\documentclass[a4paper,fleqn,usenatbib]{mn2e}

\usepackage[T1]{fontenc}
\usepackage{ae,aecompl}
\usepackage{graphicx}	% Including figure files
\usepackage{amsmath}	% Advanced maths commands
\usepackage{amssymb}	% Extra maths symbols
\usepackage[hyphens]{url}
\usepackage{pdflscape}
\usepackage{graphicx}
\usepackage{epstopdf}
\title[Unveiling  Vela-SALT ]
  { Unveiling  Vela -  Variability of Interstellar  Lines in the Direction of the Vela Supernova Remnant III. Na D and Ca II K\thanks{Based on
observations obtained with Southern African Large Telescope (SALT) and The Vainu Bappu Telescope (VBT).}}
\author[N.K.Rao et al.  ]
  {N.Kameswara Rao$^1$$^,$$^2$,David L. Lambert$^2$, Arumalla B.S Reddy$^1$, Ranjan Gupta$^3$,\\
\newauthor
        S. Muneer$^1$, Baba Varghese$^1$ \& Harinder P. Singh$^4$ \\
        \thanks{E-mail: nkrao@iiap.res.in (NKR);
  dll@astro.as.utexas.edu (DLL); muneers@iiap.res.in (SM); rag@iucaa.in (RG); singh@associates.iucaa.in
 (HPS)},\\
  $^1$Indian Institute of Astrophysics, Bangalore 560034, India\\
       $^2$The W.J. McDonald Observatory and Department of Astronomy, The University of Texas, Austin, TX 78712-1083, USA\\
       $^3$Inter-University Centre for Astronomy and Astrophysics, IUCAA Pune, 411007 India\\
       $^4$Department of Physics \& Astrophysics, University of Delhi, Delhi 110007 India\\ }
\RequirePackage[normalem]{ulem} %DIF PREAMBLE
\RequirePackage{color}\definecolor{RED}{rgb}{1,0,0}\definecolor{BLUE}{rgb}{0,0,1} %DIF PREAMBLE
\providecommand{\DIFdelbegin}{} %DIF PREAMBLE
\providecommand{\DIFdelend}{} %DIF PREAMBLE
\begin{document}

\pagerange{\pageref{firstpage}--\pageref{lastpage}} \pubyear{2014}

\maketitle

\label{firstpage}

\begin{abstract}
High-resolution optical spectra were obtained in 2017-2019 with The Southern African Large Telescope of fifteen stars in the direction of the Vela supernova remnant. Interstellar Ca\,{\sc ii} H and K and Na\,{\sc i} D lines are discussed in this paper. In particular, the line profiles are compared with profiles at a comparable spectral resolution  obtained in 1993-1996 by Cha \& Sembach. Ten of the lines of sight show changes to one or more of the components in  that line of sight. Changes include small changes (1-2 km s$^{-1}$) in radial velocity and/or increases/decreases  in equivalent width over the two decades between the periods of observation. Changes are more obvious in the Ca K line than in the Na D lines. These changes are attributed to gas disturbed by interactions between the supernova ejecta and the surrounding interstellar medium. A representative timescale may be 20-50 years. Small-scale variations in line profiles across the face of the remnant suggest, as previously remarked, that a linear scale for interactions is a small fraction of the 40 pc size of the present remnant.
\end{abstract}


\section{Introduction}

The Vela supernova remnant (SNR) is a young remnant close to the Earth. According to the VLBI parallax, the embedded pulsar is at a distance of 287$^{+19}_{-17}$ pc (Dodson et al. 2003). The SNR's age is estimated to be about 11000 yr (Reichley et al. 1970). The {\it ROSAT} X-ray image of the SNR subtends about 8 degrees (Aschenbach et al. 1995) which at the distance of 287 pc corresponds to a linear diameter of 40 pc. With its large angular size and youth, this SNR provides a fine opportunity to probe  aspects of the
  interaction of an energetic remnant with its surrounding interstellar medium (ISM). One method of probing the interaction is through absorption  spectroscopy of optical lines such as the Ca\,{\sc ii} H and K and Na\,{\sc i} D lines along  lines of sight towards stars in and behind the SNR. In particular, monitoring of the strengths and velocities of these absorption lines, especially the stronger H and K lines, appears likely to reveal details about the SNR's evolution and  its interaction with the local ISM.

The {\it ROSAT} images of the Vela SNR appear roughly circular in the plane of the sky and but slightly offset from the position of the Vela pulsar. These characteristics suggest that an assumption of a uniform spherically expanding remnant may be a fair approximation. In such case, the SNR may now extend about 40 pc along the line of sight and stars closer than about 250 pc will be between us and the SNR and stars more distant than about 330 pc will be behind the SNR. The Vela pulsar seems unlikely to have moved far from the location of the Vela supernova. With a typical `kick' velocity of about 200 km s$^{-1}$ and age of 11000 yr, the pulsar will have moved at most about 2 pc or about half a degree.  A complication is that the neighbourhood of the Vela SNR is also populated by other extended structures including the Gum nebular and the $\gamma^2$-Velorum binary which may affect the structure and evolution of parts of the Vela SNR (Cha \& Sembach 2000; Sushch, Hnatyk \& Neronov 2011).

Early studies of the interstellar absorption lines towards  stars in and around the Vela SNR
 indicated that high velocity components most commonly seen  in the Ca\,{\sc ii} profiles arise from shocked gas associated with the SNR (Wallerstein,
Silk \& Jenkins 1980; Jenkins, Wallerstein \& Silk 1984). Cha, Sembach \& Danks (1999) showed that only stars with distances greater than about 500 pc
 show  these high-velocity
components among the collection of 68 lines of sight observed by Cha \& Sembach (2000). These high-velocity components at both positive and negative velocities in excess of about 30 km s$^{-1}$ are apparently beyond the distance of the Vela pulsar at 287 pc.  In their classic study of the Ca K and Na D lines towards many B stars behind the Vela SNR, Cha \& Sembach (2000) found that  seven sight lines, some previously reported,  out of the 68 studied at the K line showed a variable high-velocity component over a three year period. High-velocity components were less common among the Na D lines but variations over the same three year period were present in some cases. Almost the entire collection of stars observed by Cha \& Sembach are behind and many considerably behind the Vela SNR.

The primary purpose of this paper is to present Ca K profiles obtained in the period 2017-2019 for 15 stars previously observed by Cha \& Sembach (2000) between 1993 and 1996 and to report on line profile variations over the approximately three decade interval. The recent spectra obtained with The Southern African Large Telescope (SALT) also provide high-resolution profiles of the Na D lines which we compare not only with the D lines observed by Cha \& Sembach but also with  the D line profiles observed with the Vainu Bappu Telescope (VBT) in primarily 2011-2012 and discussed in Paper I (Rao et al. 2016). This collection of D line profiles provides information about line profile evolution over the interval 1993-2011-2017. Three stars exhibiting extreme changes of low-velocity D line components between 1993 and 2011 were previously observed with the high-resolution spectrograph at SALT and show  no changes in their K line profiles over the 1993-2015 interval (Rao et al. 2017 - Paper II).

This paper is organized as follows. The next section discusses the observations obtained with the SALT and VBT spectrographs. Section  3 compares the K line profiles from the recent SALT observations with those reported by Cha \& Sembach and the D lines profiles from Cha \& Sembach, the VBT and the SALT. Section 4 discusses the K and D line profile variations and attempts to relate the variations to the SNR 's interactions with the surrounding ISM and related structures. Spatial variations of the K and D line profiles across the SNR are discussed in Section 5 which also discusses the equivalent ratio of the K line to the D$_2$ line within the sample of 15 stars observed by {\it SALT} and selected information from ultraviolet spectroscopy of interstellar lines. Finally, Section 6 provides a few concluding remarks.

\begin{table*}
\centering
\begin{minipage}{160mm}
\caption{\Large Observations \& Other Parameters  }
\begin{footnotesize}
\begin{tabular}{lcrcrrrrrcc}
\hline
 Star   &\multicolumn{2}{c}{$\alpha$,$\delta$(2000)}&& &  \multicolumn{3}{c}{}&&\multicolumn{2 }{c}{ Distance} \\
\cline{2-3} \cline{6-8} \cline{10-11} \\
  &    &      & & epoch& V$^{a}$   & (B-V) &E(B-V)& &d(Gaia)$^{b}$    \\
     & h m s  &  $o$ ' "  & &  & &     & & &pc   \\
\hline
HD 71302&08 24 57.2 &-42 46 11.4& &2019.10.13 & 5.95&-0.15 &0.05 && 549$^{c}$   \\
       &           &           &  &           &    &      &     &&              \\
HD 72014&08 28 52.1 &-42 35 14.9& &2018.1.05& 6.25&-0.07 &0.23 &&798$\pm$40     \\
       &           &           &  &         &     &      &     &&               \\
HD 72232&08 29 45.6 &-46 19 54.1& &2017.4.18 & 5.99&-0.15 &0.00 &&200$\pm$2     \\
       &           &           &  &          &     &      &     &&              \\
HD 72350&08 30 39.2 &-44 44 14.4& &2018.2.03& 6.30&-0.02 &0.16 &&568$\pm$14     \\
       &           &           &  &         &     &      &     &&               \\            HD 73326&08 36 02.2 &-46 30 05.9& &2017.6.02& 7.27&-0.03 &0.21 &&802$\pm$23       \\
       &           &           &  &         &     &      &     &&                \\
HD 74194&08 40 47.8 &-45 03 30.2& &2019.5.17& 7.57& 0.21 &0.50 &&2361$\pm$156   \\
       &           &           &  &         &     &      &     &&                \\
HD 74234&08 40 53.4 &-48 13 31.8& &2017.5.28& 6.95&-0.17 &0.07 &&719$\pm$27       \\
       &           &           &  &         &     &      &     &&                \\
HD 74273&08 41 05.3 &-48 55 21.6& &2017.4.18& 5.92&-0.20 &0.05 &&757$\pm$46    \\
       &           &           &  &         &     &      &     &&             \\
HD 74371&08 41 56.9 &-45 24 38.6& &2018.1.05& 5.24& 0.21 &0.28 &&1247$\pm$150    \\
       &           &           &  &         &     &      &     &&                \\
HD 74979&08 45 47.4 &-40 36 56.1& &2018.1.17& 7.24&-0.05 &0.24$^{d}$ &&937$\pm$34\\
       &           &           &  &         &     &      &     &&                \\
HD 75129&08 46 19.4 &-47 32 59.6& &2017.5.29& 6.87& 0.26 &0.35 &&1038$\pm$25     \\
       &           &           &  &         &     &      &     &&                \\
HD 75387&08 48 08.8 &-42 27 48.4& &2017.6.01& 6.42&-0.20 &0.04 &&466$\pm$12       \\
       &           &           &  &         &     &      &     &&                \\
HD 75608&08 49 21.3 &-43 22 14.4& &2017.5.07& 7.45&-0.09 &0.06 &&524$\pm$14      \\
       &           &           &  &         &     &      &     &&                \\
HD 75821&08 50 33.5 &-46 31 45.1& &2017.4.16& 5.11&-0.23 &0.07 &&655$\pm$79      \\
       &           &           &  &         &     &      &     &&                 \\
HD 76566&08 55 19.2 &-45 02 30.0& &2017.4.17 & 6.26&-0.16 &0.04 &&384$\pm$40    \\
       &           &           &  &          &     &      &     &&               \\
\hline
\end{tabular}
\\
$^{a}$:  V, B-V, and E(B-V) values are from
 Cha \& Sembach (2000) except when they are uncertain or differ too much from Simbad.
 In such cases, Simbad values are used.  \\
$^{b}$: The Gaia  parallaxes are taken from https://gea.esac.esa.int/archive/.\\
${c}$ : The parallax is from Hipparcos via Simbad. \\
$^{d}$: Cha \& Sembach (2000) give  E(B-V) as 0.00, which seem to be inconsistent
 with respect to the  colour and spectral type. Arellano Ferro \& Garrison (1979) estimate E(B-V) and E(U-B)
 as 0.24 and 0.19, respectively, which are more consistent with the spectral type.
 We adopt  0.24 as the E(B-V). \\
\end{footnotesize}
\label{default}
\end{minipage}
\end{table*}

\section{Observations}

Our high-resolution spectra of the stars discussed here were obtained at two observatories at
resolving powers $R = \lambda/d\lambda $  comparable to that of the baseline spectra
by Cha \& Sembach (2000) whose spectra from the Coud\'{e} Auxiliary Telescope at the European Southern Observatory at La Silla were at a value of $R = \lambda/d\lambda \simeq 75000 $. Thus, our comparisons of profiles for the same star do not consider corrections for different $R$ but do recognize that minor changes in line depth and profile may arise from slight differences in instrumental profiles. These differences may manifest themselves in small differences between Gaussian components fitted to apparently very similar line profiles.

New high-resolution spectra of the Ca\,{\sc ii} H \& K lines along with
  Na\,{\sc i} D lines have been acquired with the fibre-fed high-resolution spectrograph
  (HRS) on the Southern African Large Telescope   (Bramall
et al. 2010). The HRS (Crause et al. 2014) has two arms: a  blue and a red arm.   The blue arm covers the interval 3674-5490 \AA\ at a resolving power
  $R$ = 66700. This blue arm provides the H and K line profiles. The red arm which covers the interval 5490-8810 \AA\ at $R$ = 73700 provides the Na D line profiles.
   We tried to obtain a S/N ratio of several hundred in the continuum. This S/N ratios appear to be better than  values attained by Cha \& Sembach. Because of high S/N of the spectra and the use
 of Gaussian fits for estimating the equivalent widths and radial velocities the estimated
  errors in these values are often $\le$ 0.5 mA and $\le$ 1km s$^{-1}$.
Our spectra (Table 1) were obtained during the period April 2017 and October 2019. Details about the observations of the observed fifteen stars are provided in Table 1.

 IRAF routines have been used for the spectral reductions: namely, flat field
  corrections, bias subtractions,
  wavelength calibration and corrections for telluric line blending.
   We adopted the same local standard of rest (LSR) as Cha \& Sembach (2000)
  for converting  heliocentric velocities to LSR velocities.
    When necessary, the decomposition of an observed line profiles to its components was
  accomplished by fitting  Gaussian profiles. Central radial velocity
  $V_{\rm LSR} $, the full width at half maximum and the equivalent width of the
  components have thus been obtained from these fits.
  Parameters  listed in Cha \& Sembach (2000) were  used as starting values
  for the Gaussian fits and
  further changes were made so as to make  both $D_{2}$ and $D_{1}$  and independently both the H and K profiles
  yield  the same number of components, same $V_{\rm LSR} $
  and also similar full width at half-maximum. The combined Gaussian fits are
  made  to match the observed profiles such that
  no residuals are left over the noise in the surrounding continuum.
  These fits are of doubtful validity when the lines are saturated, as is the
  case for some Na D profiles.

Observations of Na D lines from 2011-2012 were obtained  with the 45 meter
   fiber-fed
 cross-dispersed echelle spectrometer at the 2.3 meter Vainu Bappu
 Telescope at the Vainu Bappu Observatory
(Rao  et al. 2005).
The  $R = \lambda/d\lambda $
 with a 60 $\micron$ slit, was 72000. The spectrum covers the wavelength range of
4000 to 10000 \AA\ with gaps.  Beyond about
5600 \AA\  the echelle orders are incompletely recorded on the E2V
2048$\times$4096 CCD chip. Note that the  Ca\,{\sc ii} H \& K lines were not provided.
 The wavelength calibration was done using
 Th-Ar  hollow cathode lamp exposures  were
  obtained soon after the exposures on the star.

The SALT Ca\,{\sc ii} H and K profiles are compared with K line profiles illustrated by Cha \& Sembach  from 1993-1996. The SALT Na D profiles are compared with profiles obtained not only by Cha \& Sembach in many cases but also with the profiles obtained at the VBT in from 2011-2012 (see Paper I). Comparison of line profiles from the different telescopes and at different times are provided in the following section.

\section{Sight lines towards the individual stars}

Ca\,{\sc ii} HK and Na\,{\sc i} D lines through the Vela SNR to stars within and behind the remnant have two striking characteristics not at all widely seen along sight lines through the general ISM. First, the SNR sight lines provide components at high velocities rarely seen outside of SNRs. Cha \& Sembach \& Danks (1999) and Cha \& Sembach (2000) adopt limits of $> +30$ km s$^{-1}$ and $< -30$ km s$^{-1}$ as  identifying gas associated with the SNR. These high velocity components are almost certainly a direct result of the interaction of the SNR with portions of the nearby ISM. Such interaction are not, however, limited to appearances at high velocities. Although the ISM in front of and behind the SNR and unaffected by the SNR is expected to provide low velocity components in the Ca\,{\sc ii} H and K and the Na\,{\sc i} D lines, not all low velocity components in these lines are may be unaffected by the SNR. Indeed, ultraviolet spectroscopy shows the presence along some lines of sight of ions not found in the quiescent ISM (e.g., Jenkins, Wallerstein \& Silk 1984). Without ultraviolet spectra, the only way to identify low velocity components as affected by the SNR may be to find temporal variations in the H and K and possibly the D lines.
Second, variations of equivalent width and velocity are rare along sight lines through the general ISM. Cha \& Sembach note one unpublished (still unpublished!) example of a variation through the general ISM (HD 28497). A careful but likely incomplete search of the literature suggests just a few other examples: $\delta$ Ori A (Price, Crawford \& Barlow 2000), HD 34078 (Rollinde et al. 2003; Boiss\'{e} et al. 2009) and $\kappa$ Vel (Smith et al. 2013). In contrast, variations on sight lines through the Vela SNR are common (Cha \& Sembach 2000; this paper).

In the following account we discuss  possible variation of the Ca H and K and Na D lines for the stars in Table 1 from the profiles published by Cha \& Sembach to those from observations reported in Table 1 and also in Paper I. (Available
spectra from the VBT and SALT also show other interstellar lines (e.g.,  Ca\,{\sc i}, CH,
  CH$^{+}$ and diffuse interstellar bands (DIBs) which will be the subject of a separate publication.)
 Although the Ca\,{\sc ii} H line occurs
  on the wing of H$\epsilon$, we extracted it whenever possible to complement the K line; if the ratio of ratio of the K to the H equivalent width ($W_{\rm \lambda}$)  is 2:1 the component is optically thin.
 In general, the spectrum around the Na D has been corrected for telluric absorption by H$_2$O lines. With the exception of HD 72232, the sight lines to the stars in Table 1 are expected to pass entirely through the portion of the SNR between the Earth and the star.

% ----------------------------------------------------------------------------------------------------------------

\subsection{HD 71302}

The line of sight to HD 71302 runs close to the north-west edge of the ROSAT contour -- see Cha \& Sembach (2000) who provide in their Figure 1 a map showing the outer {\it ROSAT} X-ray limit of the Vela SNR and the locations on the sky of their (and our) observed stars together with a simple characterization of the velocity range of their 1993-1996 K line profiles. HD 72014 is a close neighbour. The Ca K profile is almost unchanged between 1994 and 2019 with velocity components ranging between $-20$ km s$^{-1}$ and $+44$ km s$^{-1}$ (Figure 1 and Table 2). There, however, is a weak absorption component at $-201$ km s$^{-1}$ in the SALT spectrum but since this component does not appear in the Ca H profile, it is unlikely to be due to an interstellar Ca$^+$ component. The Na D lines in 1994 and recently are similar to the Ca K profiles with, perhaps, small changes in the D line equivalent widths suggesting a partial origin in the Vela SNR.

HD 71302 is an Algol-type binary (Lef\`{e}vre et al. 2009). It might also be a visual binary with a separation of $0.^"50$ (Tokovinin  et al. 2010). Our SALT spectrum shows double lines for H\,{\sc i}, He\,{\sc i} and Mg\,{\sc ii}.

 % 1
%\DIFdelbegin

%\DIFdelend
 \begin{figure*}
%\begin{minipage}{120mm}
\vspace{0.0cm}
\includegraphics[width=8cm,height=6.5cm]{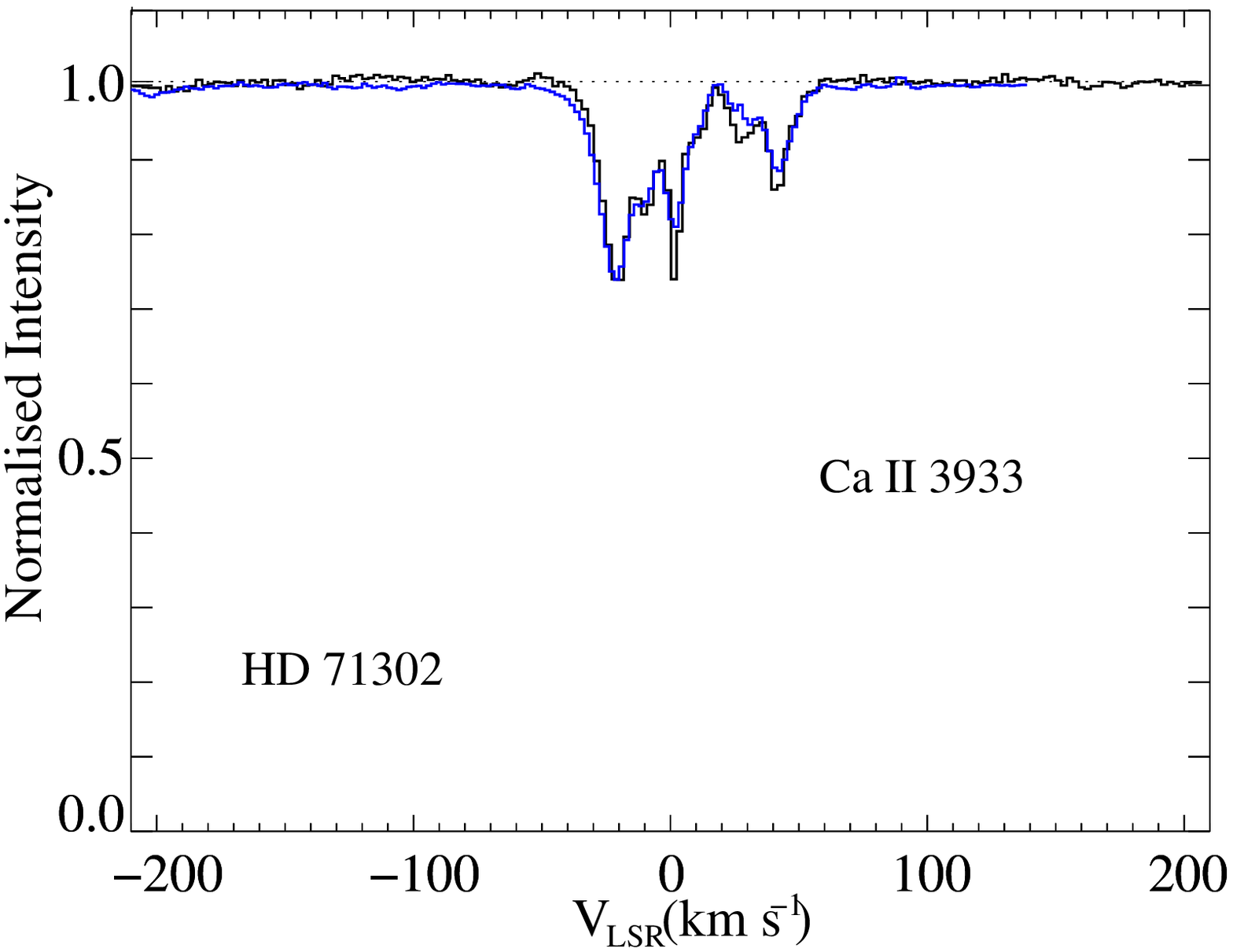}
\quad
\vspace{0.0cm}
\includegraphics[width=8cm,height=6.5cm]{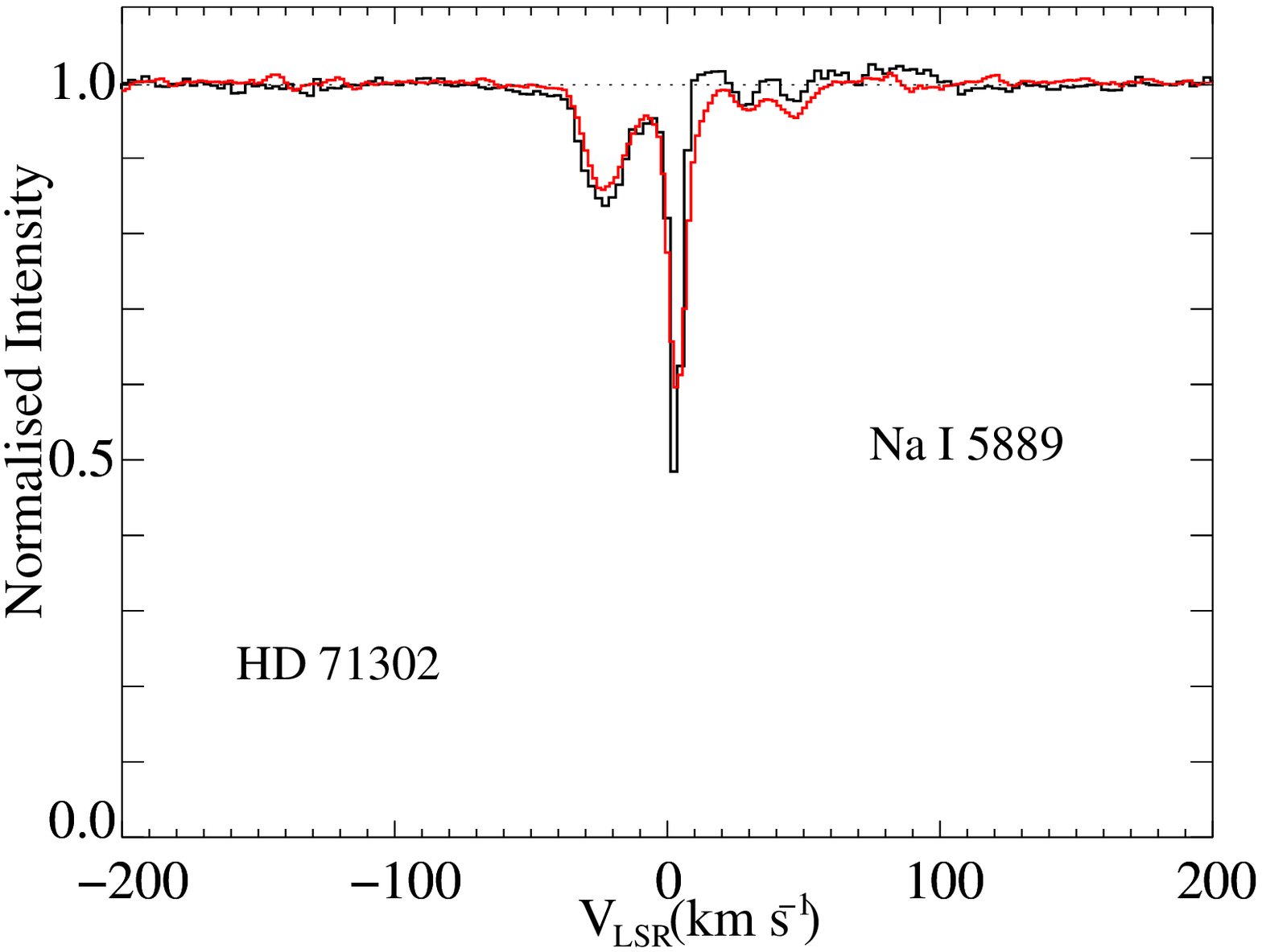}
\caption{(Left): Ca\,{\sc ii} K  profile of HD 71302 obtained with SALT on
 2019 October 13 (blue line) and the  profile (black line) obtained by Cha \& Sembach (2000) in
 1994.
 (Right): Profile of  Na\,{\sc i} D$_2$ in HD 71302 obtained by Cha \& Sembach (2000)
 (black line) shown superposed on the profile of  Na\,{\sc i} D$_2$
 observed on 2019 October 13 with SALT (red line). }
%\end{minipage}
\end{figure*}

%2
\begin{table*}
\centering
\begin{minipage}{170mm}
\caption{\Large ISM Absorption Lines of Ca\,{\sc ii} K \&  Na\,{\sc i} towards HD 71302. }
\begin{footnotesize}
\begin{tabular}{lcrrrcrrccrrrccr}
\hline
\multicolumn{1}{c}{}&\multicolumn{2}{c}{Ca\,{\sc ii} (C\&S) } &\multicolumn{1}{c}{}&\multicolumn{1}{c}{}&
\multicolumn{3}{c}{Ca\,{\sc ii} SALT} &\multicolumn{1}{c}{} & \multicolumn{3}{c}{Na\,{\sc i} (C\&S)}&\multicolumn{1}{c}{}  &\multicolumn{2}{c} {Na\,{\sc i} (SALT)}  \\
\cline{1-3} \cline{5-7} \cline{9-11} \cline{13-15}  \\
      & K &    &  & &K & H &   &    &$ D_{\rm 2}$& $D_{\rm 1}$& & &$ D_{\rm 2}$& $D_{\rm 1}$ \\
\cline{1-3} \cline{5-7} \cline{9-11} \cline{13-15}  \\
  $V_{\rm LSR}$ &$W_{\rm \lambda}$&$W_{\rm \lambda}$& &$V_{\rm LSR}$&$W_{\rm \lambda}$&$W_{\rm \lambda}$&  &$V_{\rm LSR}$&$W_{\rm \lambda}$&
  $W_{\rm \lambda}$&  &$V_{\rm LSR}$ &$W_{\rm \lambda}$&$W_{\rm \lambda}$&   \\
  km s$^{-1}$ &(mA) &(mA) &  &km s$^{-1}$ &(mA)& (mA)& &km s$^{-1}$&(mA) &(mA)& &km s$^{-1}$&(mA)& (mA)  \\
\hline
         &    &     &  &$-201$&2.5 & 1 &    &     &  &       &  &       &   &  \\
 $-20$   & 49 &     &  &$-22$&43   &22 &    &$-23$&61&24     &  &$-25.7$&29 &15    \\
         &    &     &  &     &     &   &    &     &  &       &  &$-16.7$&19 &10    \\
 $-7$    & 19 &     &  &$-10$&21   &9.5&    &     &  &       &  &$-7   $&5.5&3  \\
    3    & 19 &     &  &   2 &23   &12 &    & 3   &58&46     &  &3.6    &67 &49  \\
   11    & 10 &     &  &  11 & 6   &3  &    &     &  &       &  &11     & 7 & 2   \\
   29    & 11 &     &  &  25 &2.5  &1.3&    &     &  &       &  &29.7   & 8 & 5: \\
         &    &     &  &  31 &4.5  &2  &    &     &  &       &  &       &   &  \\
   44    & 20 &     &  &  42 &18   &9  &    &     &  &       &  &46.7   &12 &5  \\
 %        &    &     &  &     &     &   &    &     &  &       &  &       &   &  \\
  %S/N    &    &     &  &     &406  &276&    &     &  &       &  &       &300&300  \\
\hline
\end{tabular}
\\
\end{footnotesize}
\label{default}
\end{minipage}
\end{table*}

% ------------------------------------------------------------------------------------------------

\subsection{HD 72014}

HD 72014 with HD 71302 lies near the inner edge of the
  north-west portion of the SNR. HD 72014 was identified by  Cha \& Sembach as having some components at positive high velocity ($> 30$ km s$^{-1}$) gas in 1993-1996. Figure 2 shows the K line profiles (left-hand panel) from 1993 or 1996 and the SALT  profile from 2018 and the D$_2$ profiles from 1993 or 1996 and 2018. (The VBT D$_2$ profile in Paper I  is within its S/N ratio similar to the 1994 and 2018 profiles.)  The interesting feature of the SALT K profile are two clouds which were not present in 1993 and 1996. One cloud appeared at $+59$ km s$^{-1}$ and the weaker one is at +130 km s$^{-1}$. The latter  cloud appears to be present in Cha \& Sembach's illustration at $+137$ km s$^{-1}$ with an equivalent width of about 3 m\AA\ but is not listed in their Gaussian decomposition. The lower velocity parts of the K profile are unchanged between 1993-1996 and 2018. Small differences in the Na D profiles may exist also. Gaussian decomposition of the profiles is given in Table 3.

%2

%\DIFdelbegin

%\DIFdelend 
\begin{figure*}
%\begin{minipage}{120mm}
\vspace{0.0cm}
\includegraphics[trim=0.0cm 0.0cm 0.1cm 0.0cm, clip=true,width=8cm,height=6.5cm]{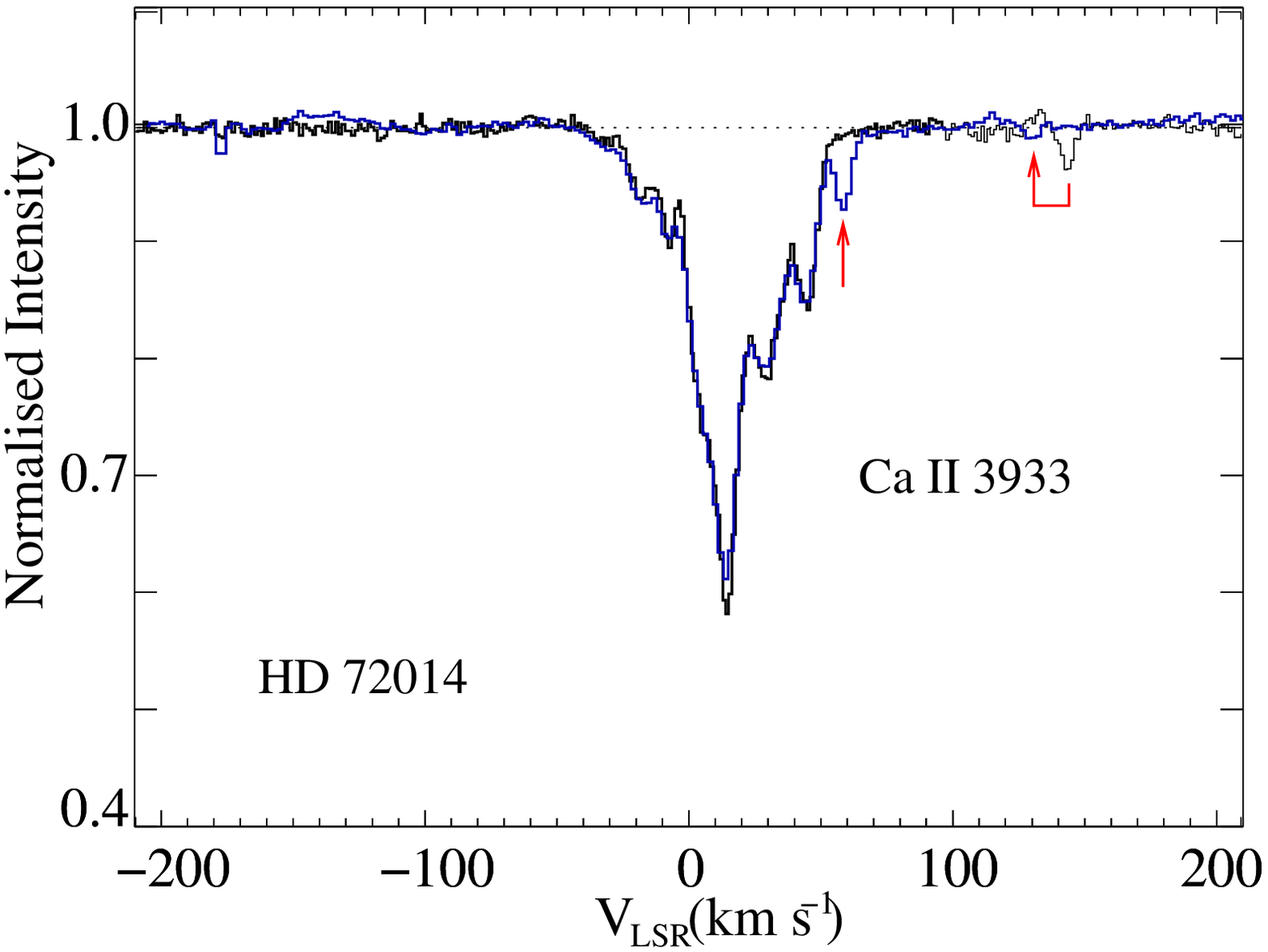}
\quad
\vspace{0.0cm}
\includegraphics[trim=0.0cm 0.0cm 0.1cm 0.0cm, clip=true,width=8cm,height=6.5cm]{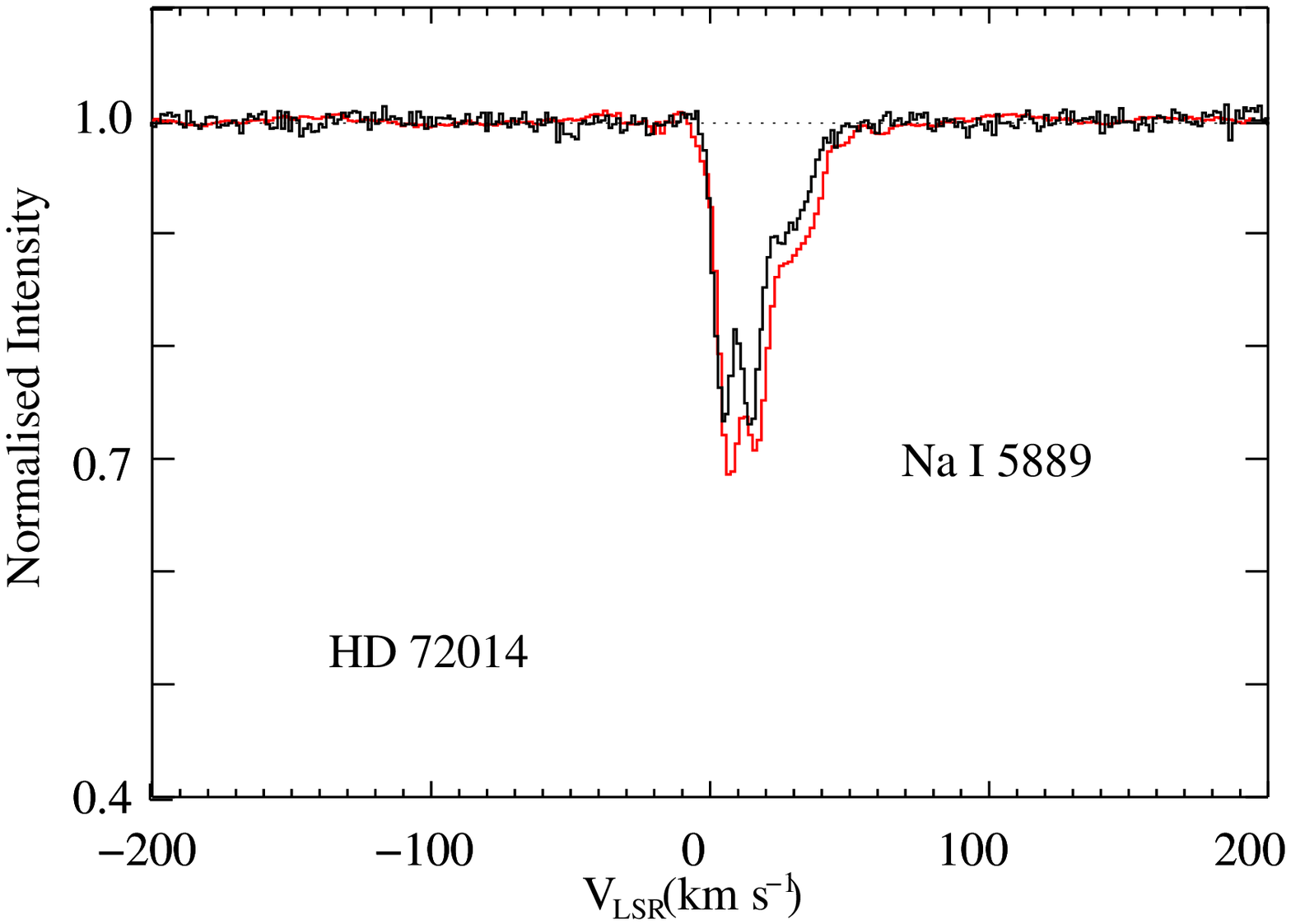}
\caption{(Left): Ca\,{\sc ii} K  profiles of HD 72014 obtained with the SALT on
 2018 January 5 (blue line) over plotted on the profile obtained by Cha \& Sembach (2000) in
 1993 or 1996.
 (Right): Profile of  Na\,{\sc i} D$_2$ in HD 72014 obtained by Cha \& Sembach (2000)
 (black line) shown superposed on the profile of the Na\,{\sc i} D$_2$
 observed in 2018 January 5 with SALT (red line). }
%\end{minipage}
\end{figure*}

%3

\begin{table*}
\centering
\begin{minipage}{170mm}
\caption{\Large ISM absorption lines of Ca\,{\sc ii} K and  Na\,{\sc i} towards HD 72014  }
\begin{footnotesize}
\begin{tabular}{lcrrrcrrccrrrccr}
\hline
\multicolumn{1}{c}{}&\multicolumn{2}{c}{C\&S Ca\,{\sc ii} K }&\multicolumn{1}{c}{}&\multicolumn{1}{c}{}&
\multicolumn{3}{c}{Ca\,{\sc ii} (SALT)} &\multicolumn{1}{c}{} & \multicolumn{3}{c}{Na\,{\sc i} (C\&S)}&\multicolumn{1}{c}{}&\multicolumn{2}{c}{Na\,{\sc i} (SALT)}&  \\
\cline{1-4} \cline{6-8} \cline{10-12} \cline{14-16}  \\
 1993  &     & 1996 &  &  & & K & H & &  & $ D_{\rm 2}$& $D_{\rm 1}$&   &&$ D_{\rm 2}$& $D_{\rm 1}$ \\
\cline{1-4} \cline{6-8} \cline{10-12} \cline{14-16}   \\
 $V_{\rm LSR}$ &$W_{\rm \lambda}$& $V_{\rm LSR}$ &$W_{\rm \lambda}$&   & $V_{\rm LSR}$ &$W_{\rm \lambda}$&$W_{\rm \lambda}$&  &$V_{\rm LSR}$  &$W_{\rm \lambda}$&$W_{\rm \lambda}$&  & $V_{\rm LSR}$&$W_{\rm \lambda}$&$W_{\rm \lambda}$   \\
km s$^{-1}$&(mA) &km s$^{-1}$&(mA)& &  km s$^{-1}$ &(mA)& (mA)& &km s$^{-1}$ &(mA) &(mA)& &km s$^{-1}$&(mA) &(mA)    \\
\hline
           &     &           &    &  &$-32$        & 3&$\le$0.8&   &            &     &    &  &   &  \\
  $-19$& 8 &$-18$ & 11 &        &$-18 $&9 &      3 &   &    &     &  &  &   &   \\
  $-8 $& 7 &$-7$ & 6  &   &$-8$ &9 &  4 &   &    &     &  &  &   &  \\
    6  & 51& 5   & 41 &   & 3  & 30  &15  &   & 5  & 55  & 21 & &5&57&25  \\
   15  & 38&15   & 45 &   &14  & 59& 34  &    &15  & 43  & 19 & &15&54&28   \\
   29  & 50&29   & 52 &   &29  & 37  & 20  &    &28  & 54  & 3  & &25&21&20 \\
       &   &     &    &   &      &     &     &    &    &     &    & &34&17  &7\\
   45  & 17&46   & 16 &   &44  & 24  & 12  &    &    &     &    & &44&4 &2 \\
       &   &     &    &   & 59 & 7 & 3&    &    &     &    & &59&3 &1 \\
 &  &     &    &   &130 & 2 & $\le$0.8 &    &     &    & &    &    &       \\
\hline
\end{tabular}
\\
\end{footnotesize}
\label{default}
\end{minipage}
\end{table*}

% ---------------------------------------------------------------------------------------------------------

\subsection{HD 72232}

HD 72232 is in the SW corner of the SNR but at a distance of 200 pc is surely in front of the Vela pulsar. Cha \& Sembach found all of the interstellar components in both the D and K lines to fall within |30| km s$^{-1}$ of the LSR velocity. Thus, the profiles would appear, as likely expected,  to be unaffected by the SNR. The K line profile is unchanged between 1994 and 2017 -- see Figure 3 where the interstellar line is superimposed on a broad stellar line. An identical conclusion applies to the Na D profiles where a stellar contribution is absent -- see also Paper I where the VBT profile from  2011 was seen to be unchanged from  1994.  Gaussian decompositions are summarized in Table 4. Along this line of sight to the near-side of the Vela SNR, there are no high-velocity components in the K line profile and the simple complex of K and D line absorbers arises from the interstellar gas between Earth and the remnant.

% 3

\DIFdelbegin

\DIFdelend \begin{figure*}
%\begin{minipage}{120mm}
\vspace{0.0cm}
\includegraphics[trim=0.0cm 0.0cm 0.1cm 0.0cm, clip=true,width=8cm,height=6.5cm]{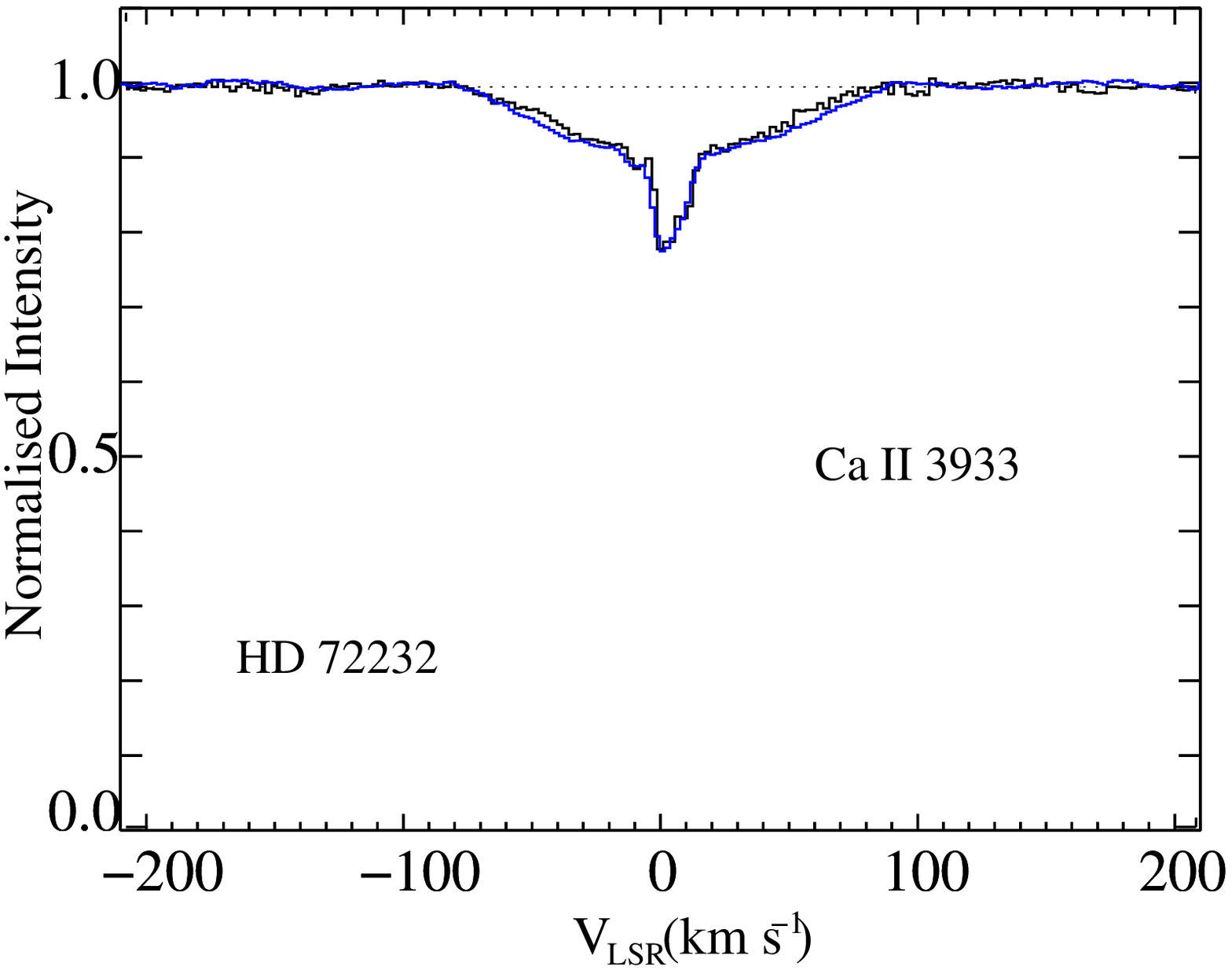}
\vspace{0.0cm}
\includegraphics[trim=0.0cm -0.4cm -0.1cm 0.0cm, clip=true,width=8cm,height=6.5cm]{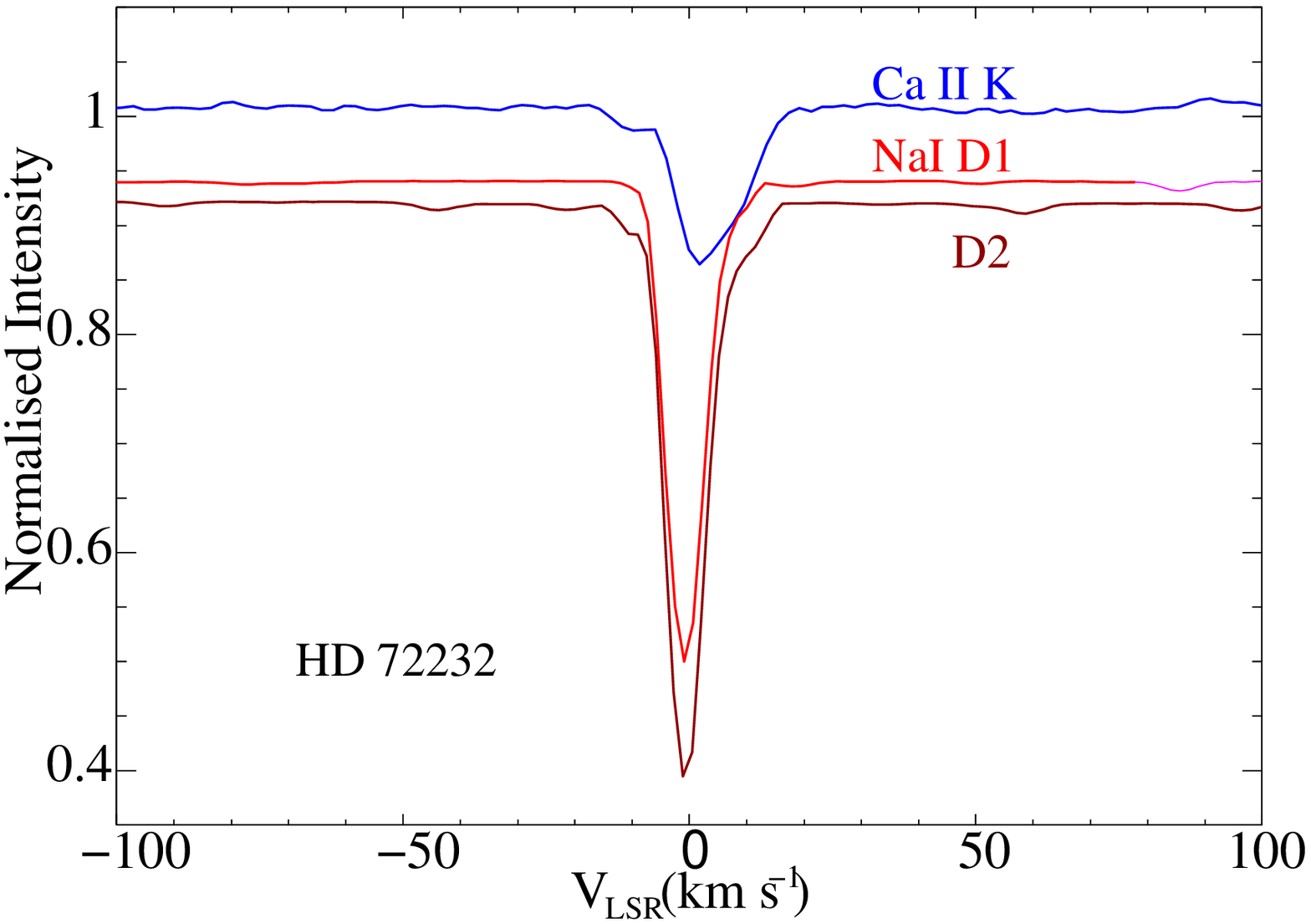}
\caption{(Left): The Ca\,{\sc ii}  K  profile of HD 72232 obtained with SALT on
 2017 May 5 (blue line) over plotted on  the profile obtained by Cha \& Sembach (2000) in
 1994.  The interstellar lines are superimposed on a broad stellar line.
 (Right): Profiles of ISM lines of   Na\,{\sc i} D$_1$, D$_2$ and  Ca\,{\sc ii} K from SALT.
 The Ca\,{\sc ii} K ISM profile is extracted by assuming the shape of stellar line  of
  HD 72232. }
%\end{minipage}
\end{figure*}

%4
\begin{table*}
\centering
\begin{minipage}{170mm}
\caption{\Large ISM  Absorption Lines of Ca\,{\sc ii} K and  Na\,{\sc i} towards HD 72232 }
\begin{footnotesize}
\begin{tabular}{lcrrrcrrccrrrccr}
\hline
\multicolumn{1}{c}{}&\multicolumn{2}{c}{Ca\,{\sc ii} K C\&S} &\multicolumn{1}{c}{}&\multicolumn{1}{c}{}&
\multicolumn{3}{c}{Ca\,{\sc ii} (SALT)} &\multicolumn{1}{c}{} & \multicolumn{3}{c}{Na\,{\sc i} (C\&S)}&\multicolumn{1}{c}{}  &\multicolumn{2}{c} {Na\,{\sc i} (SALT)}  \\
\cline{1-3} \cline{5-8} \cline{10-12} \cline{14-15}  \\
      & K&    &  & &K & H &   & &   $ D_{\rm 2}$& $D_{\rm 1}$& &  &$ D_{\rm 2}$& $D_{\rm 1}$ \\
\cline{1-3} \cline{5-7} \cline{9-11} \cline{13-15}  \\
  $V_{\rm LSR}$ &$W_{\rm \lambda}$&$W_{\rm \lambda}$& &$V_{\rm LSR}$&$W_{\rm \lambda}$&$W_{\rm \lambda}$&  &$V_{\rm LSR}$&$W_{\rm \lambda}$&
$W_{\rm \lambda}$&  &$V_{\rm LSR}$ &$W_{\rm \lambda}$&$W_{\rm \lambda}$&   \\
  km s$^{-1}$ &(mA) &(mA) &  &km s$^{-1}$ &(mA)& (mA)& &km s$^{-1}$&(mA) &(mA)& &km s$^{-1}$&(mA)& (mA)  \\
\hline
         &    &     &  &     &     &   &    &$-29$&11&$\le$6 &  &      &    &    \\
         &    &     &  & $-10$&2   &   &    &     &  &       &  &      &    &    \\
    2    &15  &     &  &  1  & 16&   &    & 1  & 84&    58 &  &$-1$&86&69 \\
    3    &117 &     &  &     &     &   &    &   &    &    &   &  &   &  \\
   10    &6   &     &  &  8 & 10 &   &    &     &  &       &  &8     &10&4  \\
\hline
\end{tabular}
\\
\end{footnotesize}
\label{default}
\end{minipage}
\end{table*}

% -------------------------------------------------------------------------------------------------------------

\subsection{HD 72350}

           The sight line to HD 72350 passes close to the Vela Pulsar and extends beyond it to a  distance  of 568 pc. Thus, it is not surprising that Cha \& Sembach noted that the D and K profiles showed components at velocities greater than $+30$ km s$^{-1}$. Figure 4 summarizes the K and D line profiles and Table 5 provides the Gaussian decomposition of the profiles. The Ca\,{\sc ii} K line contains about nine components almost all of them
  show positive radial velocities.  The Ca\,{\sc ii} and Na\,{\sc i} components have retained the same radial velocity between Cha \& Sembach's observations and our SALT observations. However, the
  Ca\,{\sc ii} K component at +39 km s$^{-1}$ was much stronger by the time of the SALT observations
  with an equivalent width  of 74 m\AA\  but it had a strength of just 48 m\AA\ in 1994. The same
  component also shows an increase in the  Na\,{\sc i} D lines: SALT observations show an equivalent width of 32 m\AA\
  for D$_2$ and 14 m\AA\ for D$_1$  but Cha \& Sembach (2000) do not list this component to their 1994 observation.
 The VBT D{$_2$} line profile from 2011 shows the strengthening of the +39 km s$^{-1}$ component (Paper I).
  Jenkins et al. (1984) noted the absence of this component in Na\,{\sc i} D lines and its
  presence in Ca\,{\sc ii} K line in their CTIO photographic spectra. Obviously this component got much stronger by 2018 February.

% % 4

\begin{figure*}
\vspace{0.0cm}
\includegraphics[trim=0.0cm 0.0cm 0.1cm 0.0cm, clip=true,width=8cm,height=6.5cm]{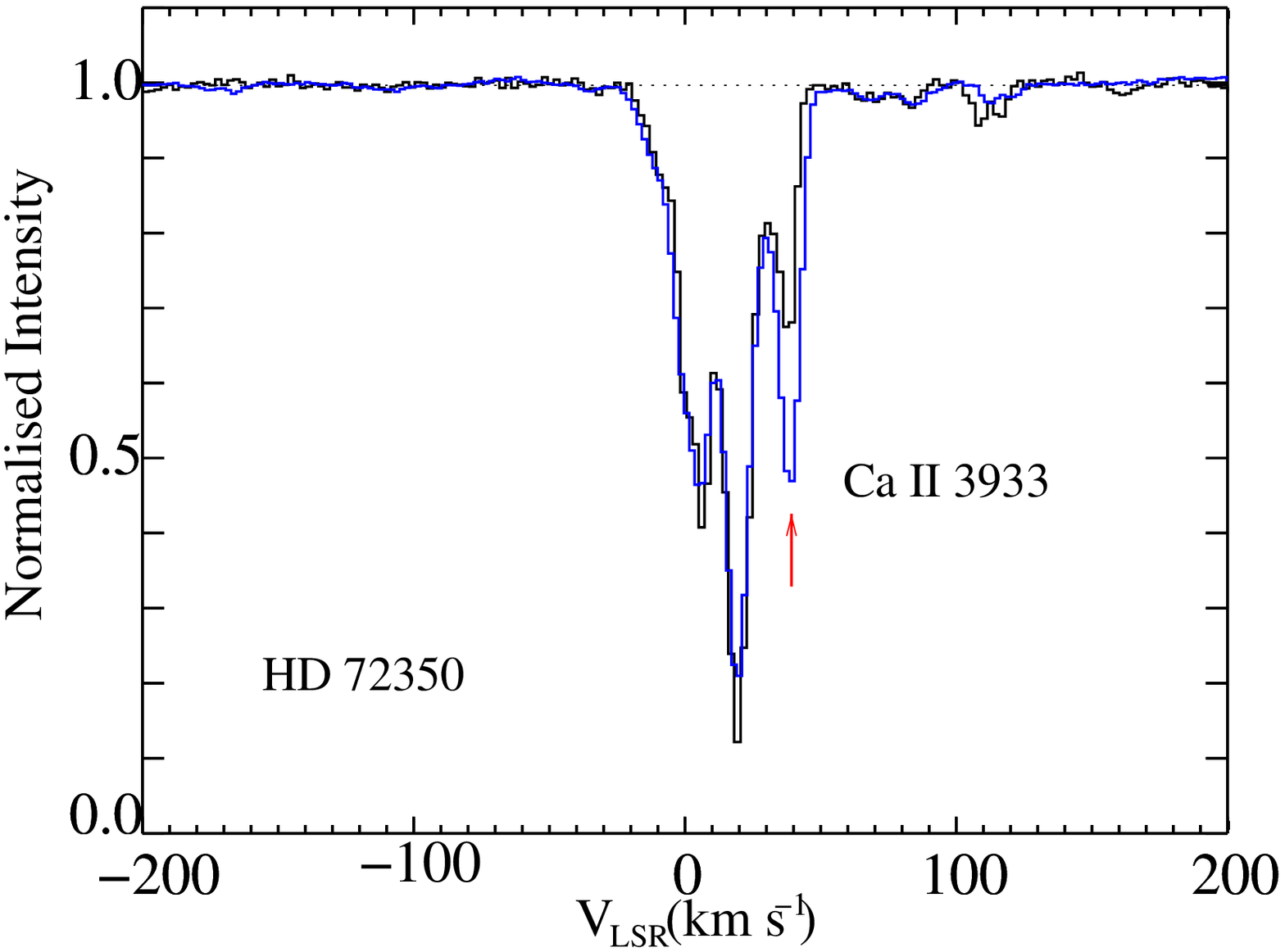}
\vspace{0.0cm}
\includegraphics[trim=0.0cm 0.0cm 0.1cm 0.0cm, clip=true,width=8cm,height=6.5cm]{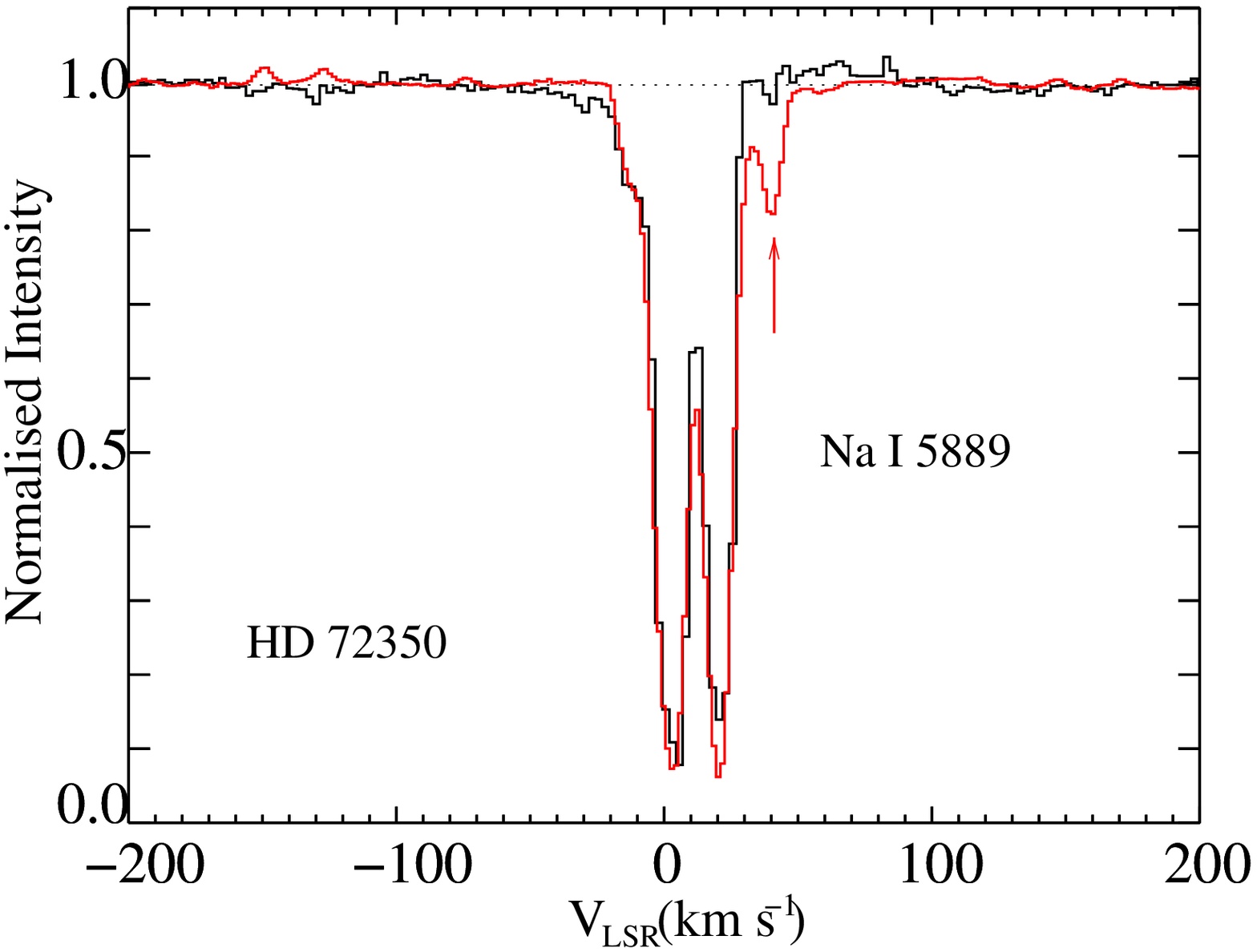}
\caption{(Left): The profiles of Ca\,{\sc ii} in  HD 72350 obtained during 1996 (black line) by Cha \& Sembach are superposed with  the profile obtained in 2018 February 3 with the SALT (blue line). (Right): Profile of  Na\,{\sc i} D$_2$ in HD 72350 obtained by Cha \& Sembach (black line) shown superposed on the profile of  Na\,{\sc i} D$_2$ observed in 2018 February 3 with the SALT (red line). Note the strengthening of the $+39$ km s$^{-1}$ component in both Ca\,{\sc ii} and Na\,{\sc i}.}
\end{figure*}

%5
\begin{table*}
\centering
\begin{minipage}{170mm}
\caption{\Large ISM Absorption Lines of Ca\,{\sc ii} K and  Na\,{\sc i} towards HD 72350.}
\begin{footnotesize}
\begin{tabular}{lcrrrcrrccrrrccr}
\hline
\multicolumn{1}{c}{}&\multicolumn{2}{c}{C\&S Ca\,{\sc ii} K }&\multicolumn{1}{c}{}&\multicolumn{1}{c}{}&
\multicolumn{3}{c}{Ca\,{\sc ii} (SALT)} &\multicolumn{1}{c}{} & \multicolumn{3}{c}{Na\,{\sc i} (C\&S)}&\multicolumn{1}{c}{}&\multicolumn{2}{c}{Na\,{\sc i} (SALT)}  &  \\
\cline{1-4} \cline{6-8} \cline{10-12} \cline{14-16}  \\
 1994  &     &    &  &  & & K & H & &  & $ D_{\rm 2}$& $D_{\rm 1}$&   &&$ D_{\rm 2}$& $D_{\rm 1}$ \\
\cline{1-4} \cline{6-8} \cline{10-12} \cline{14-16}   \\
 $V_{\rm LSR}$ &$W_{\rm \lambda}$&     &     &   & $V_{\rm LSR}$ &$W_{\rm \lambda}$&$W_{\rm \lambda}$&  &$V_{\rm LSR}$  &
$W_{\rm \lambda}$&$W_{\rm \lambda}$&  & $V_{\rm LSR}$&$W_{\rm \lambda}$&$W_{\rm \lambda}$   \\
km s$^{-1}$&(mA) &km s$^{-1}$&(mA)& &  km s$^{-1}$ &(mA)& (mA)& &km s$^{-1}$ &(mA) &(mA)& &km s$^{-1}$&(mA) &(mA)    \\
\hline
       &   &      &    &   &       &    &   &    &$-12$& 36 & 15&   &$-12.7$&20&10\\
       &   &      &    &   &$-10$  &18  &12 &    &     &    &   &   &       &  &  \\
  $-6 $& 28&      &    &   &       &    &   &    &     &    &   &   &       &  &  \\
    0  & 25&      &    &   &$-0.5$ & 30 &22 &    &     &    &   &   &       &  &  \\
       &   &      &    &   &       &    &   &    &  4  &259 &227&   & 4     &268&215\\
    7  & 67&      &    &   &  7    & 65 &38.7&   &     &    &   &   &       &   &     \\
   20  &133&      &    &   & 20.5  &125 &94 &    & 21  &206 &206&   & 21    &238&190\\
   38  & 48&      &    &   & 39    & 74.3&47& &   &    &   &   & 41    &31.5&14 \\
  114  &  4&      &    &   &113    &  3 &  2&    &     &    &   &   &       &   &    \\
  122  &  3&      &    &   &122    &  2 & 1.5&   &     &    &   &   &       &   &    \\
       &   &      &    &   &       &    &   &    &     &    &   &   &       &   &   \\
% S/N$*$& &      &    &   &       & 184&280&    &     &    &   &   &       &373&373   \\
\hline
\end{tabular} \\
\end{footnotesize}
\label{default}
\end{minipage}
\end{table*}

% ----------------------------------------------------------------------------------------------------------

\subsection{HD 73326}

HD 73326 at a distance of 802 pc lies beyond the Vela pulsar in the central region of the remnant. Cha \& Sembach noted that high-velocity components were limited to negative velocities. HD 73326 is a double-line spectroscopic binary (Garrison et al. 1977) with both
  spectra being visible on our SALT spectrum obtained on 2017 June 2. However, our VBT spectrum
  obtained on 2011 March 11 shows single lines, as do apparently the spectra available to Cha \& Sembach.  Figure 5 (left-hand panel) compares the SALT K profile  with that obtained by  Cha \& Sembach. The ISM Ca\,{\sc ii} lines are superimposed on the stellar lines during the observations of Cha \& Sembach (2000) in 1996.   Components in the SALT spectrum  and those identified by Cha \&
   Sembach match both in velocity and equivalent width except for the central $+7$ km s$^{-1}$  component for which the equivalent width of 48 m\AA\ observed by Cha \& Sembach (2000) increased to 68 m\AA\ by the time of
   SALT observations (Table 6).  The H \& K components are mostly optically thin. Cha \& Sembach did not observe the D lines. There are no real changes to the
   Na\,{\sc i} D between the VBT and the SALT observations (Figure 5 right-hand panel).

%5

%\DIFdelbegin

%\DIFdelend
\begin{figure*}
%\begin{minipage}{120mm}
\vspace{0.0cm}
\includegraphics[trim=0.0cm 0.0cm 0.1cm 0.0cm, clip=true,width=8cm,height=6.5cm]{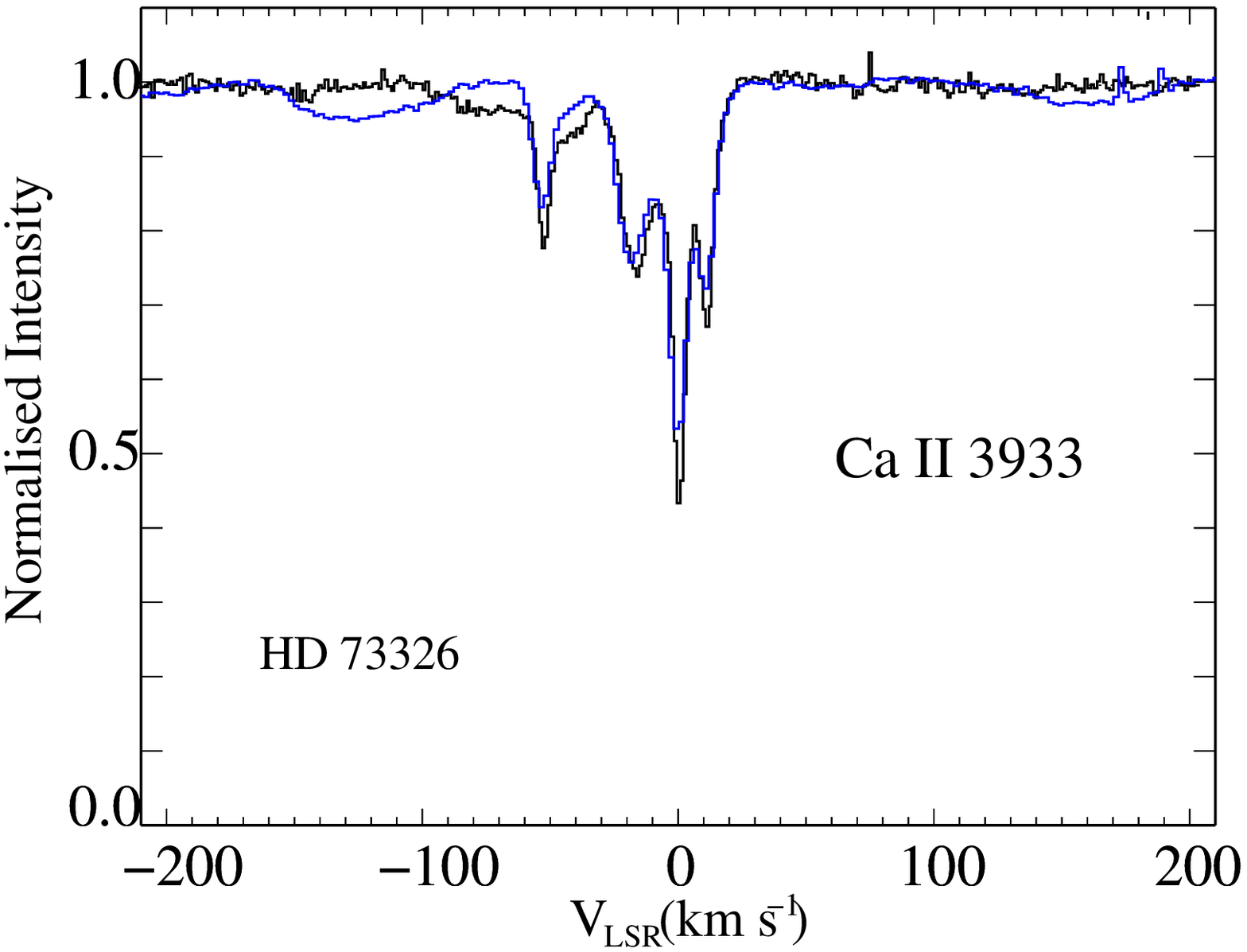}
\vspace{0.0cm}
\includegraphics[trim=0.0cm 0.0cm 0.1cm 0.0cm, clip=true,width=8.cm,height=6.5cm]{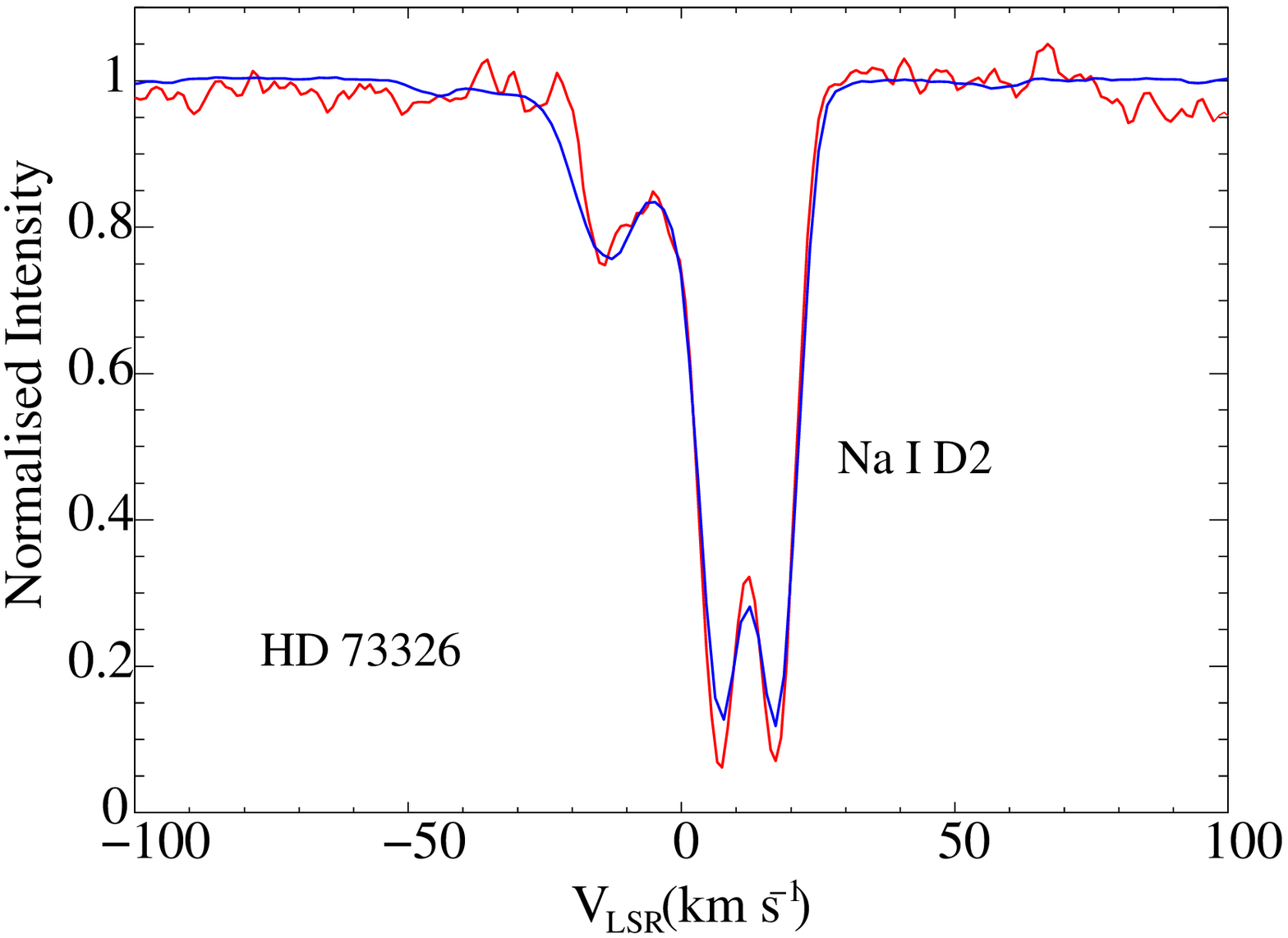}
\caption{(Left): The profile of Ca\,{\sc ii} in  HD 73326 obtained during
 1996 (black line) by Cha \& Sembach are superposed on  the profile obtained
 in 2017 June 2 with the SALT (blue line).  Two shallow stellar lines may be seen in the SALT spectrum, one to the red and one to the blue of the complex of interstellar K lines.
 (Right): Profile of  Na\,{\sc i} D$_2$ in HD 73326 obtained on
 2011 March 23 with the VBT (red line) superposed on the profile of  Na\,{\sc i} D$_2$
 observed in 2017 June 2 with the SALT (blue line). }
%\end{minipage}
\end{figure*}

%6
\begin{table*}
\centering
\begin{minipage}{170mm}
\caption{\Large ISM Absorption Lines of Ca\,{\sc ii} K and Na\,{\sc i}  towards HD 73326. }
\begin{footnotesize}
\begin{tabular}{lcrrrcrrccrrrccr}
\hline
\multicolumn{1}{c}{}&\multicolumn{2}{c}{Ca\,{\sc ii} (C\&S) } &\multicolumn{1}{c}{}&\multicolumn{1}{c}{}&
\multicolumn{3}{c}{Ca\,{\sc ii} SALT} &\multicolumn{1}{c}{} & \multicolumn{3}{c}{Na\,{\sc i} (C\&S)}&\multicolumn{1}{c}{}  &\multicolumn{2}{c} {Na\,{\sc i} (SALT)}  \\
\cline{1-3} \cline{5-7} \cline{9-11} \cline{13-15}  \\
      & K &    &  & &K & H &   &    &$ D_{\rm 2}$& $D_{\rm 1}$& & &$ D_{\rm 2}$& $D_{\rm 1}$ \\
\cline{1-3} \cline{5-7} \cline{9-11} \cline{13-15}  \\
  $V_{\rm LSR}$ &$W_{\rm \lambda}$&$W_{\rm \lambda}$& &$V_{\rm LSR}$&$W_{\rm \lambda}$ &$W_{\rm \lambda}$&  &$V_{\rm LSR}$& $W_{\rm \lambda}$  &
  $W_{\rm \lambda}$ &  &$V_{\rm LSR}$ &$W_{\rm \lambda}$&$W_{\rm \lambda}$&   \\
  km s$^{-1}$ &(mA) &(mA) &  &km s$^{-1}$ &(mA)& (mA)& &km s$^{-1}$&(mA) &(mA)& &km s$^{-1}$&(mA)& (mA)  \\
\hline
  $-46$  &14  &     &  & $-46$&17  &8.5&    &     &  &       &  & $-46$&4   &2    \\
  $-37$  &18  &     &  & $-36$& 8  &2  &    &     &  &       &  &      &    &    \\
  $-10$  &50  &     &  &$-10.6$&50 &25 &    &     &  &       &  &$-14$ &70  &26  \\
    0    &7   &     &  &      &    &   &    &     &  &       &  &$-3$  &16.5&12.5  \\
    7    &48  &     &  &  6.8 & 65 &36 &    &     &  &       &  &  7   &162 &158  \\
   17    &36  &     &  & 18.3 & 31 &18 &    &     &  &       &  & 17.5 &160 &126  \\
         &    &     &  &      &    &   &    &     &  &       &  &      &    &    \\
 %  S/N   &    &     &  &      &283 &290&    &     &  &       &  &      &588 &588 \\
\hline
\end{tabular}
\\
\end{footnotesize}
\label{default}
\end{minipage}
\end{table*}

% -------------------------------------------------------------------------------------------------

 \subsection{HD 74194}

HD 74194 lies at about the centre of the ROSAT image but at the distance of 2360 pc far behind the SNR. Its location may account for the fact that the K line profile recorded  by Cha \& Sembach and here by the SALT are among the most complex across the sample of stars in the SNR's vicinity; the ISM ahead of and also behind the SNR may contribute absorption components to the D and K lines. In particular, the K line profile is rich in high-velocity components -- all of negative velocity and none appearing in the D line profiles. Figure 6 shows the K (left-hand panel) and D (right-hand panel) line profiles obtained by Cha \& Sembach and SALT. Table 7 gives the Gaussian decompositions. (HD 74194 is a massive X-ray binary. There  were variations in the stellar
        line profiles in   spectra  obtained on successive days.)

There are some  changes between the Ca\,{\sc ii} K line profiles obtained by
     Cha \& Sembach (2000) in 1996 and the SALT observations obtained in 2019. One of the more prominent changes occurred to   the highest radial velocity component observed by Cha \& Sembach (2000) at $-141$ km s$^{-1}$
    in 1996 which was accelerated to $-146$ km s$^{-1}$
     without much  change in equivalent width by 2019.  Cha \& Sembach's components at
     -89  and $-79$ km s$^{-1}$ have lost their identity in 2019. A
      component -61 km s$^{-1}$ weakened slightly by 2019. The complex from $-30$ to $30$ km s$^{-1}$ including the deep K line at $+8$ km s$^{-1}$ is unchanged between 1996 and 2019.   A Na D component at $-25$ km s$^{-1}$ strengthened in the SALT spectrum relative to Cha \& Sembach's profile is also strengthened in the VBT spectrum from 2008.

% 6

\DIFdelbegin

\DIFdelend \begin{figure*}
%\begin{minipage}{120mm}
\vspace{0.0cm}
\includegraphics[trim=0.0cm 0.0cm 0.1cm 0.0cm, clip=true,width=8cm,height=6.5cm]{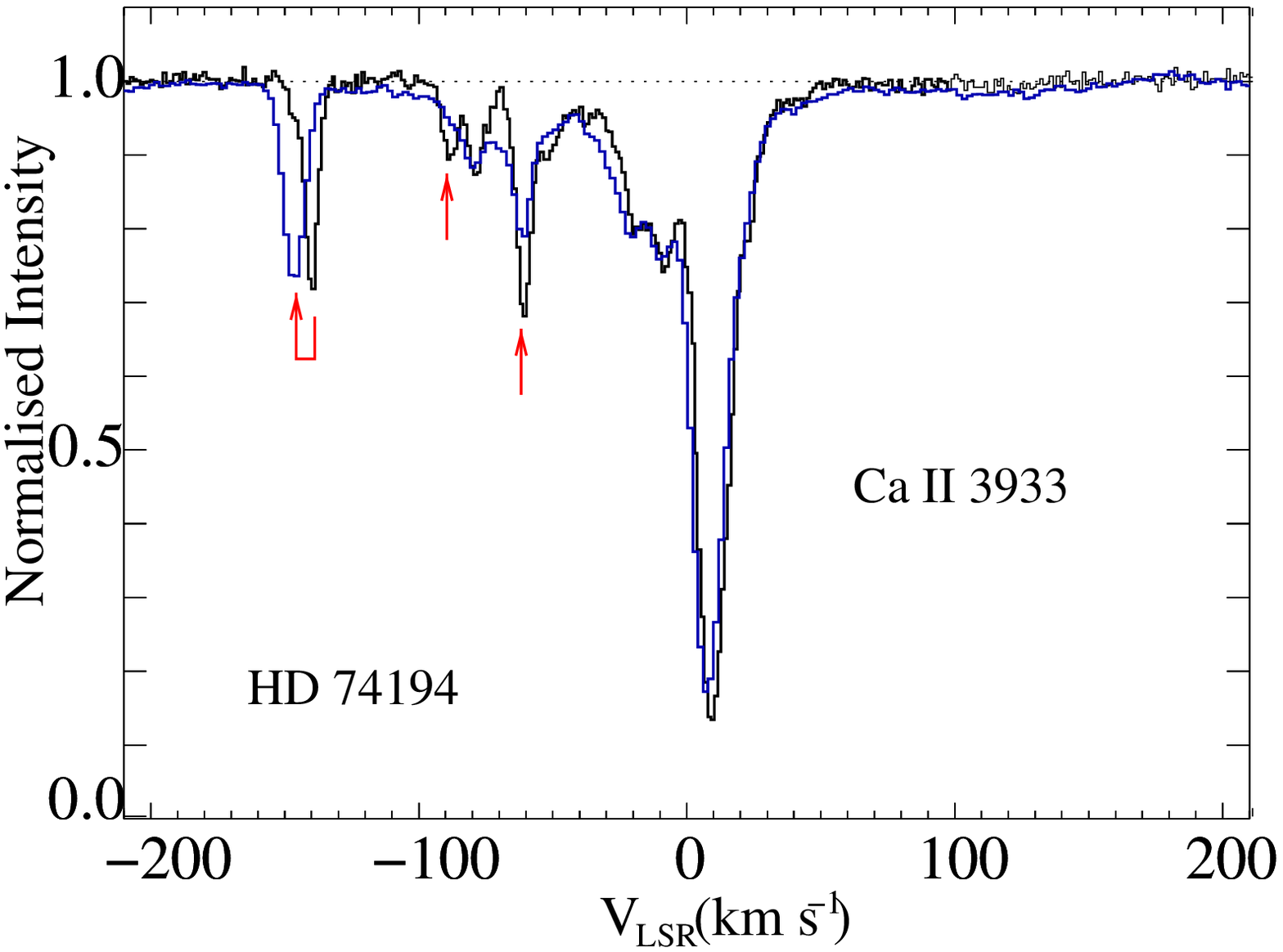}
\vspace{0.0cm}
\includegraphics[trim=0.0cm 0.0cm 0.1cm 0.0cm, clip=true,width=8cm,height=6.5cm]{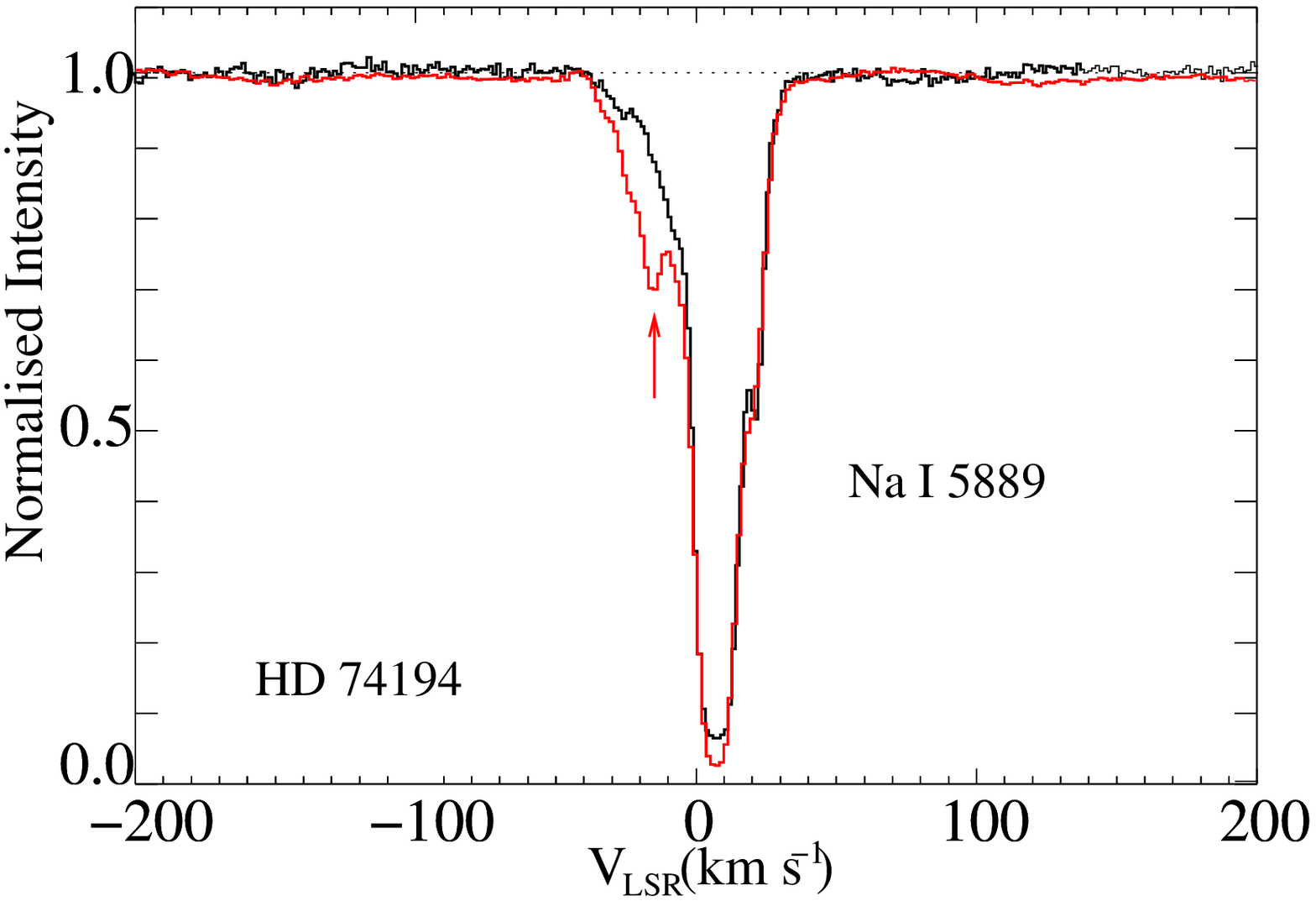}
\caption{(Left): The profile of the Ca\,{\sc ii} K line in  HD 74194 obtained during
 1993 (black line) by Cha \& Sembach is superposed on the  profile obtained
 in 2019 May 17
 with   the SALT (blue line).
 (Right): Profile of  Na\,{\sc i} D$_2$ line in HD 74194 obtained in
 1996 by Cha \& Sembach (black line) shown superposed on the profile of  Na\,{\sc i} D$_2$ line
 observed on 2019 May 17 with the SALT (red line). The red arrows indicate the changes to
 the components between Cha \& Sembach's and SALT observations.   }
%\end{minipage}
\end{figure*}

\begin{table*}
\centering
\begin{minipage}{170mm}
\caption{\Large ISM Absorption Lines of Ca\,{\sc ii} K and  Na\,{\sc i} towards HD 74194.}
\begin{footnotesize}
\begin{tabular}{lcrrrcrrccrrrccr}
\hline
\multicolumn{1}{c}{}&\multicolumn{2}{c}{C\&S Ca\,{\sc ii} K }&\multicolumn{1}{c}{}&\multicolumn{1}{c}{}&
\multicolumn{3}{c}{Ca\,{\sc ii} (SALT)} &\multicolumn{1}{c}{} & \multicolumn{3}{c}{Na\,{\sc i} (C\&S)}&\multicolumn{1}{c}{}&\multicolumn{2}{c}{Na\,{\sc i} (SALT)}  &  \\
\cline{1-4} \cline{6-8} \cline{10-12} \cline{14-16}  \\
 1993  &     & 1996 &  &  & & K & H & &  & $ D_{\rm 2}$& $D_{\rm 1}$&   &&$ D_{\rm 2}$& $D_{\rm 1}$ \\
\cline{1-4} \cline{6-8} \cline{10-12} \cline{14-16}   \\
 $V_{\rm LSR}$ &$W_{\rm \lambda}$ & $V_{\rm LSR}$ &$W_{\rm \lambda}$ &   & $V_{\rm LSR}$ &$W_{\rm \lambda}$&$W_{\rm \lambda}$&  &$V_{\rm LSR}$  &
  $W_{\rm \lambda}$& $W_{\rm \lambda}$ &  & $V_{\rm LSR}$&$W_{\rm \lambda}$&$W_{\rm \lambda}$   \\
km s$^{-1}$&(mA) &km s$^{-1}$&(mA)& &  km s$^{-1}$ &(mA)& (mA)& &km s$^{-1}$ &(mA) &(mA)& &km s$^{-1}$&(mA) &(mA)    \\
\hline
$-141$ &30 &$-140$&23 &   &$-146$&34&19 &    &    &     &    &   &    &  & \\
 $-88$ &11 & $-89$& 9 &   &      &     &     &    &    &     &    &   &    &  & \\
 $-78$ &14 &$-79$ &12 &   & $-80$& 26  & 9   &    &    &     &    &   &    &  & \\
 $-61$ &31 &$-61$ &20 &   & $-61$& 30  &10   &    &    &     &    &   &    &  &   \\
 $-49$ &10 &$-55$ &31 &   &$-49.5$&9   &1.5  &    &    &     &    &   &    &  &   \\
       &   &      &   &   &$-34$ & 13  & 5   &    &    &     &    &   &    &  &  \\
 $-17$ &50 &$-20$ &50 &   &$-21$ & 33  &17   &    &$-24$&15  &10  &   &$-25$&20&5  \\
 $-7$  &21 &$-8$  & 26&   &$-9.3$& 32  &14   &    &$-8$&58   &23  &   &$-13$&74&30 \\
   9   &133& 8    &134&   &  8   &159  &107  &    &  9 &348  &310 &   &9 &346&323  \\
       &   &     &    &   &21.5  & 24  &14   &    & 23 &55   &37  &   &22  &82 &40  \\
 %$*$   &   &     &    &   &      &     &     &    &    &     &    &   &    &   &    \\
 %      &   &     &    &   &      &     &     &    &    &     &    &   &    &   &    \\
 %S/N   &   &     &    &   &      &184  &280  &    &    &     &    &   &    &273&273   \\
\hline
\end{tabular}
\\
\end{footnotesize}
\label{default}
\end{minipage}
\end{table*}

% ---------------------------------------------------------------------------------------------------

\subsection{HD 74234}

                   HD 74234 is located near  the southern edge of the X-ray remnant at a distance of 719 pc. The K line profile consists of a complex of low velocity components and a strong complex of high velocity components (Figure 7). The Gaussian decompositions are summarized in Table 8. Unlike many high-velocity K components, the $+73$ km s$^{-1}$ has a  Na D counterpart in the SALT observation and in the VBT spectrum from 2011 (Paper I). Cha \& Sembach did not observe the Na\,{\sc i} D lines. There are major changes between the profile obtained in 1996 (Cha \& Sembach 2000) and in 2017 with the SALT. New components appeared at -40 and +47 km s$^{-1}$. The component
   at +28 km s$^{-1}$ that was present in 1996 seem to have weakened and accelerated to
  +35 km s$^{-1}$ (assuming this is the same component) and a component at +85 km s$^{-1}$
   has weakened considerably without much change in radial velocity. A more prominent change is the
   acceleration and weakening of the component where the radial velocity increased from +69 km s$^{-1}$ to +73 km s$^{-1}$.

 %7
\begin{figure*}
%\begin{minipage}{120mm}
\includegraphics[width=8cm,height=6.5cm]{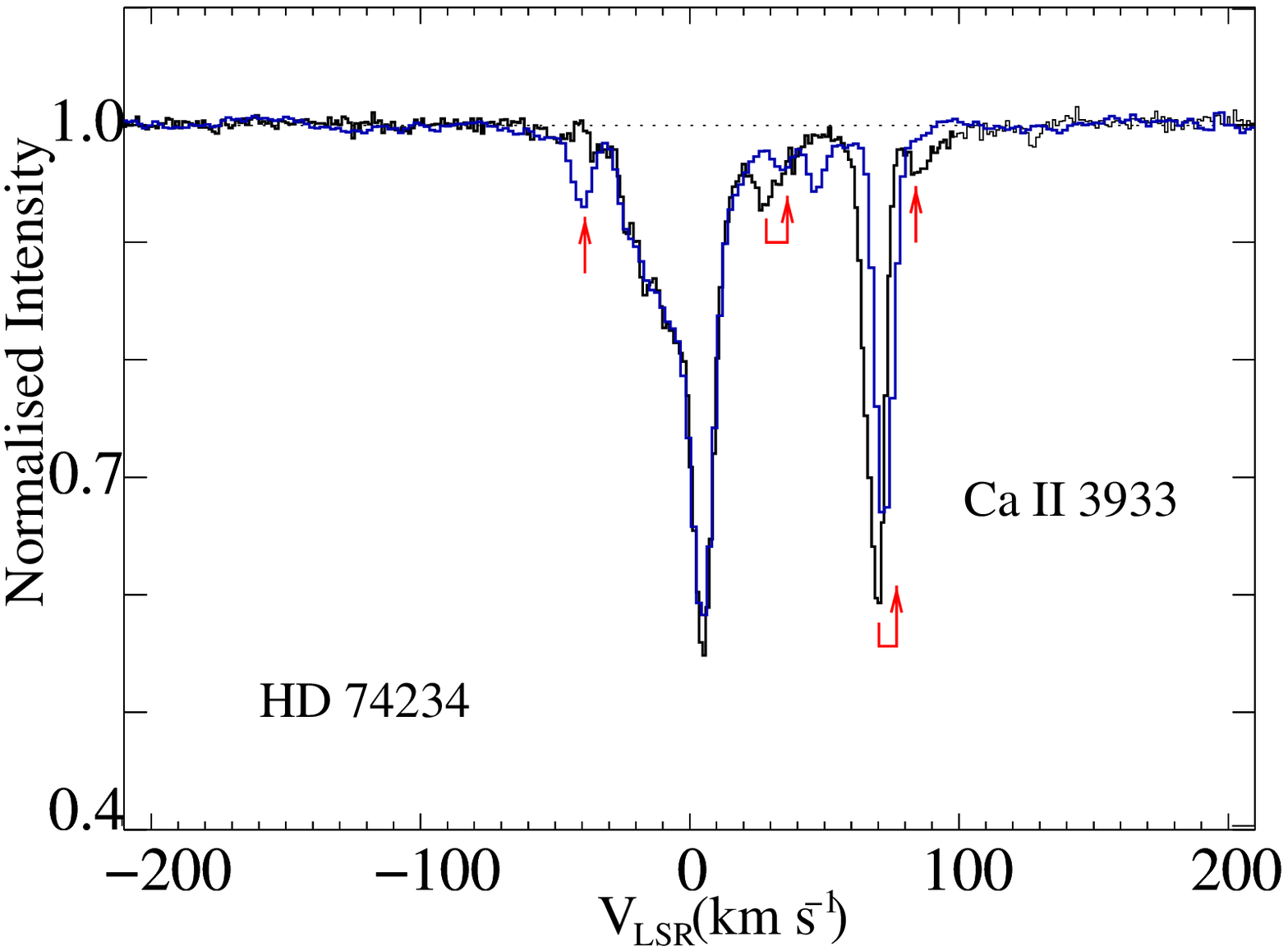}
\vspace{0.0cm}
\includegraphics[trim=0.0cm 0.0cm 0.0cm 0.0cm, clip=true,width=8cm,height=6.5cm]{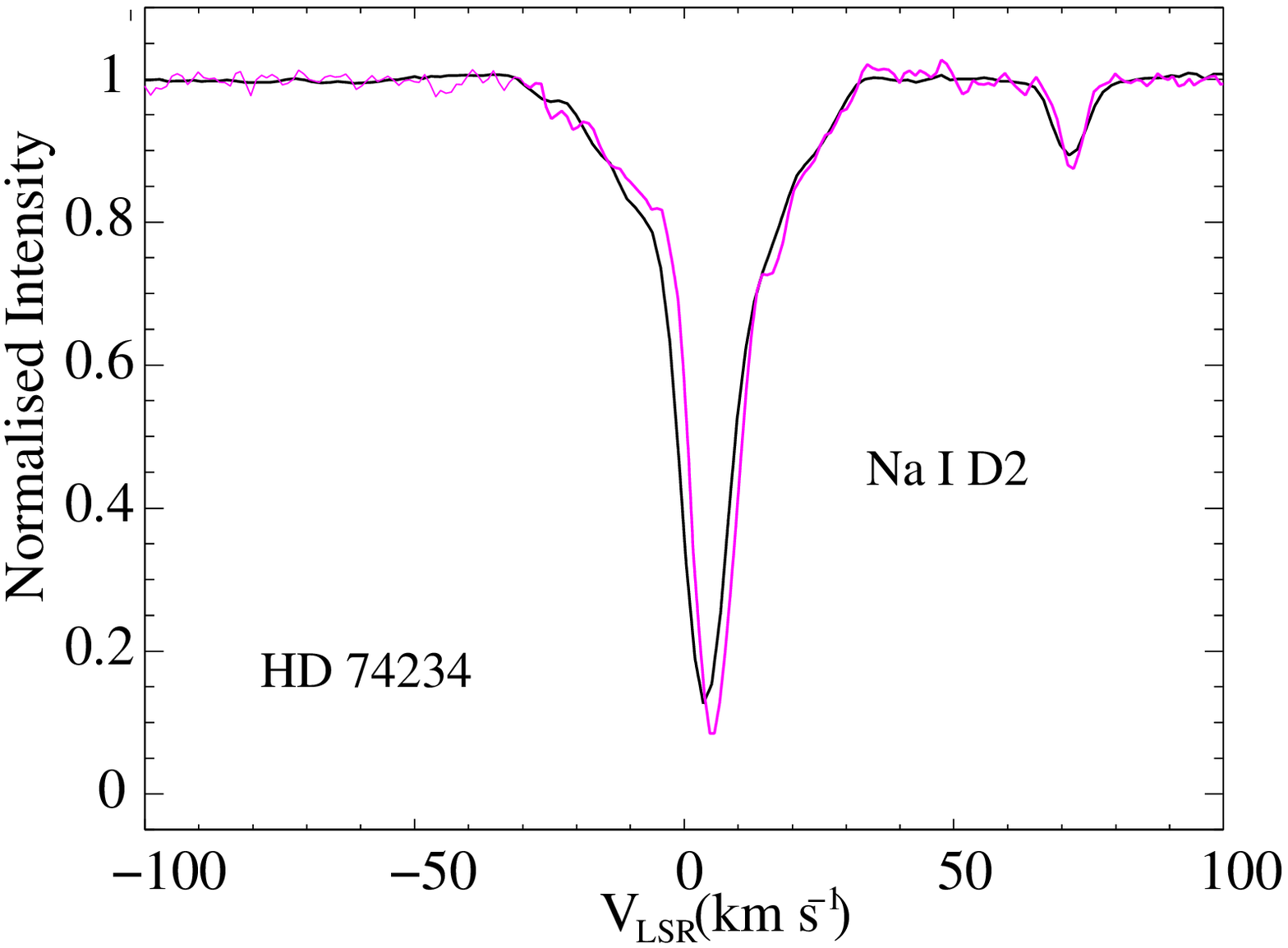}
\caption{  Profiles of  Ca\,{\sc ii} K in HD 74234 obtained in 1996 by Cha \& Sembach (black line)
 and on
 2017 May 28  with SALT (blue line).  Prominent changes in the profile between 1996
  and 2017  are  pointed out by red arrows. (Right): Na\,{\sc i} D lines from VBT (red line) and SALT (black line)
  spectra obtained in 2011 and 2017. The red arrows indicate the changes to
 the components between Cha \& Sembach's and SALT observations. }
%\end{minipage}
\end{figure*}

%8
\begin{table*}
\centering
\begin{minipage}{170mm}
\caption{\Large ISM Absorption Lines of Ca\,{\sc ii} K \&  Na\,{\sc i}  towards HD 74234 }
\begin{footnotesize}
\begin{tabular}{lcrrrcrrccrrr}
\hline
\multicolumn{1}{c}{}&\multicolumn{2}{c}{Ca\,{\sc ii} K } &\multicolumn{1}{c}{}&\multicolumn{1}{c}{}&
\multicolumn{3}{c}{Ca\,{\sc ii} SALT} &\multicolumn{1}{c}{} & \multicolumn{3}{c}{Na\,{\sc i} (SALT)}&\multicolumn{1}{c}{}    \\
\cline{1-3} \cline{5-7} \cline{9-11}   \\
      & C\&S &    &  & &K & H &   &   &$ D_{\rm 2}$& $D_{\rm 1}$&   \\
\cline{1-3} \cline{5-7} \cline{9-11}   \\
  $V_{\rm LSR}$ &$W_{\rm \lambda}$ &$W_{\rm \lambda}$& &$V_{\rm LSR}$&$W_{\rm \lambda}$ &$W_{\rm \lambda}$&  &$V_{\rm LSR}$&$W_{\rm \lambda}$  &
 $W_{\rm \lambda}$&     \\
  km s$^{-1}$ &(mA) &(mA) &  &km s$^{-1}$ &(mA)& (mA)& &km s$^{-1}$&(mA) &(mA)&    \\
\hline
         &    &     &  &$-40$  & 9    & 4 &    &     &    &     &      \\
         &    &     &  &       &      &   &    &$-24$&5 & 5   &     \\
 $-17$   & 27 &     &  &$-18$  &20.4  &11 &    &$-15$&16  & 10  &      \\
         &    &     &  &       &      &   &    &$-8$ &28  & 12  &      \\
 $-6$    & 18 &     &  &$-6.4$ &26    &13 &    &     &    &     &      \\
    5    &64  &     &  &  4.5  &56    &31 &    & 4   &185 & 158 &      \\
         &    &     &  & 17    & 8    &3.5&    &14   & 45 & 21  &       \\
   28    &15  &     &  & 34.5  & 6    & 1 &    &24   & 16 & 6   &      \\
         &    &     &  &       &      &   &    &     &    &     &       \\
         &    &     &  & 47    & 6    & 1 &    &     &    &     &       \\
 69 &47  &     &  & 72.6& 37   & 20& &71   & 17 & 7.5 &       \\
   85    &6   &     &  & 85   &  1   &0.5&    &     &    &     &      \\
\hline
\end{tabular}
\\
\end{footnotesize}
\label{default}
\end{minipage}
\end{table*}

% ---------------------------------------------------------------------------------------------------------------

\subsection{HD 74273}

           HD 74273 at 757 pc is located at the southern edge of the X-ray SNR just south of HD 74234 at 719 pc.  For HD 74273, the D and also the K line  from Cha \& Sembach in 1994 and 1996, the VBT D line profiles in 2011 and 2012 (Paper I) and the SALT spectrum in 2017 show unchanging profiles (Figure 8). Gaussian decomposition (Table 9) suggests minor changes in equivalent width in some components but these changes may arise from subtle changes in the instrumental line profiles and differences in the S/N ratios. This sight line is one  without high -- positive or negative -- velocity components in Cha \& Sembach's sample. The three low velocity components may arise from the interstellar medium between us and the SNR and, thus, the SNR may not contribute to the D and K line profiles even for this sight line extending to 757 pc.

%8

\begin{figure*}
%\begin{minipage}{120mm}
\vspace{0.0cm}
\includegraphics[trim=0.0cm 0.0cm 0.1cm 0.0cm, clip=true,width=8cm,height=6.5cm]{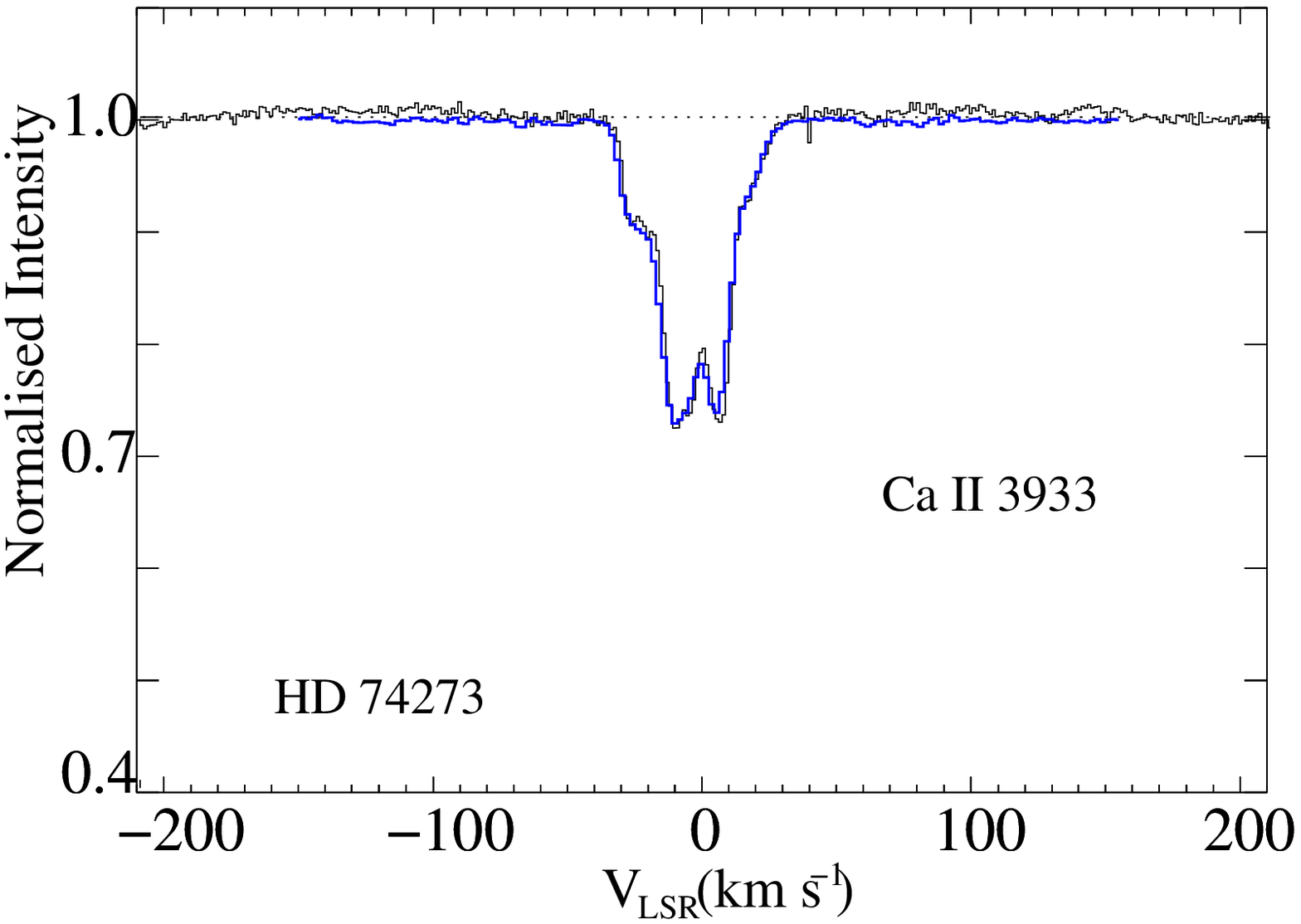}
\vspace{0.0cm}
\includegraphics[trim=0.0cm 0.0cm 0.1cm 0.0cm, clip=true,width=8cm,height=6.5cm]{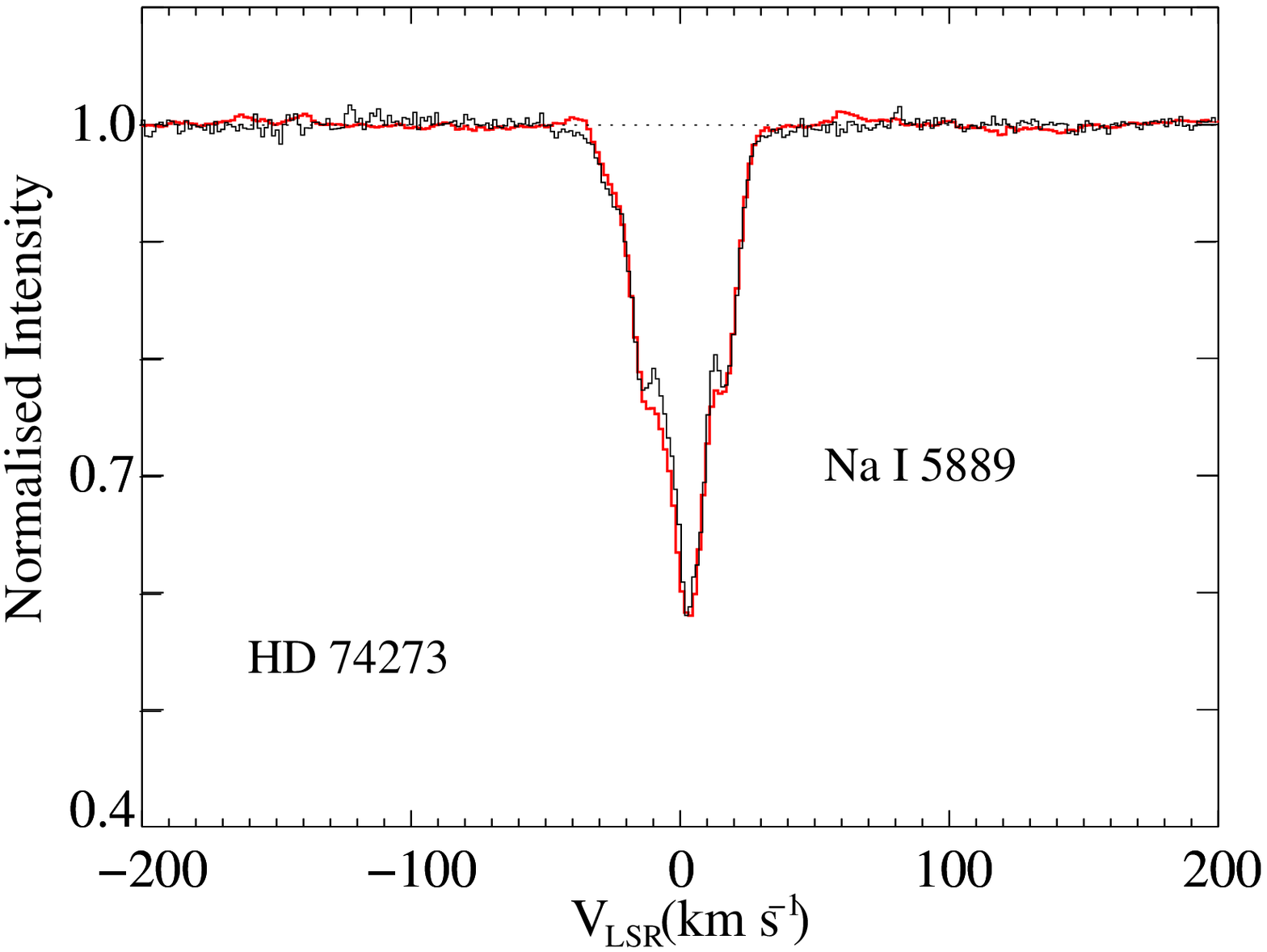}
\caption{(Left panel): Profiles of  Ca\,{\sc ii}  K in the sightline towards
  HD 74273 obtained with SALT (blue line) superposed on the K line profile obtained in 1996
  by Cha \& Sembach (2000). (Right panel):    Na\,{\sc i} D line  profiles
 of HD 74273 obtained with SALT(red) superposed on Na\,{\sc i} D2 line observed by Cha \& Sembach. }

\end{figure*}

%9
\begin{table*}
\centering
\begin{minipage}{170mm}
\caption{\Large ISM Absorption Lines of Ca\,{\sc ii} K and  Na\,{\sc i} towards HD 74273.}
\begin{footnotesize}
\begin{tabular}{lcrrrcrrccrrrccr}
\hline
\multicolumn{1}{c}{}&\multicolumn{2}{c}{C\&S Ca\,{\sc ii} K }&\multicolumn{1}{c}{}&\multicolumn{1}{c}{}&
\multicolumn{3}{c}{Ca\,{\sc ii} (SALT)} &\multicolumn{1}{c}{} & \multicolumn{3}{c}{Na\,{\sc i} (C\&S)}&\multicolumn{1}{c}{}&\multicolumn{2}{c}{Na\,{\sc i} (SALT)}  &  \\
\cline{1-4} \cline{6-8} \cline{10-12} \cline{14-16}  \\
 1994  &     & 1996 &  &  & & K & H & &  & $ D_{\rm 2}$& $D_{\rm 1}$&   &&$ D_{\rm 2}$& $D_{\rm 1}$ \\
\cline{1-4} \cline{6-8} \cline{10-12} \cline{14-16}   \\
 $V_{\rm LSR}$ &$W_{\rm \lambda}$ & $V_{\rm LSR}$ & $W_{\rm \lambda}$ &   & $V_{\rm LSR}$ &$W_{\rm \lambda}$&$W_{\rm \lambda}$&  &$V_{\rm LSR}$  &
  $W_{\rm \lambda}$& $W_{\rm \lambda}$ &  & $V_{\rm LSR}$&$W_{\rm \lambda}$&$W_{\rm \lambda}$   \\
km s$^{-1}$&(mA) &km s$^{-1}$&(mA)& &  km s$^{-1}$ &(mA)& (mA)& &km s$^{-1}$ &(mA) &(mA)& &km s$^{-1}$&(mA) &(mA)    \\
\hline
  $-25$&21 &$-27$& 11 &   &$-25$ & 14.5&7    &    &$-27$&10  &$\le$6 & &$-27$  &6.5&3  \\
       &   &     &    &   &      &     &     &    &$-11$&68  &26     & &$-12.5$&70 &33 \\
  $-9 $&51 &$-9$ & 66 &   &$-9$  & 53  & 28  &    &    &     &       & &       &   &  \\
    5  &43 &  6  & 34 &   & 5    & 39  & 21  &    & 5  &124  &70     & & 3     &121&83  \\
       &   &     &    &   &20    &  8  & 3   &    &19  & 39  & 17    & &17   &44&20   \\
       &   &     &    &   &      &     &     &    &    &     &       & &       &  &   \\
 %S/N   &   &     &    &   &      & 430 & 542 &    &    &     &       & &       &343&361\\
 %     &   &     &    &   &      &     &     &    &    &     &       & &       &   &   \\
\hline
\end{tabular}
\\

\end{footnotesize}
\label{default}
\end{minipage}
\end{table*}

% -------------------------------------------------------------------------------------------------------------

\subsection{HD 74371}

                       On the sky,  HD 74371 is located almost at the centre of the SNR,  close to HD 74194 and not far east of
     the pulsar. At the star's distance of 1250 pc, the sight line must pass right through the SNR. Unfortunately, the relatively weak interstellar K lines are superimposed on a strong and broad stellar line (Figure 9 and Table 10).  In fact, Cha \& Sembach in their Gaussian decomposition recognized two strong and roughly equal stellar K lines separated by 27 km s$^{-1}$ with ISM lines at $-10$ and $+11$ km s$^{-1}$.  Double stellar lines at different velocities were recognized earlier (Wallerstein \& Silk 1984). The SALT profile is distinctly different from that illustrated by Cha \& Sembach.  Cha \& Sembach did not observe the Na D lines. It appears that HD 74371 may be a spectroscopic binary. With the variable and strong stellar profile, it is difficult to isolate the weak interstellar profiles.  HD 74371 is, however, far from uninteresting; this sight line through the centre of the SNR provides no D or K line contributions other than those which may be attributed to the local interstellar medium. It may seem surprising that the SNR provides neither positive or negative high-velocity contributions

% %9
\begin{figure*}
 %\begin{minipage}{120mm}
\vspace{0.0cm}
\includegraphics[width=8cm,height=6.5cm]{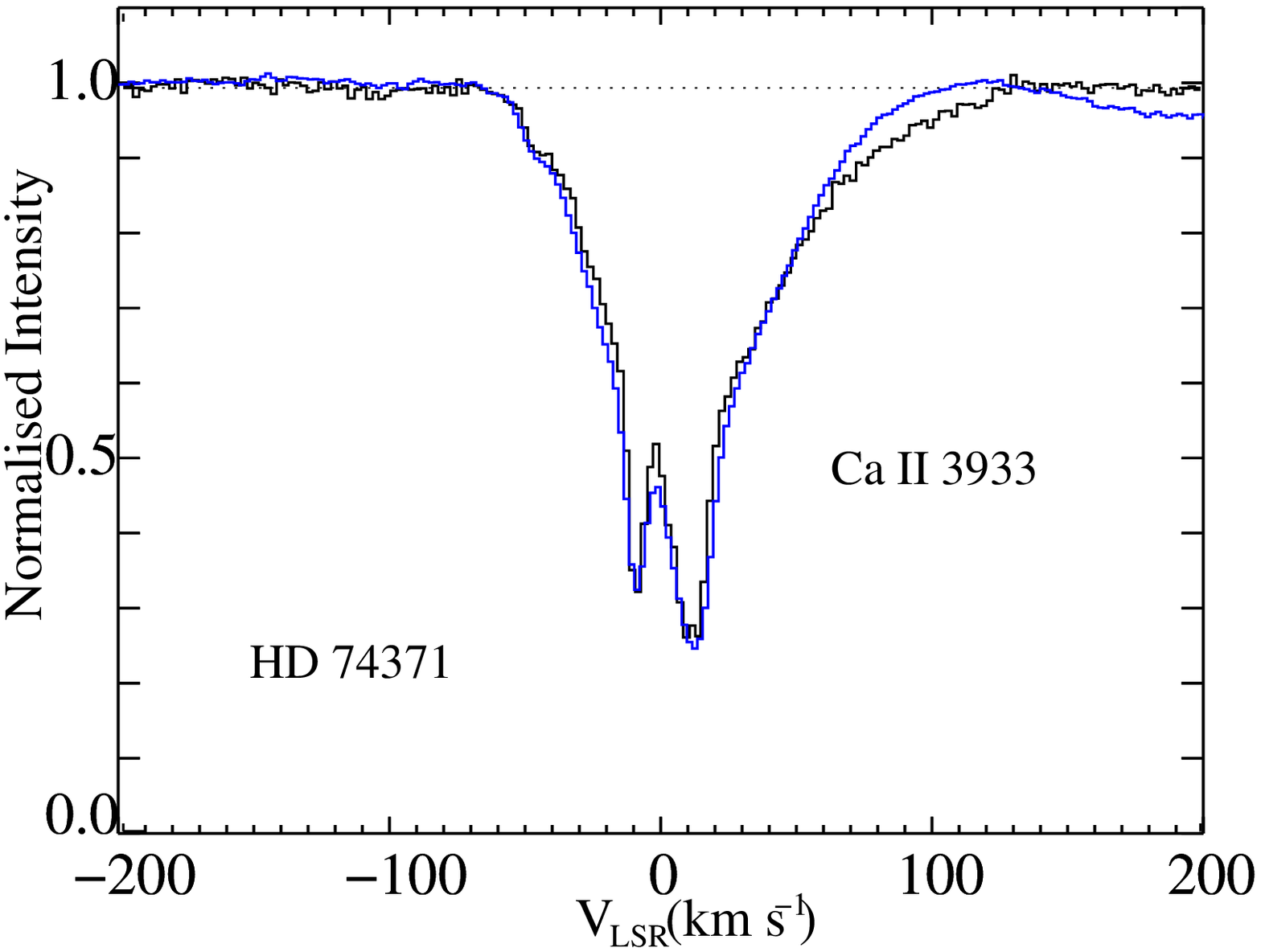}
\vspace{0.0cm}
\includegraphics[width=8cm,height=6.5cm]{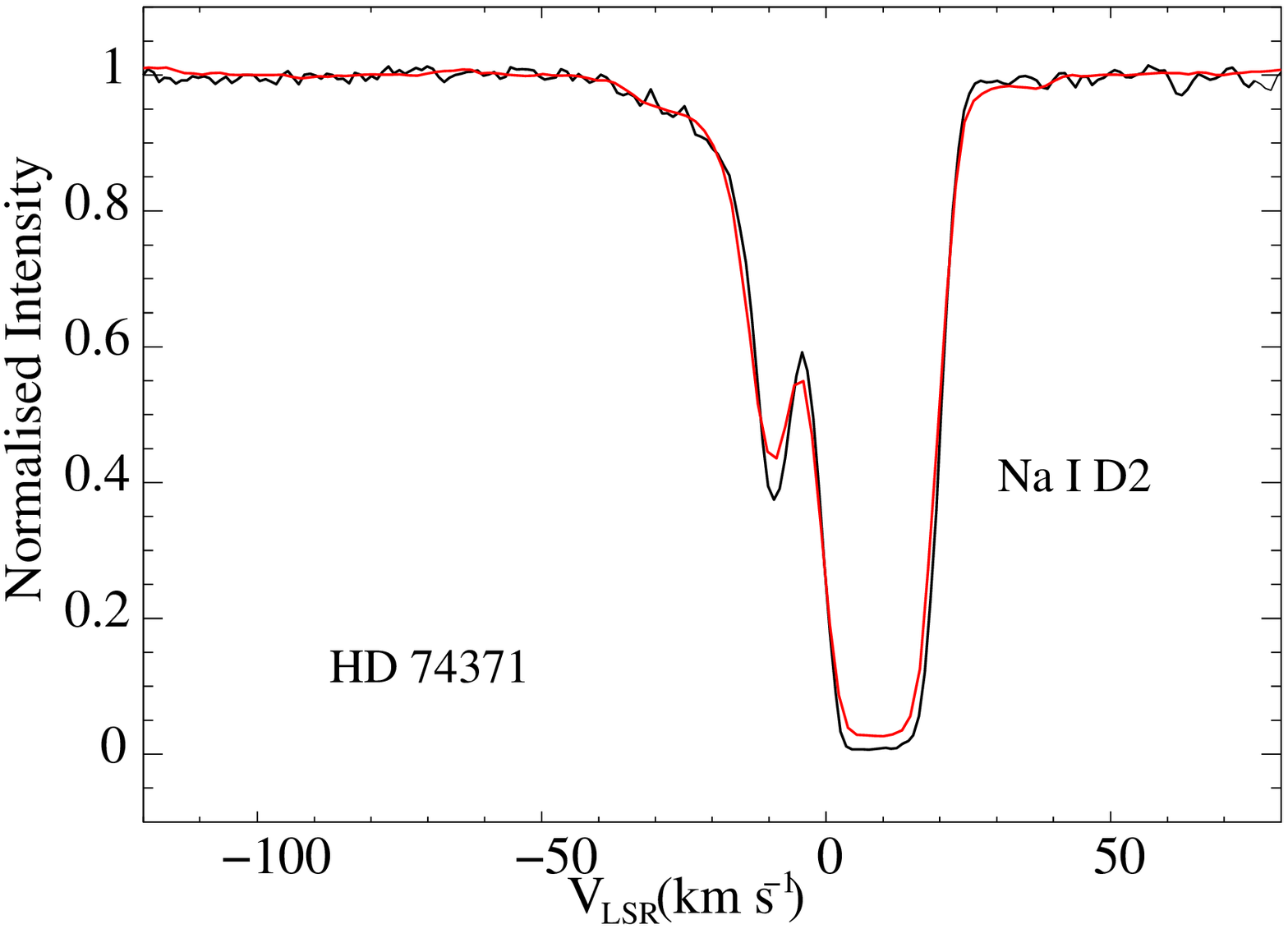}
\caption{(Left panel): Profiles of  Ca\,{\sc ii} K in the sightline towards
   HD 74371 obtained with SALT (blue line) on 2018 January 5
   is superposed on the 1996 profile obtained by Cha \& Sembach (2000)
  (black line). A strong stellar Ca\,{\sc ii} line is blended with the ISM components.
  (Right panel):   Na\,{\sc i} D$_2$  profile of HD 74371 obtained with SALT on
  2018 January 5 (red line) superposed on the VBT profile obtained on 2012 March 16 (black line).}
 %(Right panel) The SED of HD74371 showing the Far-IR
 %excess. Blackbody fits suggest a temperature of 22 K for the IR excess.
 %Observations from 2MASS, WISE and IRAS are shown along with optical.  }
 %\end{minipage}
\end{figure*}

%10
\begin{table*}
\centering
\begin{minipage}{170mm}
\caption{\Large ISM Absorption Lines of Ca\,{\sc ii} K and Na\,{\sc i}  towards HD 74371 }
\begin{footnotesize}
\begin{tabular}{lcrrrcrrccrrr}
\hline
\multicolumn{1}{c}{}&\multicolumn{2}{c}{Ca\,{\sc ii} K } &\multicolumn{1}{c}{}&\multicolumn{1}{c}{}&
\multicolumn{3}{c}{Ca\,{\sc ii} SALT} &\multicolumn{1}{c}{} & \multicolumn{3}{c}{Na\,{\sc i} (SALT)}&\multicolumn{1}{c}{}    \\
\cline{1-3} \cline{5-7} \cline{9-11}  \\
      & C\&S &    &  & &K & H &   &   &$ D_{\rm 2}$& $D_{\rm 1}$  \\
\cline{1-3} \cline{5-8} \cline{10-12}   \\
  $V_{\rm LSR}$ & $W_{\rm \lambda}$ &$W_{\rm \lambda}$& &$V_{\rm LSR}$&$W_{\rm \lambda}$ &$W_{\rm \lambda}$&  &$V_{\rm LSR}$& $W_{\rm \lambda}$  &
  $W_{\rm \lambda}$     \\
  km s$^{-1}$ &(mA) &(mA) &  &km s$^{-1}$ &(mA)& (mA)& &km s$^{-1}$&(mA) &(mA)  \\
\hline
         &    &     &  &$-42.3$& 12& 3 &    &     &    &     &   \\
         &    &     &  &       &      &   &    &$-27$&15  & 8    \\
 $-10$   & 42 &     &  &$-10$  &57    &36 &    &$-10$&126 & 68     \\
   2     &286 & &  &       &      &   &    &     &    &       \\
         &    &     &  &       &      &   &    & 4   &216 &213   \\
  11     &96  &     &  & 10    &143   &85 &    &     &    &      \\
         &    &     &  & 17.3  & 8    &3.5&    &14.5 &201 &194    \\
   29    &301 &     &  &       &      &   &    &     &    &        \\
 %        &    &     &  &       &      &   &    &     &    &      \\
 %S/N     &    &     &  &       &      &   &    &     &588 & 588  \\
         &    &     &  &       &      &   &    &     &    &      \\
\hline
\end{tabular}
\\
\end{footnotesize}
\label{default}
\end{minipage}
\end{table*}

%  ------------------------------------------------------------------------------------------------------

\subsection{HD 74979}

                         HD 74979 is located just outside the northern edge of the ROSAT contour. At a distance of 937 pc the sight line to the star crosses the entire SNR. Cha \& Sembach observed the K line but not the D lines. Their K line profile shows two strong and blended components near zero velocity and a weak high velocity component at  -87 km s$^{-1}$. The SALT Ca\,{\sc ii} K profile (Figure 10 replicates that shown by Cha \& Sembach except that the lone high velocity component is at $-89$ km s$^{-1}$) but with the same equivalent width. Examination of the SALT spectrum shows, however, that this high-velocity component is absent from the Ca\,{\sc ii} H profile. The predicted equivalent width of the H profile's $-89$ km s$^{-1}$ component is about 7 m\AA\ which is
     very easily detectable. But it is not seen. There is a possibility that the
     this  K line component  is due to an unidentified stellar line
     (most likely of O\,{\sc ii} or C\,{\sc ii}).
     The stellar lines are narrow in
     HD 74979.
    Our SALT profile of the Na D$_2$ line lines is shown in  Figure 10. The radial
      velocities of the components are marginally different to  those of the Ca\,{\sc ii} K line
     (Table 11).

%10
\DIFdelend
\begin{figure*}
%\begin{minipage}{120mm}
\vspace{0.0cm}
\includegraphics[trim=0.0cm 0.0cm 0.1cm 0.0cm, clip=true,width=8cm,height=6.5cm]{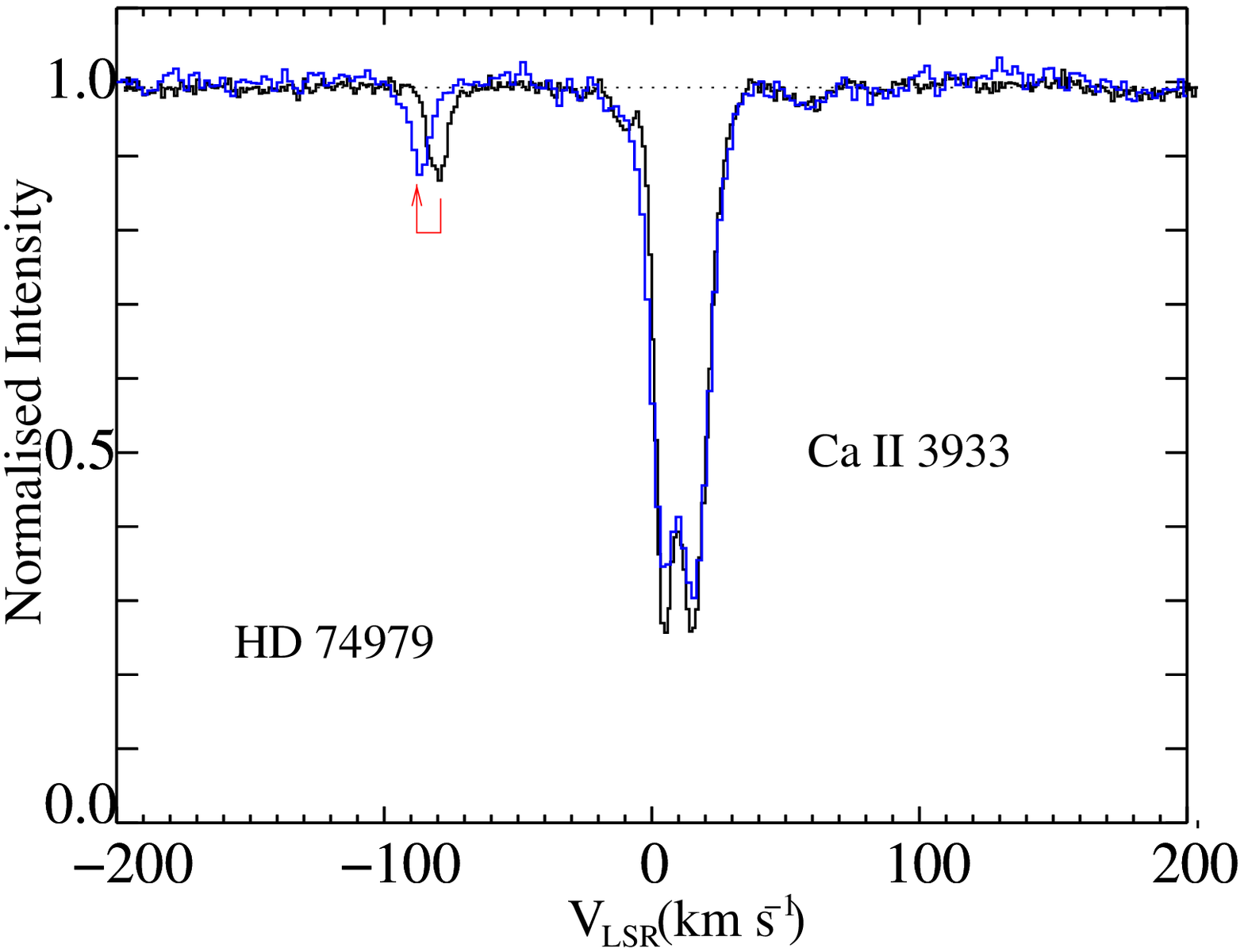}
\vspace{0.0cm}
\includegraphics[trim=0.0cm -0.1cm 0.1cm 0.0cm, clip=true,width=8cm,height=6.5cm]{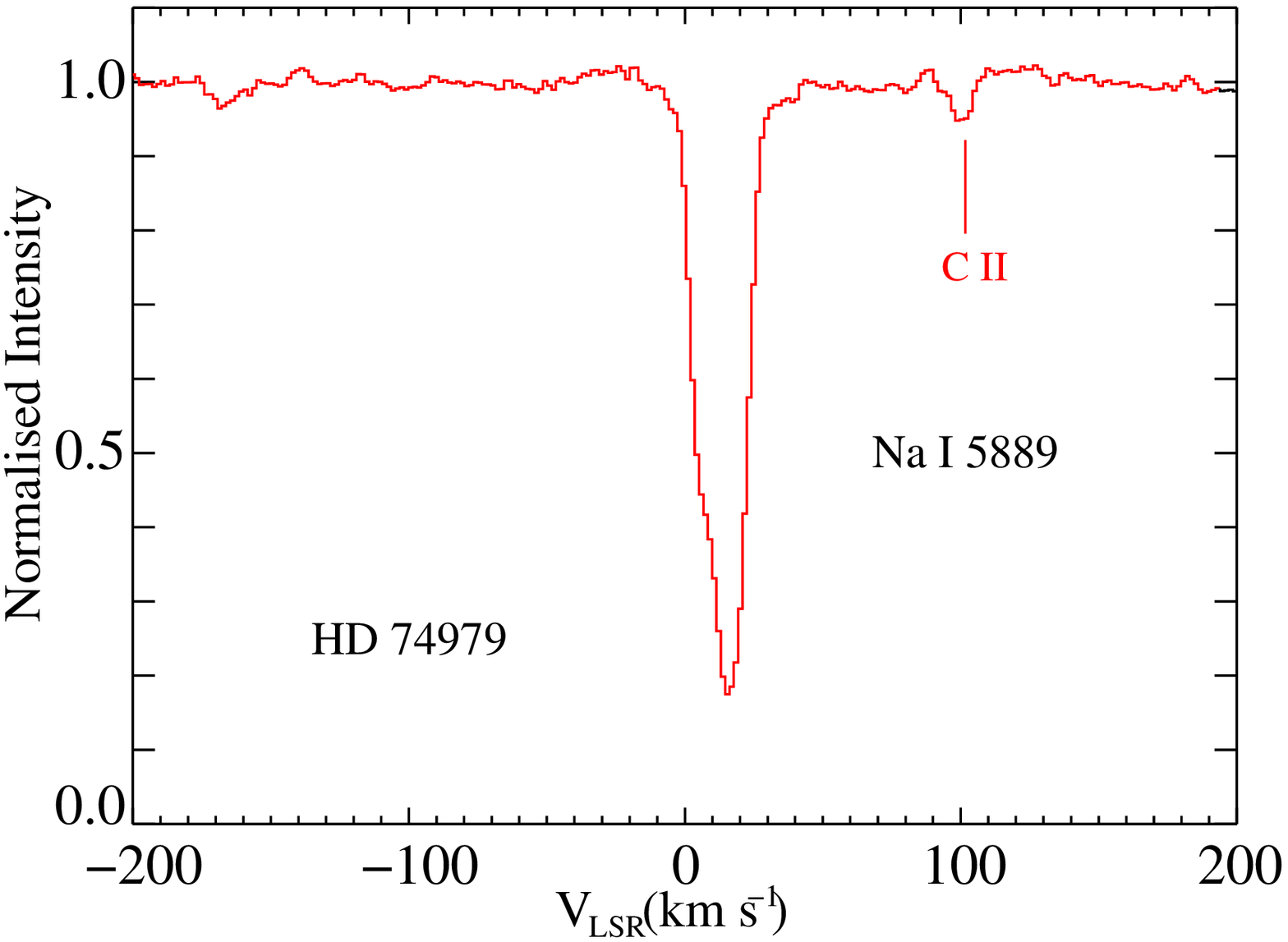}
\caption{(Left panel): Profile of  Ca\,{\sc ii} K in the sight line towards
  HD 74979 obtained with SALT (blue line) on 2018 January 17
  is superposed on the 1996 profile obtained by Cha \& Sembach (2000)
 (black line).
 (Right panel):   Na\,{\sc i} D$_2$  profile of HD 74979 obtained with SALT on
 2018 January 17 (red line). A  VBT profile (not shown here) from  2011 March 24 is noisy but there are no major differences between the SALT and VBT profiles. The absorption line at $+100$ km s$^{-1}$ in the SALT spectrum is a stellar C\,{\sc ii} line.}
%\end{minipage}
\end{figure*}

% ----------------------------------------------------------------------------------------------------

\subsection{HD 75129}

                        HD 75129 is located in the south west part of the X-ray remnant. At the stellar distance of 1038 pc, the sight line crosses the entire SNR. Cha \& Sembach's K line profile showed about  eight components   ranging in radial velocity from $-117$ to $+27$ km s$^{-1}$. Interstellar K line contributions near 0 km s$^{-1}$ are blended with the stellar K line. Our SALT K profile shows a
      dramatic change  between observations of
       Cha \& Sembach (2000) in 1996 and the SALT observations on 2017 May 29: the almost complete
       disappearance  of a strong
       absorption high-velocity component at $-67$ km s$^{-1}$ (Figure 11, Table 12).
       Cha \& Sembach (2000) did not observe the star in  Na\,{\sc i} D lines so a comparison
        with the SALT spectrum at Na\,{\sc i} D could not be made.

 %11
\begin{table*}
\centering
\begin{minipage}{170mm}
\caption{\Large ISM Absorption Lines of Ca\,{\sc ii} K and  Na\,{\sc i}  towards HD 74979 }
\begin{footnotesize}
\begin{tabular}{lcrrrcrrccrrr}
\hline
\multicolumn{1}{c}{}&\multicolumn{2}{c}{Ca\,{\sc ii} K } &\multicolumn{1}{c}{}&\multicolumn{1}{c}{}&
\multicolumn{3}{c}{Ca\,{\sc ii} SALT} &\multicolumn{1}{c}{} & \multicolumn{3}{c}{Na\,{\sc i} (SALT)}&\multicolumn{1}{c}{}   \\
\cline{1-3} \cline{5-7} \cline{9-11}   \\
      & C\&S &    &  & &K & H &   &   &$ D_{\rm 2}$& $D_{\rm 1}$  \\
\cline{1-3} \cline{5-8} \cline{10-12}   \\
  $V_{\rm LSR}$ &$W_{\rm \lambda}$ &$W_{\rm \lambda}$ & &$V_{\rm LSR}$&$W_{\rm \lambda}$&$W_{\rm \lambda}$ &  &$V_{\rm LSR}$& $W_{\rm \lambda}$   &$W_{\rm \lambda}$     \\
  km s$^{-1}$ &(mA) &(mA) &  &km s$^{-1}$ &(mA)& (mA)& &km s$^{-1}$&(mA) &(mA)&    \\
\hline
 $-87$   & 15 &     &  & -89.3&15&  &    &     &    &         \\
 $-13$   &  9 &     &  &$-12$&13      &5  &    &    &     &         \\
   1     & 73 &     &  &  1  &76     &45   &    &   &     &         \\
         &    &     &  &       &      &   &    & 4  &116  &60       \\
  12     &146 &     &  & 13    &135   &80 &    & 14 &185  &195      \\
         &    &     &  &       &      &   &    & 27 & 6   &11       \\
         &    &     &  &       &      &   &    &    &    &          \\
 %S/N     &    &     &  &       &79    &87 &    &     &130 &130      \\
         &    &     &  &       &      &   &    &     &    &          \\
\hline
\end{tabular}
\\
\end{footnotesize}
\label{default}
\end{minipage}
\end{table*}

% 11

\DIFdelend \begin{figure*}
%\begin{minipage}{120mm}
\vspace{0.0cm}
\includegraphics[trim=0.0cm 0.0cm 0.1cm 0.0cm, clip=true,width=8.cm,height=6.5cm]{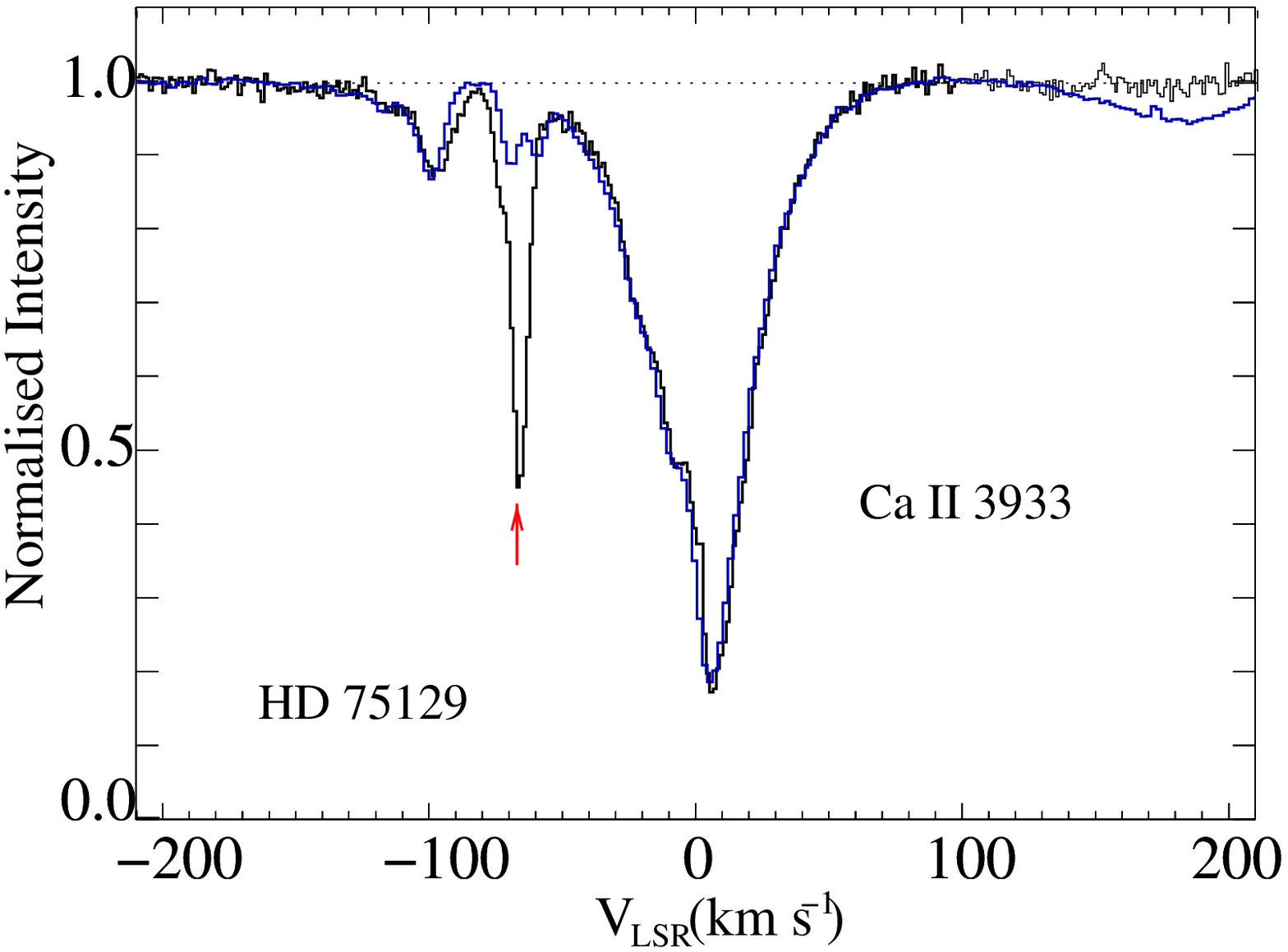}
\vspace{0.0cm}
\includegraphics[trim=0.0cm 0.0cm 0.0cm 0.0cm, clip=true,width=8.cm,height=6.5cm]{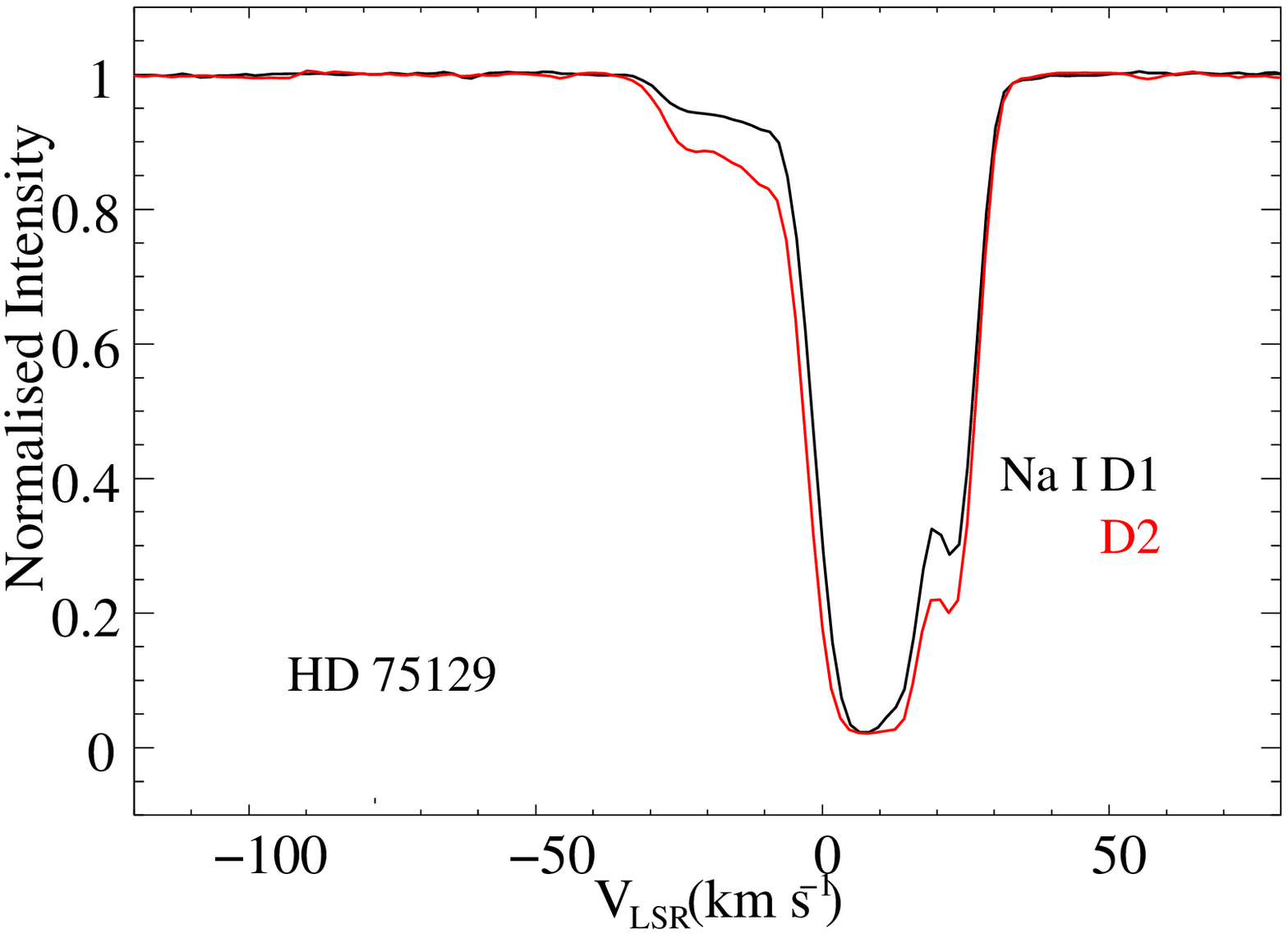}
\caption{(Left panel): Profiles of  Ca\,{\sc ii} K along the sight line towards
  HD 75129 obtained with SALT (blue line) on 2017 May 29 and
 the 1996 profile obtained by Cha \& Sembach (2000)
 (black line). The component at -67 km s$^{-1}$ has weakened considerably by 2017
 (Right panel):   Na\,{\sc i}  D$_1$ and D$_2$ profiles of HD 75129 obtained with the SALT on 2017 May 29.}
%\end{minipage}
\end{figure*}

%12
\begin{table*}
\centering
\begin{minipage}{170mm}
\caption{\Large ISM Absorption Lines of Ca\,{\sc ii} K and  Na\,{\sc i} to wards HD 75129.}
\begin{footnotesize}
\begin{tabular}{lrrcrrccrrrccr}
\hline
\multicolumn{2}{c}{C\&S Ca\,{\sc ii} K }&\multicolumn{1}{c}{}&\multicolumn{1}{c}{}&
\multicolumn{3}{c}{Ca\,{\sc ii} (SALT)} &\multicolumn{1}{c}{} & \multicolumn{3}{c}{}&\multicolumn{1}{c}{}&\multicolumn{2}{c}{Na\,{\sc i} (SALT)}    \\
\cline{1-2} \cline{4-6} \cline{8-10} \cline{12-14}  \\
   1996 &  &  & & K & H & &  & $ D_{\rm 2}$& $D_{\rm 1}$&   &&$ D_{\rm 2}$& $D_{\rm 1}$ \\
\cline{1-2} \cline{4-6} \cline{8-10} \cline{12-14}   \\
 $V_{\rm LSR}$ &$W_{\rm \lambda}$&   & $V_{\rm LSR}$ &$W_{\rm \lambda}$ &$W_{\rm \lambda}$ &  &$V_{\rm LSR}$  &
$W_{\rm \lambda}$&$W_{\rm \lambda}$&  & $V_{\rm LSR}$&$W_{\rm \lambda}$ &$W_{\rm \lambda}$    \\
km s$^{-1}$&(mA)& &  km s$^{-1}$ &(mA)& (mA)& &km s$^{-1}$ &(mA) &(mA)& &km s$^{-1}$&(mA) &(mA)    \\
\hline
       &    &   &$-117$ & 5  &4    &    &    &     &       & &       &   &   \\
  $-98$& 26 &   &$-99.3$&24  &14   &    &    &     &       & &       &   &   \\
  $-67$& 62& &$-70$& 13& 7   &    &    &     &       & &       &   & \\
       &    &   &$-60$&11  & 4   &    &    &     &       & &       &   &   \\
       &    &   &$-34$ &32   & 12  &    &    &     &       & &       &   &   \\
       &    &   &      &     &     &    &    &     &       & &$-22$  &24 &12   \\
       &    &   &$-15$ &90   &65  &    &    &     &       & & $-13$  &20 &11 \\
    2  &396 &   &      &     &     &    &    &     &       & &       &   &   \\
    7  &109 &   & 6.5  &280  &138  &    &    &     &       & & 5   &300&267   \\
       &    &   &      &     &     &    &    &     &       & &15   &125&128   \\
       &    &   & 27   &50   & 35  &    &    &     &       & & 23.5  &127&48   \\
       &    &   &      &     &     &    &    &     &       & &       &   &   \\
 %S/N  &    &   &      &368  & 308 &    &    &     &       & &       &430&430   \\
\hline
\end{tabular}
\\
 \end{footnotesize}
\label{default}
\end{minipage}
\end{table*}

 % -----------------------------------------------------------------------------------------------

\subsection{HD 75387}

     HD 75387 is in the northern part of the X-ray outer contour and at a distance of 466 pc probably lies behind the bulk of the SNR. Profiles of the K and the D$_2$ line obtained by SALT in 2017 and by Cha \& Sembach in 1994 are compared in Figure 12 and  Table 13. Apart from minor changes to the wings of the principal K blend near $0$ km s$^{-1}$, the 1994 and 2017 K line profiles agree with respect to velocity and equivalent width.   A distinct component at $-39$ km s$^{-1}$ unusually  in both the D and K lines is unchanged between 1994 and 2017.                 Our measurement
    of the Na\,{\sc i} D$_2$ line is affected by the stellar C\,{\sc ii}  lines blending with it
    (see red arrows in Figure 11). This line was also present in the VBT spectrum from 2011 (Paper I).
    We suspect that even Cha \& Sembach (2000)'s measurements of this line may have been
    affected by these stellar lines.
 There is not much of a change in the profiles between  1996 to 2017.

% 12
\begin{figure*}
%\begin{minipage}{120mm}
\includegraphics[trim=0.0cm 0.0cm 0.1cm 0.0cm, clip=true,width=8cm,height=6.5cm]{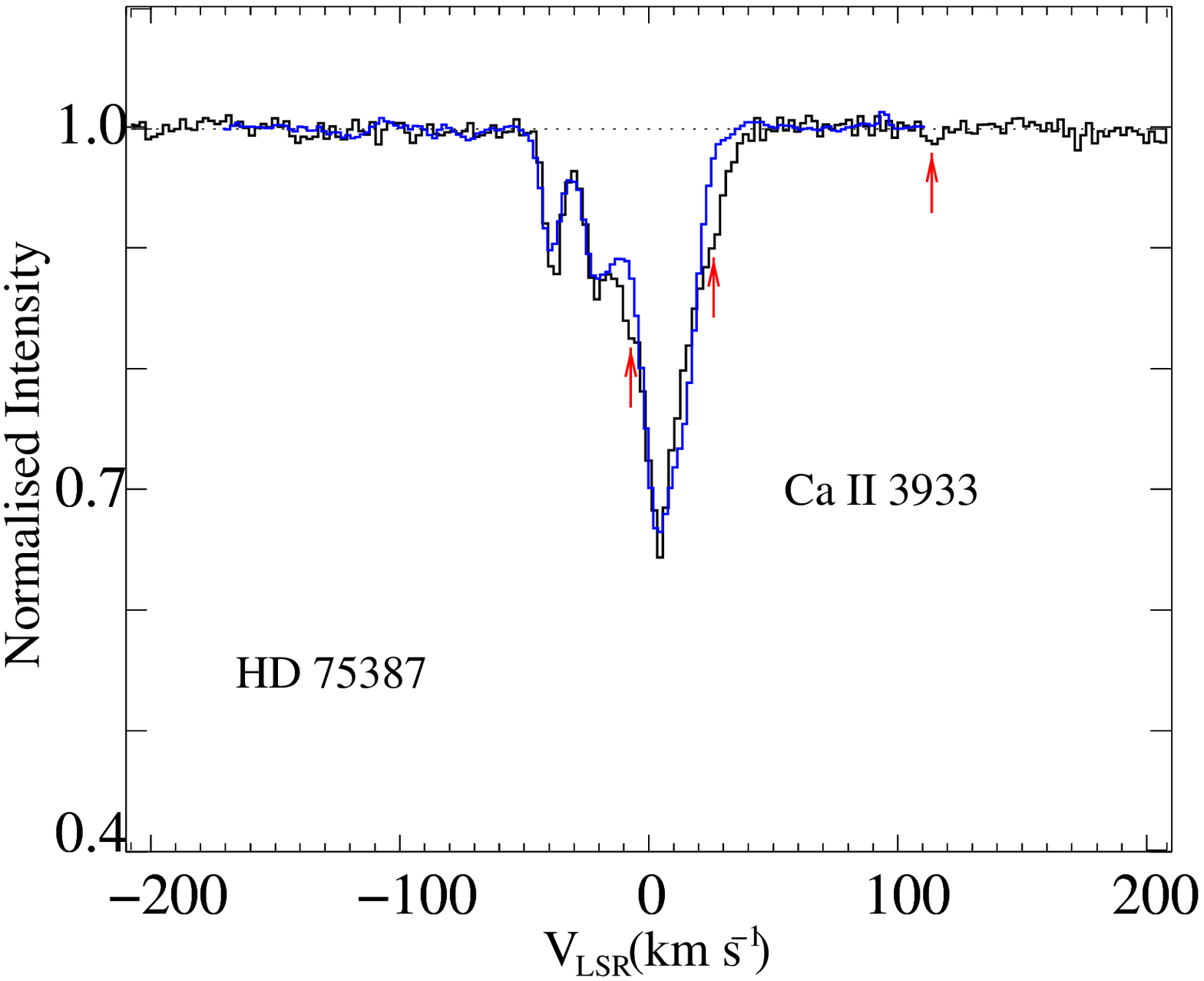}
\includegraphics[trim=0.0cm 0.0cm 0.3cm 0.0cm, clip=true,width=8cm,height=6.5cm]{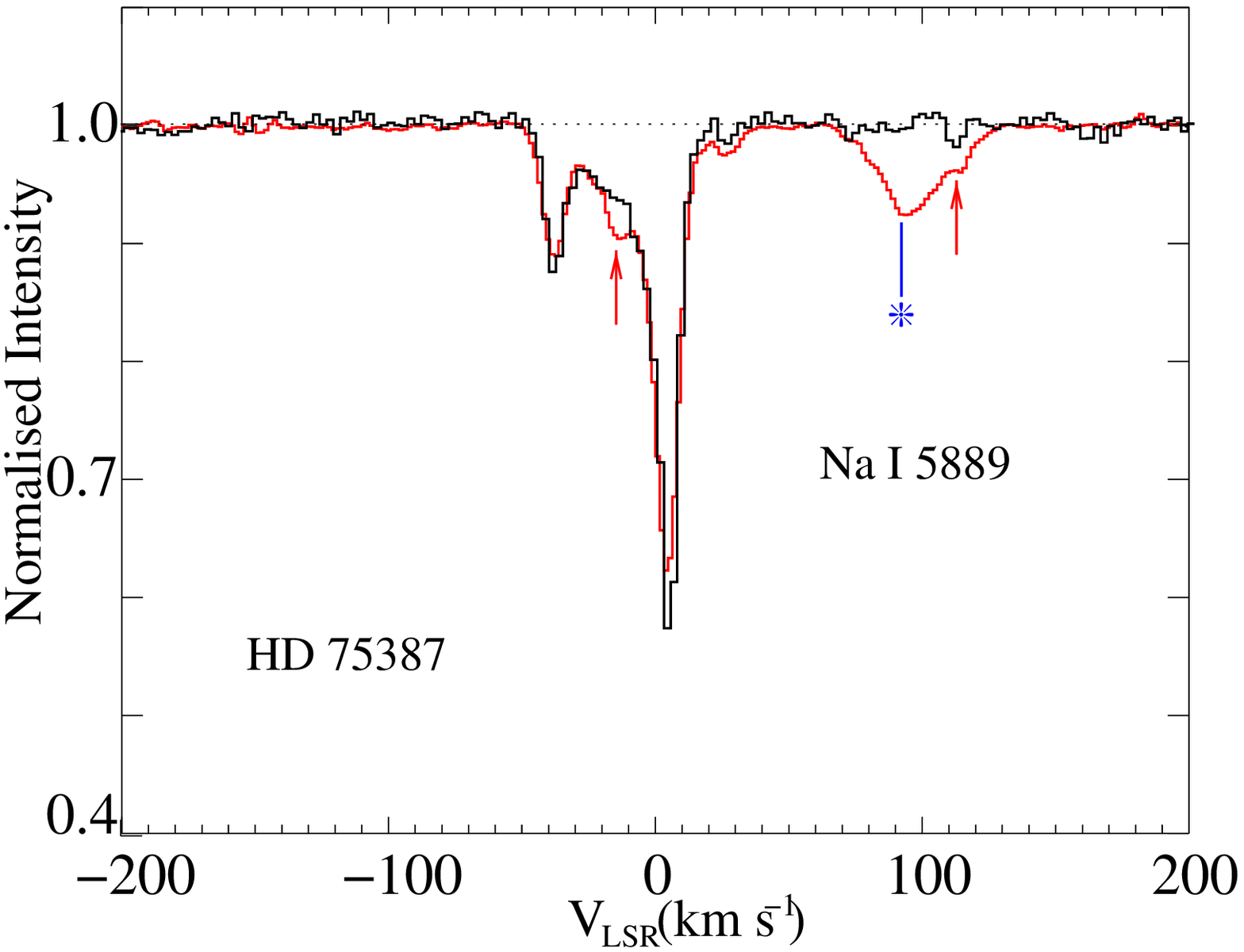}
\caption{(Left):  Ca\,{\sc ii} K   profile of HD 75387
 obtained in 1994 by Cha \& Sembach (2000) is compared with the profile obtained on
 2017 June 1  with SALT (blue line). The profiles are very similar indicating no
 significant change occurred between 1994 and 2017. (Right): Na\,{\sc i} D$_2$   profile
 obtained in 1996 by Cha \& Sembach (2000) is compared with the profile obtained on
 2017 June 1  with SALT (red line). The profiles  indicate
 significant change are present between 1996 and 2017. A stellar line is indicated in
 the SALT spectrum. The red arrows indicate the changes to
 the components between Cha \& Sembach's and SALT observations.  A red arrow suggests
 a weak Na D component is present at +115 km s$^{-1}$ in 2017.}
%\end{minipage}
\end{figure*}

%13
\begin{table*}
\centering
\begin{minipage}{170mm}
\caption{\Large ISM Absorption Lines of Ca\,{\sc ii} K and  Na\,{\sc i} towards HD 75387.}
\begin{footnotesize}
\begin{tabular}{lcrrrcrrccrrrccr}
\hline
\multicolumn{1}{c}{}&\multicolumn{2}{c}{C\&S Ca\,{\sc ii} K }&\multicolumn{1}{c}{}&\multicolumn{1}{c}{}&
\multicolumn{3}{c}{Ca\,{\sc ii} (SALT)} &\multicolumn{1}{c}{} & \multicolumn{3}{c}{Na\,{\sc i} (C\&S)}&\multicolumn{1}{c}{}&\multicolumn{2}{c}{Na\,{\sc i} (SALT)}  &  \\
\cline{1-4} \cline{6-8} \cline{10-12} \cline{14-16}  \\
 1994      &     &  &  &  & & K & H & &  & $ D_{\rm 2}$& $D_{\rm 1}$&   &&$ D_{\rm 2}$& $D_{\rm 1}$ \\
\cline{1-4} \cline{6-8} \cline{10-12} \cline{14-16}   \\
 $V_{\rm LSR}$ &$W_{\rm \lambda}$ & $V_{\rm LSR}$ & $W_{\rm \lambda}$ &   & $V_{\rm LSR}$ &$W_{\rm \lambda}$&$W_{\rm \lambda}$&  &$V_{\rm LSR}$  &
 $W_{\rm \lambda}$ & $W_{\rm \lambda}$ &  & $V_{\rm LSR}$&$W_{\rm \lambda}$& $W_{\rm \lambda}$  \\
km s$^{-1}$&(mA) &km s$^{-1}$&(mA)& &  km s$^{-1}$ &(mA)& (mA)& &km s$^{-1}$ &(mA) &(mA)& &km s$^{-1}$&(mA) &(mA)    \\
\hline
  $-39$&14 &     &    &   &$-39$ & 14  &6    &    &$-39$& 21 &11     & & $-39$ &24 &12   \\
  $-20$&25 &     &    &   &$-20.7$&22.5&3.5  &    &    &     &       & & $-22.5$&11& 4  \\
  $-9$ &14 &     &    &   &$-8.2$ &15& 4.5   &    &$-4$& 12  &$\le$ 6  & &       &   &  \\
   3   &60 &     &    &   & 3     &55  &22.5 &    & 4  & 68 & 35     & &  2    &88 & 52.5  \\
  17   &44 &     &    &   & 14    &36  & 11.5&    &    &     &       & & 25    &5  & 2.5  \\
 115   &   &     &    &   &      &     &     &    &115 &     &       & & 115      &   &   \\
 %S/N   &   &     &    &   &      &430  & 542 &    &    &     &       & &       &343&361 \\
\hline
\end{tabular}
\\

\end{footnotesize}
\label{default}
\end{minipage}
\end{table*}

%----------------------------------------------------------------------------------

\subsection{HD 75608}

HD 75608 at 524 pc lies slightly south of the HD 75387 at a similar distance. The SALT spectrum reveals that HD 75608 is a previously undiscovered shell star. Comparison of the K line profiles from 1996 by Cha \& Sembach and from SALT in 2017 is made in Figure 13. The SALT Na D$_2$ profile is also shown in this figure. Cha \& Sembach did not observe the Na D lines. Extraction of the components from the ISM and the SNR is complicated by the likely presence of a shell component and its likely variability. (In the SALT spectrum, the shell lines have a LSR velocity of $+51$ km s$^{-1}$.)

%13
\begin{figure*}
%\begin{minipage}{120mm}
\includegraphics[trim=0.0cm 0.0cm 0.1cm 0.0cm, clip=true,width=8cm,height=6.5cm]{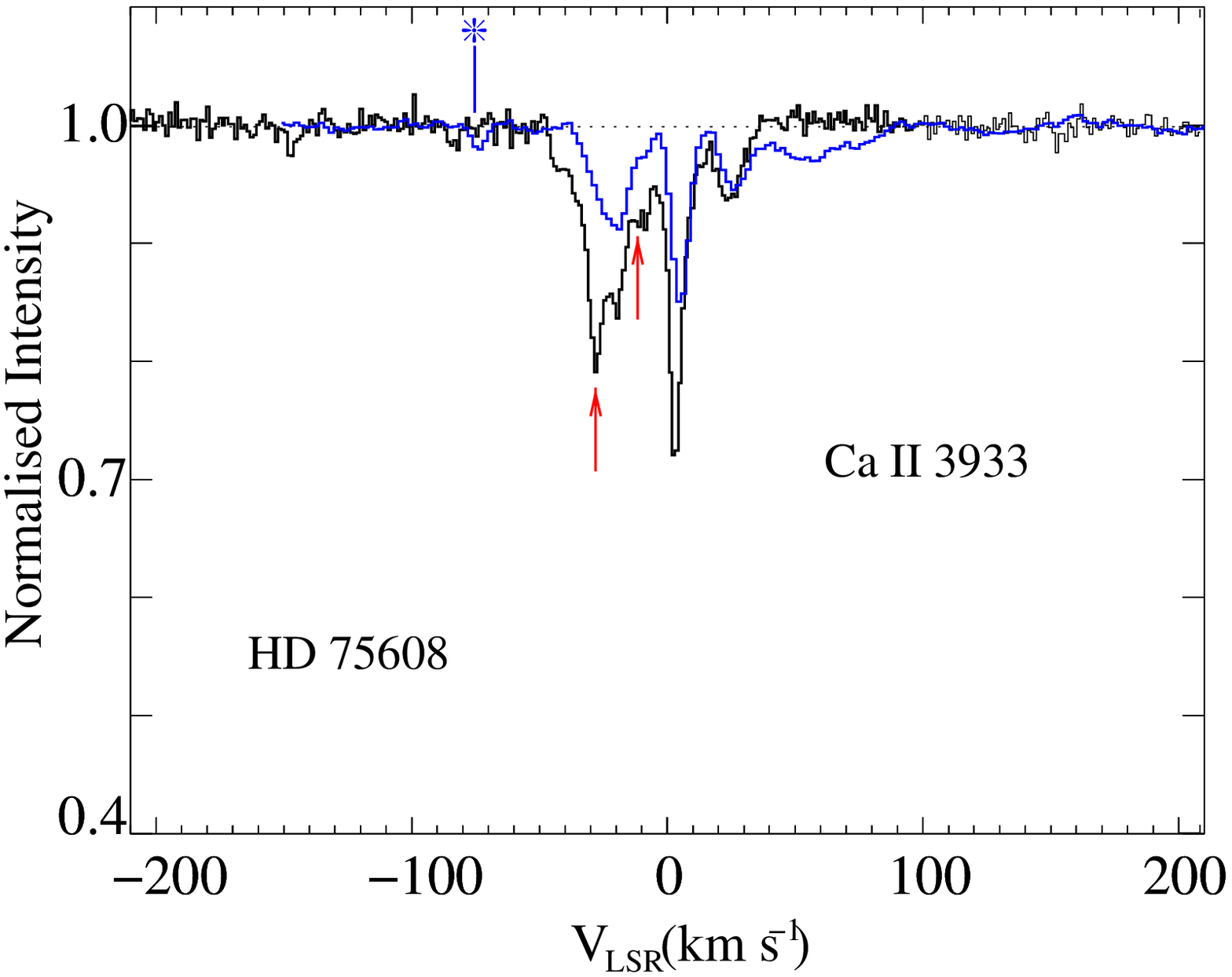}
\vspace{0.0cm}
\includegraphics[trim=0.0cm 0.0cm -0.4cm 0.0cm, clip=true,width=8cm,height=6.5cm]{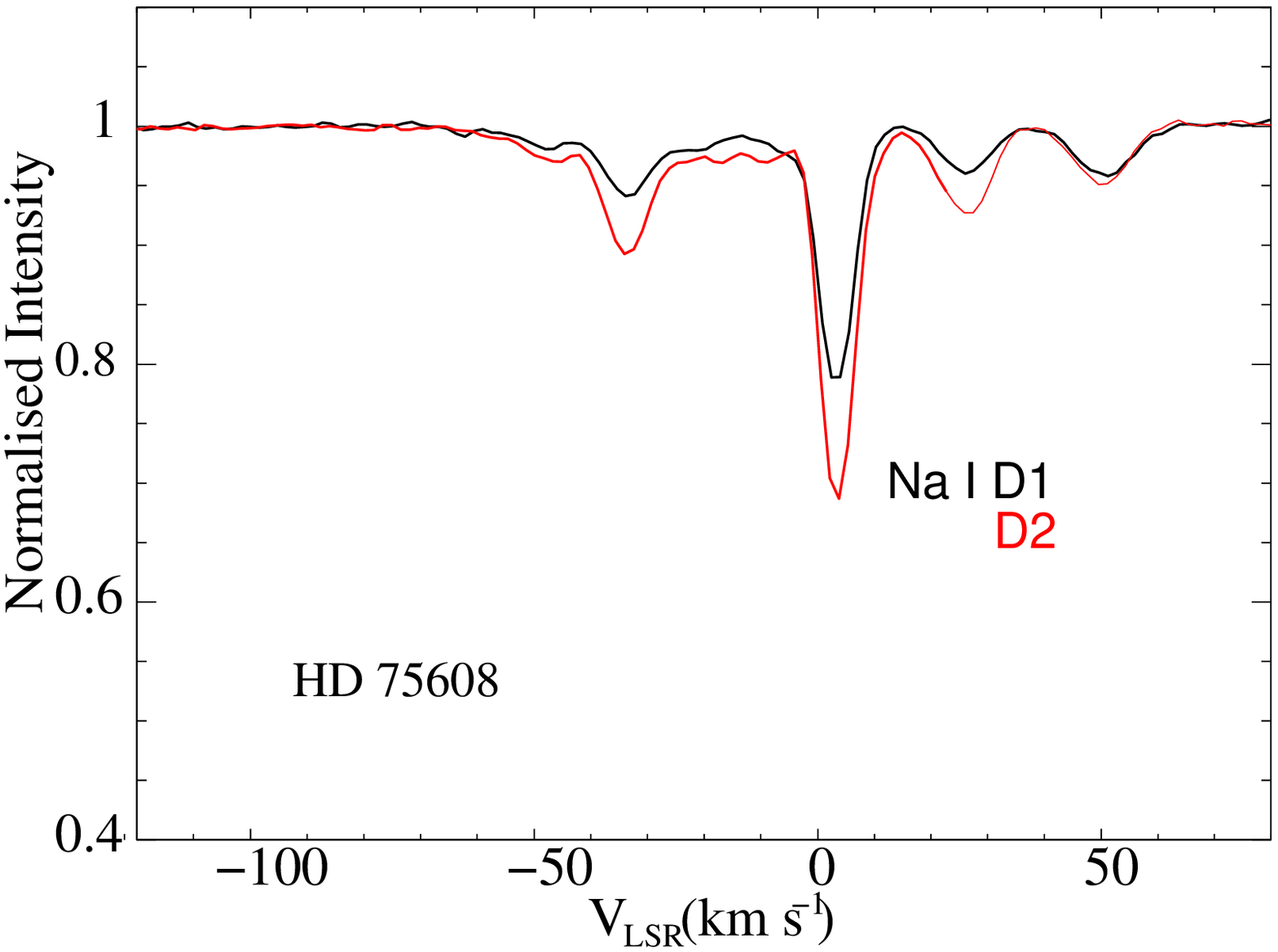}
\caption{(Left):  Ca\,{\sc ii} K   profile
 obtained in 1996 by Cha \& Sembach (2000) of HD 75608 is compared with the profile obtained on
 2017 May 7  with SALT (blue line).  Significant changes occurred between 1996 and 2017. HD 75608  appears to have a
 circustellar  gas shell/disk. (Right): SALT profiles of  Na\,{\sc i} D lines. No significant
 change between SALT and VBT profiles (Paper I) of D lines of  HD 75608.}
%\end{minipage}
\end{figure*}

%14
\begin{table*}
\centering
\begin{minipage}{170mm}
\caption{\Large ISM Absorption Lines of Ca\,{\sc ii} K and Na\,{\sc i}  towards HD 75608 }
\begin{footnotesize}
\begin{tabular}{lcrrrcrrccrrr}
\hline
\multicolumn{1}{c}{}&\multicolumn{2}{c}{Ca\,{\sc ii} K } &\multicolumn{1}{c}{}&\multicolumn{1}{c}{}&
\multicolumn{3}{c}{Ca\,{\sc ii} SALT} &\multicolumn{1}{c}{} & \multicolumn{3}{c}{Na\,{\sc i} (SALT)}&\multicolumn{1}{c}{}   \\
\cline{1-3} \cline{5-7} \cline{9-11}  \\
      & C\&S &    &  & &K & H &   &   &$ D_{\rm 2}$& $D_{\rm 1}$  \\
\cline{1-3} \cline{5-8} \cline{10-12}   \\
  $V_{\rm LSR}$ &$W_{\rm \lambda}$&$W_{\rm \lambda}$ & &$V_{\rm LSR}$&$W_{\rm \lambda}$ &$W_{\rm \lambda}$&  &$V_{\rm LSR}$& $W_{\rm \lambda}$ &  $W_{\rm \lambda}$     \\
  km s$^{-1}$ &(mA) &(mA) &  &km s$^{-1}$ &(mA)& (mA)& &km s$^{-1}$&(mA) &(mA)  \\
\hline
         &    &      &  &       &      &   &    &$-48$&6   &3.5       \\
 $-37$   & 8  &      & &       &      &   &    &$-33.5$&22&12          \\
 $-27$   & 21 &      & &  $-27$&4.3   &3  &    &     &     &           \\
 $-19$   & 14 &      & &  $-20$&11.5  &8  &    &$-19$& 5   &3         \\
 $-9$    & 12 &      & & $-13$ &4     & 3 &    &     &     &           \\
   4     & 26 &      & &   5   &17    & 9 &    & 3.3 & 46  &32        \\
  23     & 12 &      & &  26   &10    & 3 &    & 26  & 16  &8         \\
         &    &      & &       &      &   &    & 50  & 11  &9         \\
         &    &      & &       &      &   &    &     &    &            \\
 %S/N     &    &     & &       &300   &328&    &     &522 &522       \\
  %       &    &      & &       &      &   &    &     &    &          \\
\hline
\end{tabular}
\\
\end{footnotesize}
\label{default}
\end{minipage}
\end{table*}

% ------------------------------------------------------------------------------------------------------

\subsection{HD 75821}

                    The sight line to HD 75821 at a distance of 655 pc crosses the SNR to the south east of the central line.  HD 75821 = KX Vel is a double line spectroscopic eclipsing binary of high eccentricity with a period of 26.3 days  and masses of
    16.8 and 9.5 M$_{\odot}$ (Mayer et al. 2014). The radial velocity amplitudes are
    92.9 and 163.9 km s$^{-1}$) and the $\gamma$-velocity  of the system is
   V$_{\rm lsr}$ = 14.5 km s$^{-1}$.  Our SALT observations in 2017 correspond to the 0.20 phase of the binary system and
    the heliocentric radial velocities from four
    He\,{\sc i} lines are  $-35.0$ km s$^{-1}$ for the primary and +132.1 km s$^{-1}$ for
    the secondary.

The  K line in the SALT spectrum consists of a complex at about 0 km s$^{-1}$ and another  near $-100$ km s$^{-1}$ and a pair of weak lines near $+100$ km s$^{-1}$ (Figure 14 and Table 15). The D line profiles contain almost all of the K components but with different relative strengths. Of particular interest is the complexity of the profile near 0 km s$^{-1}$; this complexity  is rare across Cha \& Sembach's sample. One may speculate that the complexity results from the addition of one or more sharp lines from the interactions with the SNR to the common simpler profile from the local ISM.
Changes in the profiles were first noted by Cha \& Sembach who illustrate changes in the high-velocity components of the Ca\,{\sc ii} H \& K, and the Na\,{\sc i} D lines
    obtained in 1993 and 1996. In particular,  they show that the -98 km s$^{-1}$ component appears to have been accelerated with a slight change in equivalent width whereas the component at
   -85 km s$^{-1}$  decreased in equivalent width  from 66 m\AA\ in 1993 to 53 m\AA\ in 1996
    without a change in radial velocity. This trend is confirmed by Mayer et al. (2014)
    from spectra obtained in early 2005 and late 2011. Our SALT spectrum shows a continuation
   of the trend, namely the -98 km s$^{-1}$ component is accelerated by 2017 to -102 km s$^{-1}$ with no change in equivalent width and  the -85 km s$^{-1}$
   component became still weaker dropping from 53 m\AA\ in 1996 to 27 m\AA\ by 2017. The $-100$ km s$^{-1}$ is present and also variable in the D lines (Paper I).
   No trace of the -84 km s$^{-1}$ component is present  in the Na\,{\sc i} D lines.

                    The two weak components  at
    +84 and +106 km s$^{-1}$ seen in the 2011.9  spectrum from the VBT and the 2017 spectrum from the SALT  were not mentioned by
    Cha \& Sembach (2000) although a weak +84 km s$^{-1}$ component appears    in the illustration of their 1996 spectrum. However,  Danks \& Sembach (1995) do
    show the component at +83 km s$^{-1}$ in their tables; it is possible
    that  +106 km s$^{-1}$ component strengthened after 2005.
The Na\,{\sc i} D profiles  show six components fewer than in Ca\,{\sc ii} K profile.
    The -98 km s$^{-1}$ component shows the same amount of acceleration as for
    Ca\,{\sc ii} K line.   The equivalent width ratio of  Ca\,{\sc ii} K to Na\,{\sc i} D$_2$ is
    different for different components. An intriguing aspect of the Ca\,{\sc ii} H \& K profiles is that
    there are four pairs of negative and positive components equally spaced within $\pm$2 km s$^{-1}$) with respect to a central component at 0.4 km s$^{-1}$      as though they form a
 set of rings or a whirlpool around a central point.
                    The line of sight to HD 75821 provides a very puzzling situation.

% 14
\begin{figure*}
%\begin{minipage}{120mm}
\vspace{0.0cm}
\includegraphics[trim=0.0cm 0.cm 0.0cm 0.cm, clip=true,width=8cm,height=6.5cm]{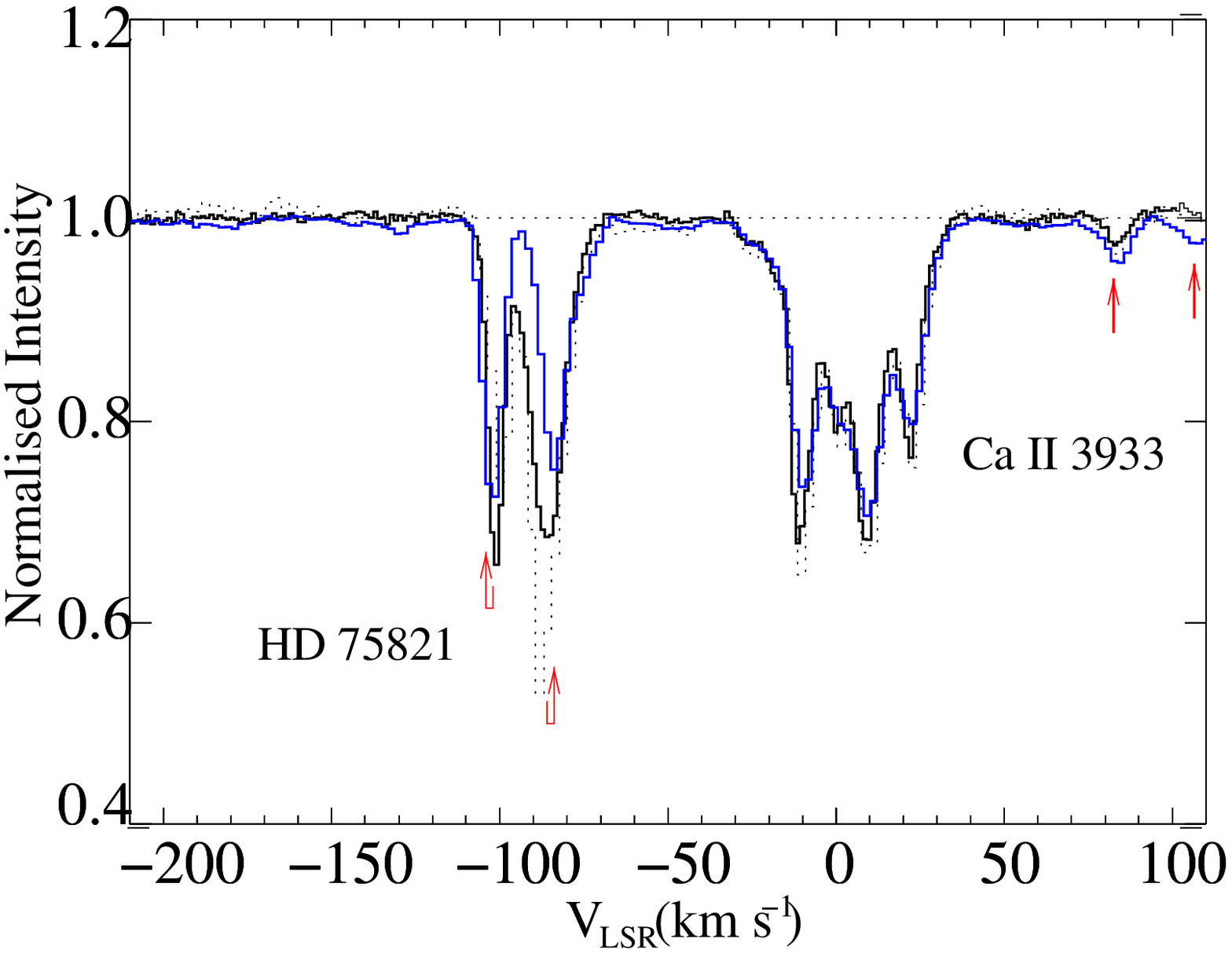}
\vspace{0.0cm}
\includegraphics[trim=0.0cm -0.2cm 0.0cm 0.0cm, clip=true,width=8.cm,height=6.5cm]{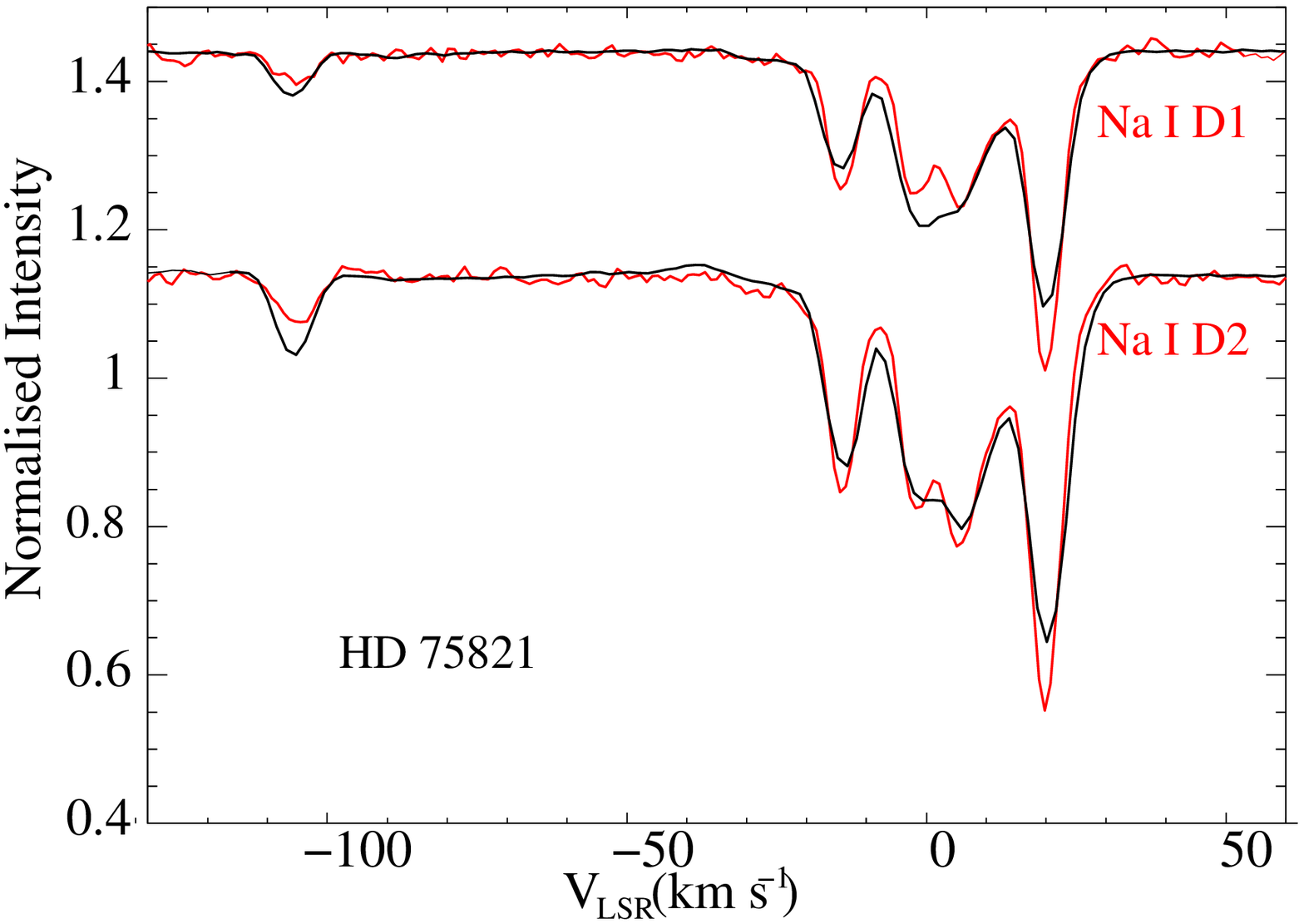}
\caption{(Left): The Ca\,{\sc ii} K   profile from the HD 75821 sight line obtained with SALT
  in 2017 is superposed on the profile obtained by Cha \& Sembach in 1993.
 (Right): The Na\,{\sc i} D$_1$ and D$_{2}$   profiles from the  HD 75821 sight line
 obtained in 2012 with the VBT (Paper I, red line) are compared with the profiles obtained on
 2017 April 17  with the SALT (black line). }
%\end{minipage}
%\endfigure=\end@float
\end{figure*}

% 15
\begin{figure*}
%\begin{minipage}{120mm}
\vspace{0.0cm}
\includegraphics[trim=0.0cm 0.0cm 0.0cm 0.cm, clip=true,width=13cm,height=9cm]{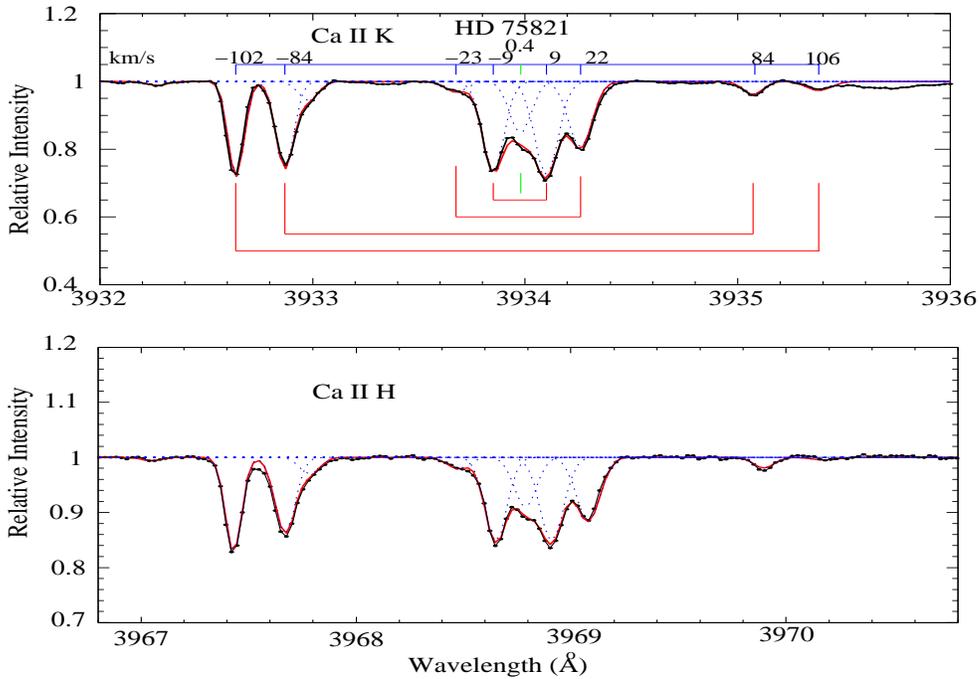}
\caption{  Ca\,{\sc ii}    profiles of HD 75821
 obtained in 2017 with SALT are shown with Gaussian fits to various components.
 Note the symmetric velocity spacing of components from the central component
 at 0.4 km s$^{-1}$ (marked by green line). Four pairs of equal negative and positive
 radial velocity components.
 2017 April 17  with SALT (blue line). }
%\end{minipage}
%\endfigure=\end@float
\end{figure*}

%15
\begin{table*}
\centering
\begin{minipage}{170mm}
\caption{\Large ISM Absorption Lines of Ca\,{\sc ii} K and  Na\,{\sc i} towards HD 75821  }
\begin{footnotesize}
\begin{tabular}{lcrrrcrrccrrrccr}
\hline
\multicolumn{1}{c}{}&\multicolumn{2}{c}{C\&S Ca\,{\sc ii} K }&\multicolumn{1}{c}{}&\multicolumn{1}{c}{}&
\multicolumn{3}{c}{Ca\,{\sc ii} (SALT)} &\multicolumn{1}{c}{} & \multicolumn{3}{c}{Na\,{\sc i} (C\&S)}&\multicolumn{1}{c}{}&\multicolumn{2}{c}{Na\,{\sc i} (SALT)}  &  \\
\cline{1-4} \cline{6-8} \cline{10-12} \cline{14-16}  \\
 1993  &     & 1996 &  &  & & K & H & &  & $ D_{\rm 2}$& $D_{\rm 1}$&   &&$ D_{\rm 2}$& $D_{\rm 1}$ \\
\cline{1-4} \cline{6-8} \cline{10-12} \cline{14-16}   \\
 $V_{\rm LSR}$ &$W_{\rm \lambda}$  & $V_{\rm LSR}$ &$W_{\rm \lambda}$ &   & $V_{\rm LSR}$ &$W_{\rm \lambda}$& $W_{\rm \lambda}$&  &$V_{\rm LSR}$  &
 $W_{\rm \lambda}$   & $W_{\rm \lambda}$ &  & $V_{\rm LSR}$&$W_{\rm \lambda}$&$W_{\rm \lambda}$   \\
km s$^{-1}$&(mA) &km s$^{-1}$&(mA)& &  km s$^{-1}$ &(mA)& (mA)& &km s$^{-1}$ &(mA) &(mA)& &km s$^{-1}$&(mA) &(mA)    \\
\hline
       &   &     &    &   &$-130$  &1.4&0.7  &    &    &     &    & &    &    &   \\
$-98$  &21 &$-99$& 26 &   &$-102$  &27 &16.5 &    &$-99$& 11 & 6  & &$-103$&16& 8  \\
$-85$  &66 &$-85$& 53 &   &$-84$   &27 &16   &    &    &     &    & &    &    &   \\
       &   &     &    &   &$-75  $ &6.5&3.5  &    &    &     &    & &    &    &   \\
       &   &     &    &   &$-23$   &4  &2.6  &    &    &     &    & &$-23$& 2 &2.5 \\
$-9$   &30 &$-10$&30  &   &$-9.6$  &33.5&19.5&    &$-11$&42  &28  & &$-12 $&44&27 \\
   1   &23 &  0  &22  &   & 0      &19.5&11.5&    & 0  &30   & 20 & & 0.5&46.5&37   \\
 10    &41 &  9  &39  &   & 10     &38 & 21  &    & 9  &78   & 43 & & 9  &62  &39   \\
 23    &28 & 22  &26  &   & 22     &27 & 16  &    &23  &76   & 51 & &22  &86  &60   \\
       &   &     &    &   & 84     &4.5& 2   &    &    &     &    & &    &    &   \\
       &   &     &    &   & 106    &4  & 1   &    &    &     &    & &    &    &   \\
       &   &     &    &   &        &     &     &    &    &     &    & &    &    &   \\
%S/N    &   &     &    &   &        &378  &     &    &    &     &    & &    &276&276   \\
%       &   &     &    &   &      &     &     &    &    &     &    & &    &    &   \\
%a\,{\sc i}& &   &    &   &      &     &     &    &    &     &    & &$-11.1$&1.3&   \\
%S/N    &   &     &    &   &      &     &     &    &    &     &    & &       &576&   \\
 %      &   &     &    &   &      &     &     &    &    &     &    & &    &    &   \\
\hline
\end{tabular}
\\
\end{footnotesize}
\label{default}
\end{minipage}
\end{table*}

% -----------------------------------------------------------------------------------------------------------

\subsection{HD 76566}

                 The line of sight to HD 76566 passes about the midpoint of the X-ray emission east of the Vela pulsar. With HD 76566 at 384 pc, the sight line probably extends through the SNR.
                  The  profile of Ca\,{\sc ii} K obtained with the SALT in 2017
    agrees with the 1994 observation of Cha \& Sembach (2000) except for
     the component at +9 km s$^{-1}$ (+6 km s$^{-1}$ according to Cha \& Sembach) which strengthened
     from 68 to 110 m\AA\ between 1994 and 2017 (Figure 16).  The Na\,{\sc i} D lines also show similar components with
     unchanged equivalent width between 194 and 2007.  Gaussian components are shown in the Table 16. This sight line is one of the few without high-velocity (positive or negative) components.

% 16
\begin{figure*}
% \begin{minipage}{120mm}
\vspace{0.0cm}
\includegraphics[trim=0.0cm 0.0cm 0.1cm 0.0cm, clip=true,width=8cm,height=6.5cm]{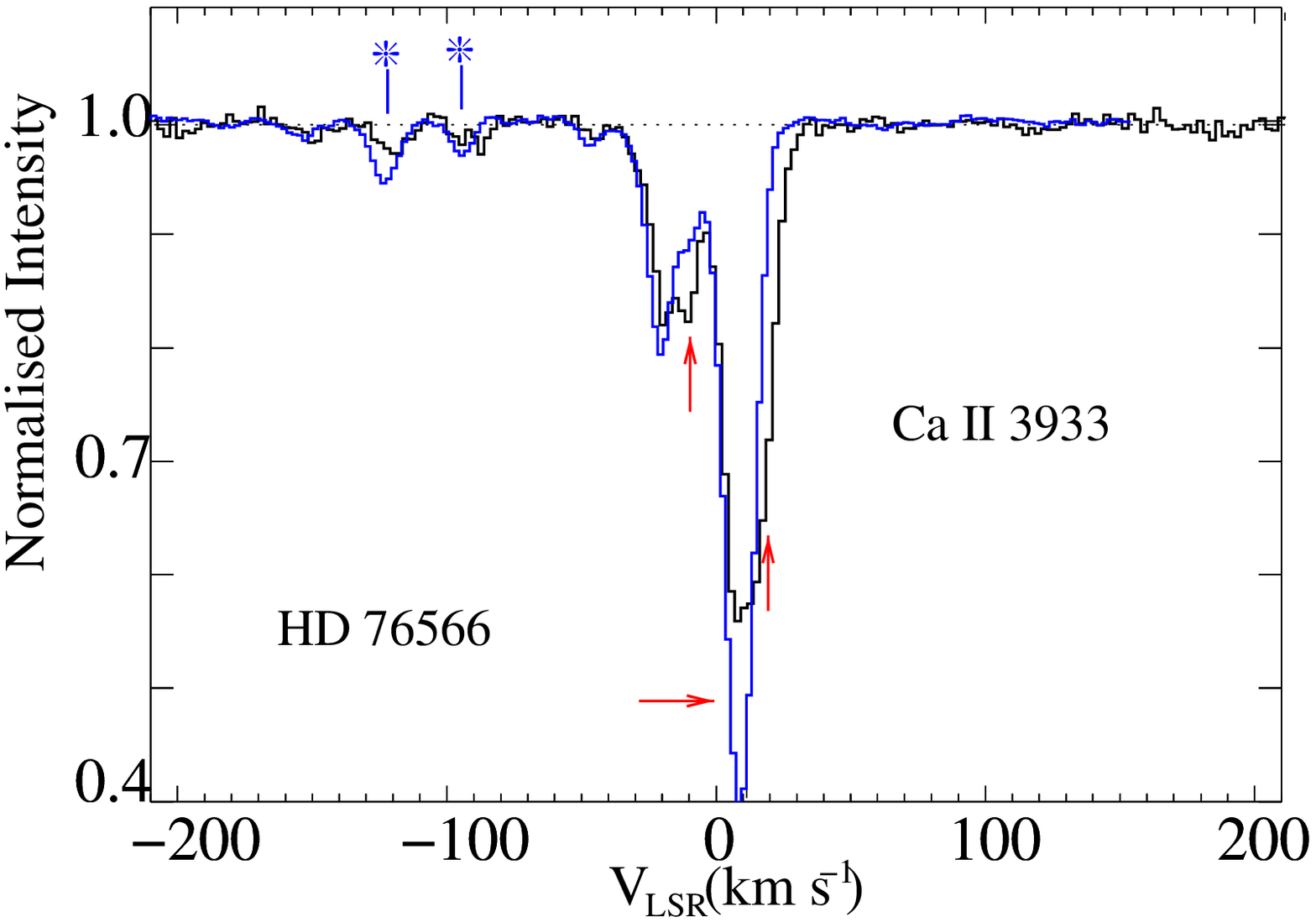}
\vspace{0.0cm}
\includegraphics[trim=0.0cm -0.2cm 0.1cm -0.5cm, clip=true,width=8cm,height=6.7cm]{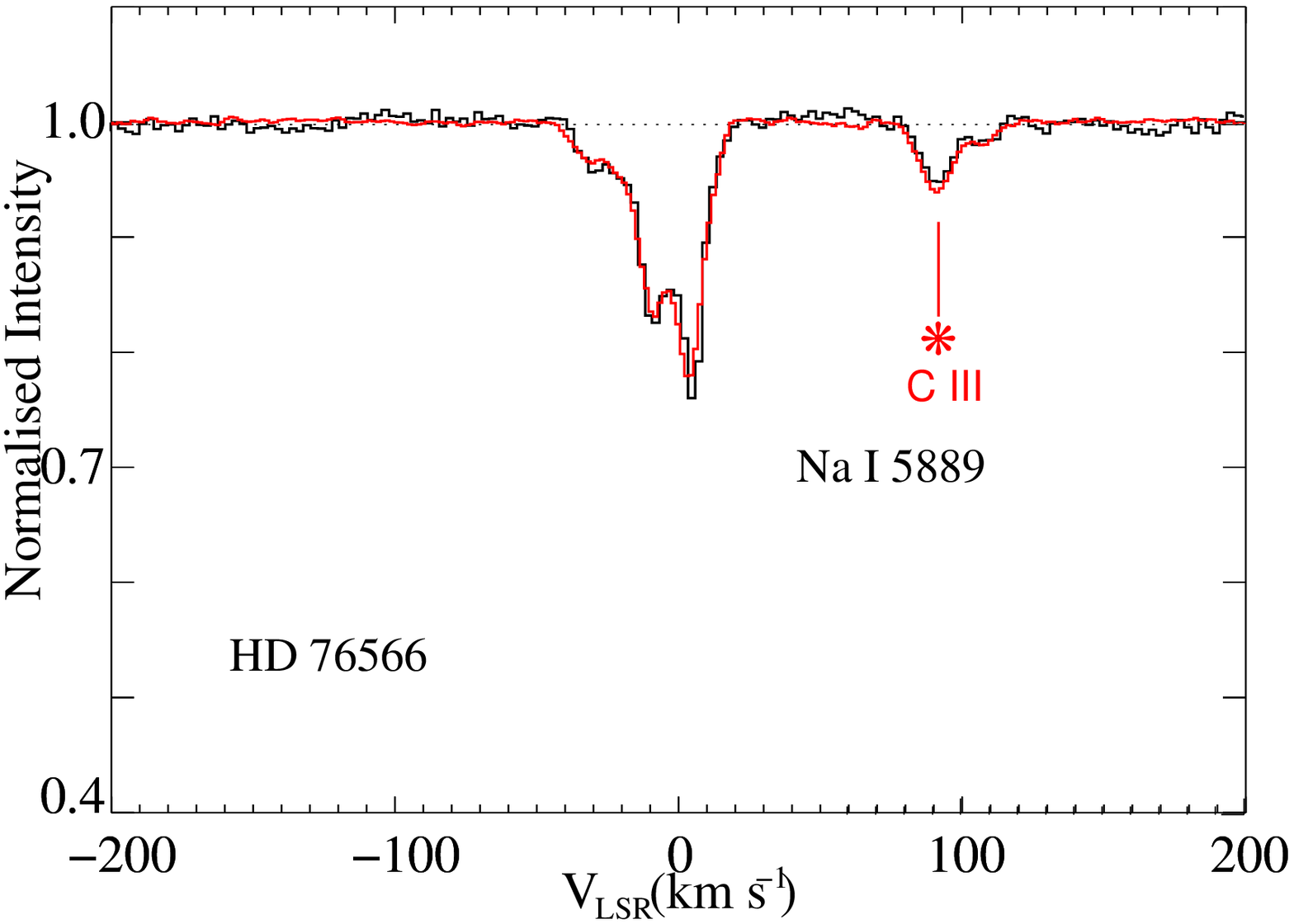}
\caption{(Left): The Ca\,{\sc ii} K   profile (black line) of HD 76566
  obtained in 1994 by Cha \& Sembach (2000) is compared with the profile obtained on
  2017 April 17  with SALT (blue line). The two features marked with stars are stellar
  S\,{\sc ii} lines. The component at 8 km s$^{-1}$
  strengthened considerably between 1994 and 2017.  A blended feature at about $-20$ km s$^{-1}$ weakened between 1994 and 2017. (Right): Na\,{\sc i}  D$_2$ profiles from SALT and VBT obtained
  on 2011 Apr 02.}
%\end{minipage}
\end{figure*}

%16
\begin{table*}
\centering
\begin{minipage}{170mm}
\caption{\Large ISM Absorption Lines of Ca\,{\sc ii} K and  Na\,{\sc i} towards HD 76566  }
\begin{footnotesize}
\begin{tabular}{lcrrrcrrccrrrccr}
\hline
\multicolumn{1}{c}{}&\multicolumn{2}{c}{C\&S Ca\,{\sc ii} K }&\multicolumn{1}{c}{}&\multicolumn{1}{c}{}&
\multicolumn{3}{c}{Ca\,{\sc ii} (SALT)} &\multicolumn{1}{c}{} & \multicolumn{3}{c}{Na\,{\sc i} (C\&S)}&\multicolumn{1}{c}{}&\multicolumn{2}{c}{Na\,{\sc i} (SALT)}  &  \\
\cline{1-4} \cline{6-8} \cline{10-12} \cline{14-16}  \\
 1993  &     & 1996 &  &  & & K & H & &  & $ D_{\rm 2}$& $D_{\rm 1}$&   &&$ D_{\rm 2}$& $D_{\rm 1}$ \\
\cline{1-4} \cline{6-8} \cline{10-12} \cline{14-16}   \\
 $V_{\rm LSR}$ & $W_{\rm \lambda}$ & $V_{\rm LSR}$ &$W_{\rm \lambda}$ &   & $V_{\rm LSR}$ &$W_{\rm \lambda}$&$W_{\rm \lambda}$&  &$V_{\rm LSR}$  &
 $W_{\rm \lambda}$   & $W_{\rm \lambda}$ &  & $V_{\rm LSR}$&$W_{\rm \lambda}$&$W_{\rm \lambda}$   \\
km s$^{-1}$&(mA) &km s$^{-1}$&(mA)& &  km s$^{-1}$ &(mA)& (mA)& &km s$^{-1}$ &(mA) &(mA)& &km s$^{-1}$&(mA) &(mA)    \\
\hline
       &   &     &    &   &$-45$   &3.5 &0.8 &    &    &     &    &  &    &    &   \\
       &   &     &    &   &        &    &    &    &    &     &    &  &$-30.5$&9.5&3   \\
 $-20$ &31 &     &    &   &$-20$   &31  &4   &    &$-25$&13  &$\le$6& &$-20$ & 5 &1   \\
 $-10$ &15 &     &    &   &$-9$    &14  &6   &     &$-9$ &39 &11  &  &$-10$&38 &14  \\
  6&68 &     &    &   & 8.7    &110 &50  &    & 5  &51   &20  &  &4    &53&32  \\
   15  &53 &     &    &   &        &    &    &    &    &     &    &  &     &  &    \\
       &   &     &    &   &        &    &    &    &    &     &    & &    &    &   \\
%S/N    &   &     &    &   &        &373&334  &    &    &     &    & &    &607&607   \\
\hline
\end{tabular}
\\\end{footnotesize}
\label{default}
\end{minipage}
\end{table*}

% -----------------------------------------------------------------------------------------------------------

\section{Temporal  variations of the line profiles}

Our collection of SALT spectra provides an opportunity to assess the frequency of profile variations over the several decade interval from 1993-1996 and 2017-2019. Our report supplements the examples of line profile variations discussed by Cha \& Sembach from spectra gathered over a much shorter time interval. Of our 15 stars, five are judged to have K line profiles unchanged over the interval 1993-1996 and 2017-2019: HD 71302. 72232, 74223, 74371 and 75608. HD 72232 is too close to Earth to be affected by the SNR and HD 75608 is a newly-discovered shell star with a K line profile too complex to securely isolate the SNR's contributions. The remaining ten stars show variations -- simple to complex - in the radial velocity and/or equivalent width of interstellar K line components at low and/or high-velocity.  An example of a line of sight with changes to several K line components is provided by HD 74234. At the other extreme is  HD 74979 and, of course, the few lines of sight where the K line profile is unchanged over the 20 year interval. Changes to the K lines and to the D lines for these ten lines of sight are discussed in detail in the next two sections. Table 17 summarizes the  principal changes to the K line profiles over the two decade interval.

Just five of our 15 stars provide Ca K and Na D profiles identical to those reported by Cha \& Sembach. Three of the four were noted by Cha \& Sembach as having K (and D) profiles with exclusively low velocity ($< |30| $ km s$^{-1}$) components. The exception in the quartet is HD 71302 with a red-shifted component at $+44$ km s$^{-1}$. One other star - HD 72232 -  at 200 pc is almost certainly projected in front of the SNR and, thus, the two K and the single D components originate in the ISM most likely  unaffected by the SNR. The remaining  two -- HD 74273 and 74371 -- have weak K profiles suggesting two or three low velocity components which could arise from the local ISM.

Two other stars in our sample of 15 were classified as lacking high-velocity components to the K line profile: HD 75608 and HD 76566. HD 75608's SALT spectrum shows it to be a shell star and apparently has K line profile differences between 1996 and 2017 with several components weakening considerably (Figure 13). HD 76566 lacks high-velocity  components in the K (and D) lines but shows  clear profile differences at the K line from 1994 to 2017 but no discernible changes in the D lines (Figure 16 and Table 16): the dominant component of the K line almost doubled in equivalent width and two minor components weakened between 1994 and 2017. The K line changes are attributed to an evolving interaction with the SNR. This small sample of stars lacking high-velocity components in the K lines  suggests that a lack of high-velocity components in K line profiles may not be assumed to represent absorption by the ambient ISM alone but are (or will be) subject to crafting by interactions between the ISM and the SNR.

High-velocity components to the K (and D) line profiles may be plausibly attributed only to interactions between the ISM and  the SNR. Four stars in our sample of 15 showed positive high-velocity components and nine showed negative high-velocity components according to Cha \& Sembach. (Five stars in Cha \& Sembach's large sample showed both positive and negative high-velocity components but none of these five were observed by SALT.)

Three of the four stars -- HD 72014, 72350 and 74234 -- in Table 1 for which Cha \& Sembach reported positive high-velocity K components exhibited  differences in the K line profiles when reobserved with SALT. HD 71302 completing this quartet showed essentially no change when reobserved with SALT. HD 72014  -- showed new K components at $+59$ and $+130$ km s$^{-1}$ with its low velocity complex unchanged between 1994 and 2018 with its maximum K line depth at about $+14$ km s$^{-1}$ (Figure 2). These `new' components do not appear in the D lines. For HD 72350 (Figure 4)  a K component at $+47$ km s$^{-1}$ approximately doubled in strength by 2018 with hints of additional fresh high-velocity but weak components at 70 to 120 km s$^{-1}$. As with HD 72014, low velocity absorption in the K profile from about 20 km s$^{-1}$ is unchanged between 1994 and 2018. HD 74234's K line profiles (Figure 7) are striking for the strong component at  about 70 km s$^{-1}$ which moved  7 km s$^{-1}$ to the red and weakened between 1996 and 2017. In addition, new weak K components appeared by 2017 at about $-40$ and $+47$ km s$^{-1}$. As for HD 72014 and 72350, a strong K blend reaching maximum depth at 5 km s$^{-1}$ is unchanged between 1996 and 2017. These common features near 0 km s$^{-1}$ are presumably dominated by contributions  from the local ISM unaffected by the SNR but some contribution here and in other lines of sight from the SNR may be present at low velocity too.

Six stars in Table 1 were noted by Cha \& Sembach  as having K line components at high negative velocities.  HD 73326 (Figure 5) has a lone high-velocity K line at $-46$ km s$^{-1}$ which with the complex near 0 velocity has remain unchanged between 1996 and 2017. HD 74194 showed a prominent K component at $-141$ km s$^{-1}$ in 1996 which was accelerated to $-146$ km s$^{-1}$ by 2019 with some increase in strength. Other changes were noted for the SALT spectrum (Figure 6) in components with velocities between $-60$ and $-80$ km s$^{-1}$ but  the very strong low-velocity complex around  $+8$ km s$^{-1}$ is unchanged from 1996. HD 74979 has a simpler K line profile than HD 74194: a strong unchanging blended double near 0 km s$^{-1}$ and a single high-velocity line with its velocity changing from $-87$ km s$^{-1}$ in 1996 to $-89$ km s$^{-1}$ in 2018.
 HD 75129 provides the most striking change to a high-velocity K line component: a strong
 line at $-67$ km s$^{-1}$ in 1996 has almost vanished by 2017 (Figure 11) while a weaker
 component at $-99$ km s$^{-1}$ is unchanged between 1996 and 2017. The 67 km s$^{-1}$ K line component is not seen in the D lines.  Such a drastic
 change in a K line might result from either a change in the degree of ionization of the gas or
 the gas cloud moving out of the line of sight during the 21 year period. A major decrease in
 ionization would likely result in an increase of the neutral atoms such as and Ca\,{\sc i} and Na\,{\sc i} I but  an absorption component in the  Na I D lines at about the same radial
  velocity of $-67$ km s$^{-1}$  is not observed. Increased ionization converting Ca$^{+}$ to Ca$^{++}$ might require local shock  heating, a possibility that
  can not be ruled out. It is also possible   that the gas cloud moved off the line of sight.

 HD 75387's high-velocity component at $-39$ km s$^{-1}$ in both the K and the D lines is unchanged over the 21 years. HD 75821 is the star in Table 1 with the most high-velocity (negative and positive) components (Figure 14). The component at $-85$ km s$^{-1}$, which was noted by Cha \& Sembach as weakening clearly between 1993 and 1996, was even weaker in 2017 but the component at $-100$ km s$^{-1}$ has kept an approximately constant equivalent width from 1996 to 2017.

Of the seven sight lines in Table 1 which  show Ca K line profiles differences between 1993-1996 and 2017-2019, all show high-velocity components -- positive or negative  -- and  differences found among some of these high-velocity components refer to combinations of changes of radial velocity and equivalent width. In brief, lines of sight showing high-velocity components to the K line appear to evolve on a time scale of roughly about 50 years, a time scale estimate which attempts to account for the observation that lines of sight lacking high-velocity components in 1993-1996 may also lack these components now.

This report on variable K line profiles with some accompanying parallel changes to the D lines is not the first for sight lines through the Vela SNR. Hobbs, Wallerstein \& Hu (1982) described a new installer component appearing in the spectrum of HD 72127A. From spectra covering a span of a few years, Danks \& Sembach (1995) identified two lines of sight and Hobbs et al. (1991) reported on one line of sight with variable K line profiles. Among their sample of 68 sight lines observed for their K lines in the 1993-1996 interval, Cha \& Sembach  (2000) found seven sight lines including confirmation of the three previous discoveries. Of the 13 variable components across the seven sight lines, five sight lines provided  only  high velocity variable components. Variable low velocity components ($< |30| $ km s$^{-1}$)  were found for two sight lines. Such variations at low radial velocity may result from energetic interactions within the SNR but observed at a high angle of inclination or from weak interactions occurring in or nearly along the line of sight.

High-velocity components have distinctive values of the ratio of the Ca$^+$ to the Na column densities. This was first shown by Jenkins, Silk \& Wallerstein (1976) who studied ultraviolet lines along four lines of sight with the {\it Copernicus} satellite and demonstrated that the high-velocity components were the result of shocks and suggested that interstellar grains in these shocked regions were destroyed by sputtering to release Ca atoms to drive the Ca$^+$/Na ratio to levels not seen in the diffuse ISM. Exploration of this ratio among high-velocity components in lines of sight through the SNR was continued by Danks \& Sembach (1995) who examined seven sight lines covered in the Cha \& Sembach (2000) catalog of 68 lines of sight.

These high velocity components in the Vela stars, often variable, differ dramatically from the
 three low-velocity components discussed in Paper II. In Paper I providing 2011-2012 VBT
 spectra of the Na D lines for many of the lines of sight observed in 1993-1996 by
Cha \& Sembach, three lines of sight were discovered in which  strong low-velocity D line
 components reported by Cha \& Sembach had weakened greatly in the VBT spectra. For these
three lines of sight, there are no high-velocity components. Paper II showed that in each
 of the three cases with dramatic changes to the D lines the Ca K line profile is weak and
 with a profile from SALT spectra obtained in 2015 unchanged from that reported
by Cha \& Sembach. Paper I did not explain satisfactorily the large changes to the Na column
 density with essentially no change to the Ca$^+$ column density. Such changes are the reverse
 of the behaviour among high-velocity components where changes to the K line are often
 accompanied by an unchanged D line profile. It is likely important to note that two of the
 three lines of sight in question fall outside the ROSAT contour used to define the SNR.
 Indeed, one (HD 63578) is associated with the $\gamma^2$ Vel wind bubble.  In the case of the
 third line of sight: the star HD 76161  at a distance of 271 pc may be located at the outer
 edge of  the SNR and in proximity to the region of interaction (shocked) of SNR and
  ISM. It is possible that the Lyman $\alpha$ generated in this recombining region could have
 ionized   Na\,{\sc i} in the sightline towards HD 76161, leaving Ca\,{\sc ii} unaffected.

\begin{table*}
\centering
\begin{minipage}{160mm}
\caption{\Large Highlights of the evolution of ISM Ca\,{\sc ii} K line profiles between 1993-1996 and
 2017-2019  }
\begin{footnotesize}
\begin{tabular}{|ll|}
\hline
\multicolumn{1}{c}{Star}   & \multicolumn{1}{c}{Comment}  \\

\hline
 HD 72014    & - New cloud at $+59$ km s$^{-1}$. High velocity component at $-137$ km s$^{-1}$
                is now at $-130$ km s$^{-1}$. \\
             &              \\
 HD 72350    & - Cloud at$+39$ km s$^{-1}$ has strengthened. \\
             &               \\
 HD 73326    & - Low velocity cloud at $+7$ Km s$^{-1}$ is stronger but this is a SB2. \\
             &              \\
 HD 74194    &  - Among several changes to high-velocity components one at $-141$ km s$^{-1}$ is
                 now at $-146$ km s$^{-1}$. \\
             &  ~ Decrease in strength of $-50$ km s$^{-1}$ component.  \\
             &          \\
 HD 74234    & - Complex profile changes at +ve and -ve high-velocities. The component at $+69$ Km s$^{-1}$ \\
             &  ~ has weakend and accelerated to $+73$ km s$^{-1}$. \\
             &         \\
HD 74979    & - The component at $-87$ km s$^{-1}$ has accelerated to $-89$ km s$^{-1}$. \\
             &           \\
 HD 75129    & - A component at $-67$ km s$^{-1}$ had almost vanished by 2017.  \\
             &              \\
 HD 75387    & - Clear changes to the principal K components near 0 km s$^{-1}$.  \\
             &            \\
 HD 75821    & - Profile variations near $-100$ km S$^{-1}$ between 1993 and 1996 have continued. \\
             & ~ $-98$ km s$^{-1}$ component has accelerated to $-102$ km S$^{-1}$ by 2017. \\
             &             \\
 HD 76566    & - Component near $+9$ km s$^{-1}$ has strengthened.  \\
             &            \\
\hline
\end{tabular}
\\\end{footnotesize}
\label{default}
\end{minipage}
\end{table*}

\section{Spatial variations of the line profiles}

Inspection of the 68 Ca K line profiles obtained by Cha \& Sembach (2000) reveals a variety of
 profiles without a clear pattern across the face of the {\it ROSAT} image or with distance to the
 observed star. The outline of the  {\it ROSAT} image and locations in the plane of the sky are shown in Figure 17 from Cha \& Sembach. Cha \& Sembach classified the profiles according to the presence of
 high-velocity  components. Four categories were provided according to the velocity of
 components in the line of sight: (i) [red filled circles] one or more components with a velocity
  $\geq +30$ km s$^{-1}$,  (ii) [blue squares] one or more components with a velocity
  $\leq -30$ km s$^{-1}$, (iii) [green triangles]  high-velocity components covering both intervals $\geq +30$ and
 $\leq -30$ km s$^{-1}$ and (iv) [plus signs] lines of sight with all components within the velocity
 limits $-30$ to $+30$ km s$^{-1}$.  All of the lines
 of sight ((i) - (iv)) contain  a blend of absorption components centred about
 on 0 km $^{-1}$ of varying degrees of complexity. The high-velocity components and  some
of the low-velocity components are attributed to interactions between the supernovae  and the ambient ISM. Sixteen of the 68 lines of sight are in group (i), 17 in group (ii),
 5 in group (iii) and 30 in group (iv). All but 8 of the 68 lines of sight are to stars
 within the ROSAT contour selected by Cha \& Sembach as defining the boundary of the Vela SNR.
 The ISM along the line of sight ahead of and beyond the SNR is responsible for other
 low-velocity components which might be expected to vary smoothly across the face of the SNR
 and likely to be at their simplest where the components from within the SNR are at a minimum
 or absent.

The SNR's contribution should be at a minimum, probably negligible, to the lines of sight
  to the five stars whose {\it GAIA} parallax indicates the star may be closer than the
 estimate to the SNR's  outer boundary. The quintet are HD 68243 at 281 pc (also outside
 the {\it ROSAT} contour), HD 71459 at 246 pc, HD 72232 at 200 pc and in Table 1, HD 74650 at 185
 pc towards the southern age of the {\it ROSAT} contour and HD 76161 near the southwest edge of the
 {\it ROSAT} contour.  Four of the five have K profiles consisting of two low-velocity components
 separated by about 10 km s$^{-1}$ which are plausibly attributed to the ISM ahead of the SNR.
 The exception is HD 71459 near the northwest edge of the {\it ROSAT} contour  with a K profile of
 three roughly equal components at $-11$, $+5$ and $+30$ km s$^{-1}$, the latter high-velocity
 component appears to differ from the low-velocity components attributed to the foreground
 ISM and is plausibly attributed to interaction with the SNR.

 Another set of Cha \& Sembach's stars are not projected on the {\it ROSAT} contour. Ten including
 HD 68243 above are to the west of the contour and of these three and possibly the entire set
  are so far beyond the contour that they are very likely to be   unrelated to the SNR.
 Two of these western outliers -- HD 63578 and HD 68217 -- were featured in Paper II  and
 shown to have similar low-velocity splitting of their Ca K profiles which were little changed
 between 1993-1994 and 2015 despite large weakening of the Na D profiles over the same interval.
 Several other western outliers show low-velocity K profiles quite similar to the pair from
 Paper II. A minority include a higher velocity component at about $+35$ km s$^{-1}$. The
 lone star HD 79275 well outside the eastern edge of the {\it ROSAT} contour also has a K profile
 suggesting two low-velocity components and a high-velocity component at $-28$ km s$^{-1}$. All
 of these outliers except one (HD 65814) are at distances of 330 to 620 pc, i.e., behind but
 close to the SNR. The exception is HD 65814 at 2030 pc. The Ca K profiles comprised of
 low-velocity components are plausibly attributed to the quiescent ISM and to lines of sight
which do not intersect the SNR.  By extension of the plausibility argument, the low-velocity
 components are assumed to be representative of lines of sight which do intersect the SNR; the
 depth of the SNR appears to be too shallow to erase by ionization or to strengthen by
 destruction of grains an ISM low-velocity component. An occasional high-velocity component
 may result from an interaction with the SNR. Our assumption is that K line profiles for lines
 of sight crossing the SNR and unaffected by interactions with the SNR will on average resemble
 the profiles of the forefront stars and those stars outside the {\it ROSAT} contours, i.e., two
 low-velocity components separated by about 10 km s$^{-1}$ and possibly a higher velocity
 components approaching $\pm$30 km s$^{-1}$.

The Ca K profiles from lines of sight through the SNR lacking in high-velocity components
 (plus signs in Figure 17) include contributions from the quiescent ISM including a fairly frequent appearance of a component near $-10\pm5$ km s$^{-1}$ and another
 around $+28\pm5$ km s$^{-1}$.\footnote{Several such lines of sight include components
 outside the limits defining this class according to Cha \& Sembach's Table 2. The most
 notable is HD 75241  whose K profile was resolved into ten components including two with
 a velocity less than $-30$ km s$^{-1}$ and two greater than $+30$ km s$^{-1}$ which would
 suggest a class (iii) line of sight.} This pair of `extreme' interstellar components may also
 result from interactions with the SNR. Final identification of their origin may require  a
 search for temporal variations. Unfortunately, our present collection of SALT observations
 of the Ca K line includes only three of these  lines of sight: HD 74273, 74371 and
 76566. Of this trio only HD 76566 shows changes in the K line profile  and  the major
 change  between 1994 and 2017 occurs in the principal component at $+9$ km s$^{-1}$ which
 strengthened after 1994 and such a change indicates that this component has a significant
 contribution from interactions with the SNR.

% 17
\begin{figure*}
% \begin{minipage}{120mm}
\vspace{0.0cm}
\includegraphics[trim=0.0cm -.0cm 0.0cm 0.cm, clip=true,width=13cm,height=9cm]{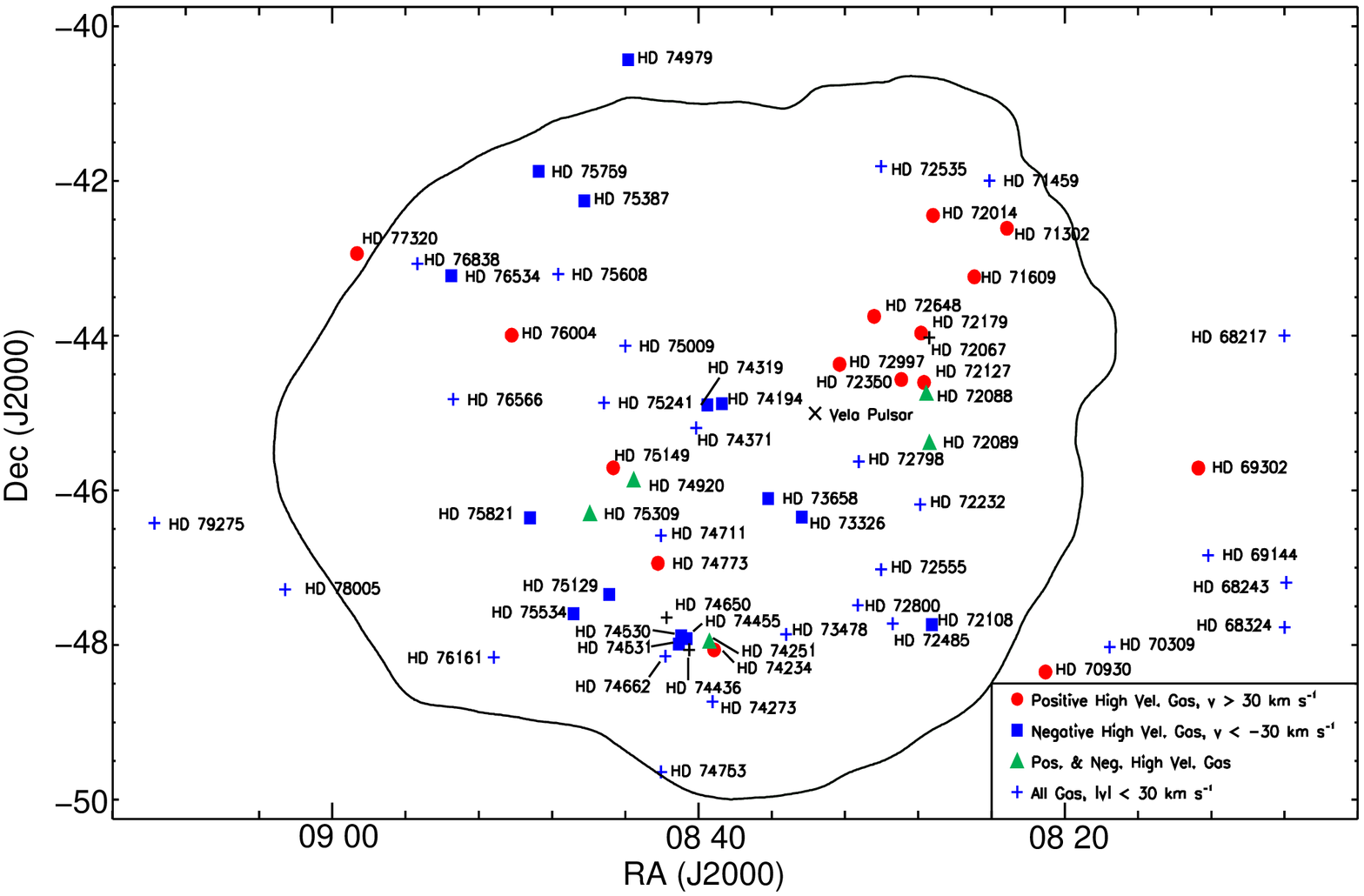}
\caption{ The location of the stars observed by Cha \& Sembach (2000) as projected on the sky. The contour is from {\it ROSAT}. The different symbols -- see key on the figure -- show the different velocity ranges found in the Ca\,{\sc ii} K profiles.  The cross refers to the
 location of Vela Pulsar.  }
%\end{minipage}
\end{figure*}

% 18
\begin{figure*}
% \begin{minipage}{120mm}
\vspace{0.0cm}
\includegraphics[trim=0.0cm -.0cm 0.0cm 0.cm, clip=true,width=13cm,height=9cm]{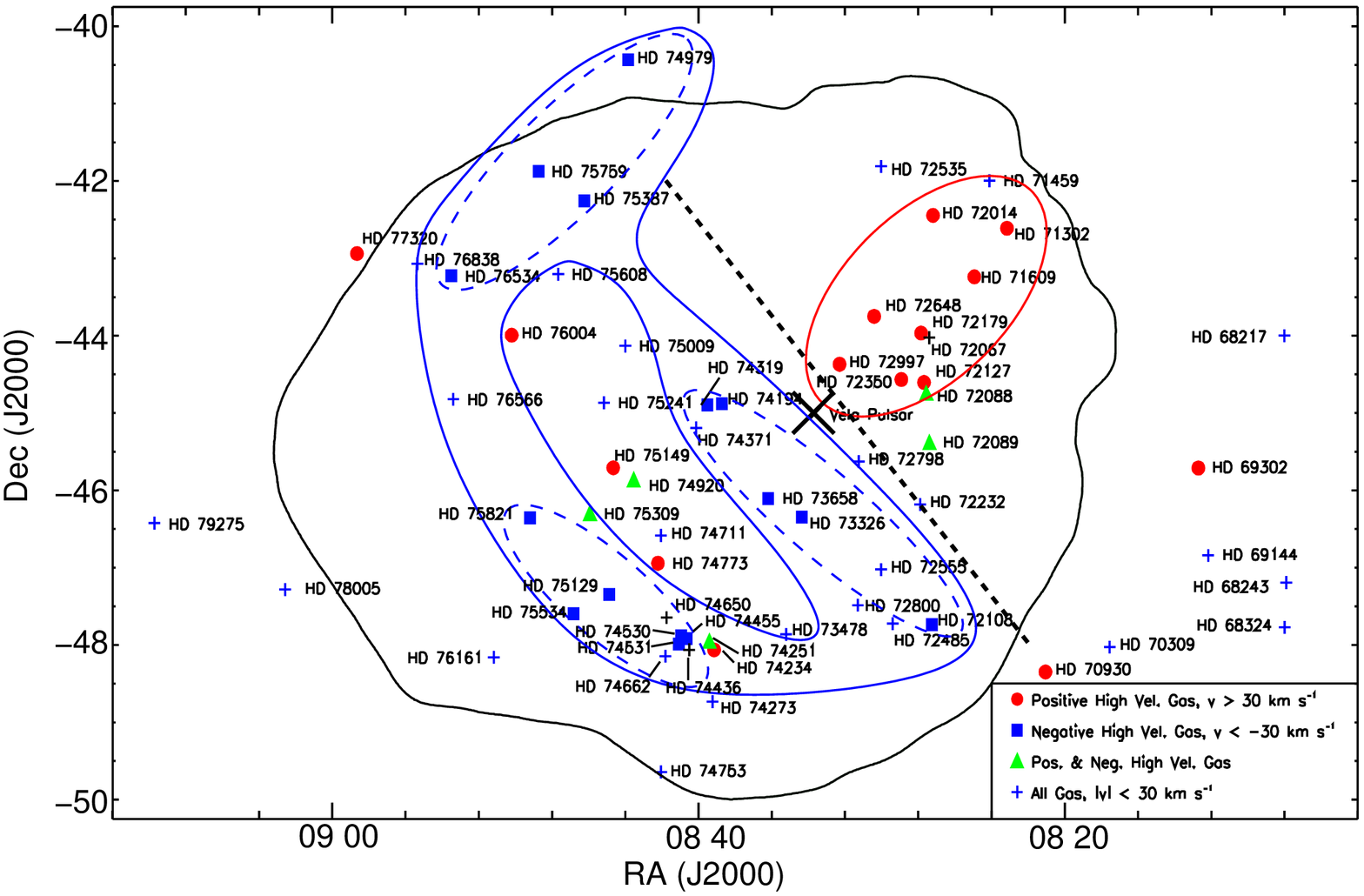}
\caption{ Location of the stars (Figure 17) observed by Cha \& Sembach (2000) with groupings of stars showing receding high-velocity clouds inside the red line and groupings of stars showing approaching high-velocity clouds inside the blue `donut'. The  dashed demarcation line through the
 pulsar  separates approaching and receding high-velocity clouds. }
%\end{minipage}
\end{figure*}

Examination of the set of Ca K (and Na D) line profiles across and beyond the {\it ROSAT} contour
 should on the large scale reveal an influence of the supernova's geometry on the dynamics
 of the remnant. Inspection of Figure 18 shows that the locations of
 the high positive (red dots) and high negative (blue squares) velocity appear to fall into
 regions bounded by either the red  line or situated within the blue `donut'.  The obvious red contour
 lies to the north-west of the pulsar. There are no lines of sight with strong velocities of
 approach in this area but there are several nearby lines of sight with an absence of high-velocity
 components. Lines of sight
 with a component of high velocity of approach are distributed in three zones  shown by dashed lines and together approximating a
 ring or a distorted `donut'. One is led to suggest that these areas of high positive and negative velocity  may
originate from the supernova (i.e., the pulsar which should not have moved over from the supernova's
location). Perhaps, the ejecta did not enter the surrounding ISM  in  a spherically symmetric
 fashion. A caveat: there is a spread in  velocity within each zone including lines of sight
 lacking any high velocity components suggesting that there may be a local component to the
 radial velocity in addition to a projection effect such that a radial velocity is not the
 total velocity of a component within the SNR.

 The observation that the spread of velocities may differ between nearby stars suggests that
 local components are significant and may dominate over a large-scale expansion driven by the
 blast from the supernova. To investigate these local components, we examine a few samples of
 stars appearing in close proximity on the map of the SNR (Figure 17). Some samples
 combine stars in Table 1 with stars taken from Cha \& Sembach while  others are drawn entirely
 from observations by Cha \& Sembach. Given that differences between contemporary observations
 of stars in Table 1 and generally very similar to those for the same stars by Cha \& Sembach
  from 1993-1996 our conclusions are unlikely to be seriously affected by combining observations
 separated by several decades.

    Our first sample includes HD 71302 and HD 72014 in Table 1 and three other stars in the
 north-west red ring (Figure 18). Three of the stars -- HD 71302, 71609 and 72014 --
 have a high-velocity recession component with a very similar velocity ($+44$, $+48$ and $+45$
 km s$^{-1}$, respectively) and consistent equivalent widths. Moreover,  Cha \& Sembach
 resolved the three K line profiles into Gaussian components finding that six components were
 demanded in each case with a close correspondence across the sight lines with velocities from
 about $-20$ km s$^{-1}$ to $+46$ km s$^{-1}$.  With distance of 549 to 1475 pc the lines of
 sight sample the entire SNR as well as the ISM ahead of and beyond the SNR. Angular distances
 from HD 71302 are 0.7 degrees for both stars.  The next two neighbouring lines of sight with
 high-velocity components and probing the entire SNR are to HD 72179 and HD 72648 at angular
 distances of 1.6 and 1.8 degrees from HD 71302. For HD 72648, the K line profile's components
 are similar to the three other stars but lacks the component at $-20$ km s$^{-1}$. HD 72179's
 line profile is strikingly different from that displayed by the other stars: Cha \& Sembach
 resolve it into 11 components between $-33$ and $+127$ km s$^{-1}$. Two lines of sight laking
 high-velocity components run just north of HD 71302 and HD 72014. One -- HD 71459 -- at 246 pc
 likely is in the SNR foreground. The other HD 72535 at 770 pc and 1.5 degrees from HD 71302
 has a line of sight sampling the SNR and the ISM and shows four of the velocity components
 from $-9$ to $+29$ km s$^{-1}$ seen in the HD 71302, 71609 and 72014 suggesting that the trio's
 components at $-20$ and $+46$ km s$^{-1}$ arise from interactions within the SNR. The sample
 is small but suggests that paths through the SNR in this part of the SNR may be similar at
 angular distances of up to about 0.7 degrees.  Obviously, a more thorough sampling of the
region is desirable.

In other parts of the SNR, angular separations smaller than 0.7 degrees correspond to major
 changes in the K line profiles but in some cases differences in the length of the lines of
 sight may be a contributing factor. In Figure  17  HD 74194 at 2360 pc and HD 74319 at
 543 pc are separated by just 0.15 degrees but their K line profiles are quite different.
 HD 74194's is shown in Figure 6  to include about five components with velocities less
 than $-30$ km s$^{-1}$ but  HD 74319's profile, as shown by Cha \& Sembach,  consists of only
 components at $-10$ and $+11$ km s$^{-1}$. A similar pair -- HD 73326 at 802 pc and HD 73658
 at 2318 pc -- separated by 0.35 degrees has a weak high velocity ($-126$ km s$^{-1}$)
 component for  the more distant star but otherwise the component structure of the K lines is
 quite similar. In a contrasting pair -- HD 75129 at 1038 pc and HD 75534 at 2065 pc -- the
line of sight to the closer star has the component with the extreme velocity of
 $-98$ km S$^{-1}$ and a component at $-67$ km s$^{-1}$ which shows a dramatic weakening
between 1996 and 2017.

A region of the Vela SNR in the southern part of the {\it ROSAT} contour contains eight stars
 within about 0.3 degree. One star -- HD 74650 at 185 pc -- is in the SNR's foreground and as
 expected its K line profile comprises just components at $-1$ and $+6$ km s$^{-1}$ from
the ISM. Six of the other seven stars are at distances of 730 to 830 pc. The eighth star
 at 1360 pc is well behind the SNR. All K lines show a comparable component at
about $+6$ km s$^{-1}$ and also at about $-6$ km s$^{-1}$, both of which may be confidently
 identified with the undisturbed ISM of which parts contributed to the line profile seen in
 the nearby star HD 74650. The striking feature of this octet is that the collections of
high-velocity components seen in five of the seven lines of sight are not duplicated in another
 star even though the angular separations are less than 0.3 degrees and distances to six of the
 seven stars are within $\pm50$ pc. And similarly, there are within this small area,
 two stars -- HD 74436 at 1362 pc and HD 74662 at 761 pc -- lacking high-velocity components.
 Thus, this closely packed octet confirm the earlier suggestions that the SNR's interactions
 with the ISM results generally in small-scale disturbances of the ISM. In optical spectra, such interactions are most easily detected  as high-velocity absorption components within the Ca\,{\sc ii} H and K line profiles but components from these interactions may also exist at low velocities where the H and K profiles are dominated by contributions from the unaffected ISM.

\subsection{N(Ca\,{\sc ii})/N(Na\,{\sc i}) ratios}

Previous studies of the high-velocity components along lines of sight through the Vela SNR
 have shown that the ratio of the K line's equivalent width to the D lines equivalent widths
 is higher than in the low-velocity components arising from the undisturbed
 ISM -- Sembach \& Danks (1994) who considered spectra included in the
 Cha \& Sembach (2000) study  including several lines of sight reobserved here. For
 high-velocity components, the ratio of the Ca\,{\sc ii} to the Na\,{\sc i} column densities
 was about 20 times that for low-velocity components with quite a scatter from one line of
 sight to another. -- see also  Jenkins et al. (1984) and  Jenkins \& Wallerstein (1995).
 A possible explanation but not the sole explanation is that the supernova blast wave's
 interaction with the ISM destroys dust grains and releases Ca (and other non-volatile
 elements) into the gas and so increases the Ca/Na ratio in the gas. This mode of enhancement
 likely competes with an enhancement  of ionization converting neutral Na  to Na$^+$ at a
 greater rate than Ca$^+$ is converted to Ca$^{2+}$. Here, the key point is to search among
 low-velocity components within the K and D lines for ones indicating a high ratio of
 Ca\,{\sc ii}/Na\,{\sc i} indicative of gas which has interacted with the SNR.

       Inspection of Figures 1 to 16 and Tables 2 to 16 shows, as anticipated, that while
 the high-velocity components are detected in the K line few  are seen in the D lines.
 For
 the 15 lines of sight in Table 1 and for each of the components listed in the Table
 corresponding to a given line of sight, we estimated the ratio of the equivalent
 widths $W_\lambda$ of the K and D$_2$ line which, if the line of sight is optically thin,
 is also the ratio of the column density of Ca\,{\sc ii} to Na\,{\sc i}. Optical depth is
 estimated from the equivalent width ratio of the K to the H line for Ca$^+$ and the ratio
 of the D$_1$ to the D$_2$ line for neutral Na.  For a few lines of sight, the Na D lines
 are saturated and decomposition within the ISM complex near 0 km s$^{-1}$ is difficult.
 For a few lines of sight, a stellar Ca\,{\sc ii} K line dominates the K line. For a majority
 of the lines of sight, high-velocity K line components, the matching D lines are not
 detectable. An estimate of the upper limit for the D$_2$ line is obtained following
 Hunter et al. (2006) who provide the 3$\sigma$ detection limit
 as 3$\Delta\lambda_{inst}$/(S/N)
 where $\Delta\lambda_{inst}$ is the instrumental  FWHM resolution which for SALT at the
 D lines is 0.08\AA. Then for a representative S/N our SALT spectra provide a detection
 limit for a D line of  $\leq$ 1 m\AA.

                   Our appraisal of the the K/D$_2$ equivalent width ratio was conducted
 with two goals: first, to note the change in the ratio for those high-velocity components
 whose K line equivalent width changed appreciably between values reported by Cha \& Sembach
 and the value provided by our SALT spectrum; second, to search for K line components at
 low velocity with a K/D$_2$ equivalent width ratio representative of the high-velocity
 components. For the undisturbed ISM, the K/D$_2$ equivalent width ratio is generally less
 than one. For high-velocity components, the ratio is invariably greater than two and very
 much higher with a clear detection of the K line and  undetectable D lines.

                                Notable changes in high-velocity components, as set out in
 Section 3, occurred for the lines of sight to the following stars and velocities
 (in parentheses following the star's name): HD 72014 ($+130, +59$ km s$^{-1}$),
  72350 ($+39$ km s$^{-1}$), 74194 $-61$ km s$^{-1})$, 74234 ($+69, -40$ km s$^{-1}$),
  75129 ($-67$ km s$^{-1}$) and 75821 ($-84$ km s$^{-1}$). Changes include both increases
  and decreases in the K line's equivalent width between 1993, 1996 and 2017-2019. An
  increase in one component may be paired with a decrease in another high-velocity component
  in the same line of sight (HD 74234).

                      Among low-velocity components, the clearest example of a change
     between 1933-1996 and 2017-2019 is seen for the dominant contribution to the K line
    in HD 76566 (Figure 16) whose profile narrowed and deepened from 1994 to 2017. The total
     equivalent width seems not have changed. The velocity change is larger than shifts
     reported here among high-velocity components but may be related because the component
    which shifted to the blue lacked a Na D component as shown by the lack of a change in
   the D line between 1994 and 2017 (Table 16). Other but minor changes to the K line at
    low velocity may have occurred at low velocity for HD 74234 and HD 75387 and at a
       velocity just in excess of $+30$ km s$^{-1}$, the  boundary between low and high
       velocity, for HD 72350 and HD 74234. Given that the arbitrary definition of a
        low-velocity component spans only 60 km s$^{-1}$ and  high-velocity components
         span at a minimum an interval of 140 km s$^{-1}$, the detection of one
     strong K-- weak D at low-velocity  among our 15 lines of sight may not be at odds
     with an expectation that components resembling the high-velocity components are present
      among the complex lines arising at low velocities from undisturbed parts of the ISM.
      Most of the variable components show highier values of column density ratio
        of Ca\,{\sc ii} to Na\,{\sc i}  thus confirming that they are affected by
      supernova shock interactions.
     Improved understanding of the variety among the low-velocity components and thorough
    definition of conditions in the high-velocity components must await  high signal-to-noise     high-resolution ultraviolet spectra.

\section{Concluding Remarks}

                 The present study has shown that variations in the interstellar Ca\,{\sc ii}
 H and K line profiles and to a lesser extent in the Na\,{\sc i} D lines are a common
 occurrence in stars situated behind the Vela SNR. A typical timescale for variations is a
 few years and small angular separations between stars may result in appreciable profile
 differences. There is now evidence for clouds moving into a line of sight or being created.
 Acceleration of clouds has been observed, particularly near the edge of the SNR.
  The present study also shows  a puzzling set of absorption components along the
  sightline to HD75821 suggesting a possible  whirlwind.
  Detailed
 and  repeated observations of  more lines of sight covering  the 8 degree wide SNR  would
 be of interest: extending our sample of 15 of the 68 sight lines covered by
 Cha \& Sembach (2000) is likely needed to understand the the mode of ejection of the
 supernova
 material  and kinematics of the SNR. But optical
 absorption line spectroscopy provides few probes of an SNR where temperatures and pressures
 may be considerably higher than in the undisturbed ISM. A thorough application of
 ultraviolet spectroscopy will be necessary. Insights into the SNR have already been provided
 by rather limited ultraviolet spectroscopy. Among highlights from this spectroscopy, we may
 mention a rich collection of ions from \emph {IUE} spectra (Jenkins, Silk \& Wallerstein 1984),
 exceptionally strong lines from excited terms of the C\,{\sc i} $^3$P ground term
 (Jenkins et al. 1984: Nichols \& Slavin 2004; Jenkins \& Wallerstein 1995;
 Jenkins et al. 1998) in low-velocity and high-velocity gas, and detections of C\,{\sc iv}
 from \emph {IUE}
 and O\,{\sc vi} lines from {\it FUSE} spectra (Slavin, Nichols \& Blair 2004) again in
 low and high-velocity components. A clear impression gained from such observations and
 theoretical analysis is of gas shocked by the supernova. Although description of the
 physical conditions demands additional ultraviolet spectroscopy illumination of the
 temporal and spatial variations is better pursued with optical spectroscopy where telescope
 time is more freely (and more cheaply) available than from space.

\section{Acknowledgements}

We thank the SALT astronomers for their considerable help in conducting the HRS
 observations for us. We also would like to express our appreciation of the
 helpful and encouraging comments by the referee Dr. Ken Sembach.
 This research has made use of the SIMBAD database, operated
at CDS, Strasbourg, France.



\begin{thebibliography}{99}
\bibitem[\protect\citeauthoryear{Author}{2012}]{Author2012} 
Arellano Ferro, A., Garrison, R.F., 1979, RMxAA, 4, 351
\bibitem[\protect\citeauthoryear{Author}{2012}]{Author2012} 
Aschenbach, B., Egger, R., Trumper, J., 1995, Nature, 373, 587
\bibitem[\protect\citeauthoryear{Author}{2012}]{Author2012} 
Boiss\'{e}, P., Rollinde, E., Hily-Blant, P., et al., 2009, A\& A, 501, 221
\bibitem[\protect\citeauthoryear{Author}{2012}]{Author2012} 
Bramall, D. G., Sharples, R., Tyas, L., et al. 2010, in Society of Photo-Optical Instrumentation Engineers (SPIE) Conference Series, Vol. 7735,  4R
\bibitem[\protect\citeauthoryear{Author}{2012}]{Author2012} 
Cha, A.N., Sembach, K.R., 2000,  ApJS, 126, 399
\bibitem[\protect\citeauthoryear{Author}{2012}]{Author2012} 
Cha, A.N., Sembach, K.R.,  Danks, A.C., 1999, ApJ, 515, L25
\bibitem[\protect\citeauthoryear{Author}{2012}]{Author2012} 
Crause, L.A., Sharples, R., Bramali, D., et al. 2014, SPIE, 9147E, 6
\bibitem[\protect\citeauthoryear{Author}{2012}]{Author2012} 
Danks, A.C., Sembach, K.R., 1995, AJ, 109, 2627
\bibitem[\protect\citeauthoryear{Author}{2012}]{Author2012} 
Dodson, R., Legge, D., Reynolds, J. E, McCulloch, P. M., 2003, ApJ, 596, 1137
\bibitem[\protect\citeauthoryear{Author}{2012}]{Author2012} 
Garrison, R.F., Hiltner, W.A., Schild, R.E., 1977, ApJS, 35, 111
\bibitem[\protect\citeauthoryear{Author}{2012}]{Author2012} 
Hobbs, L.M., Ferlet, R., Welty, D.E., Wallerstein, G., 1991, ApJ, 378, 586
\bibitem[\protect\citeauthoryear{Author}{2012}]{Author2012} 
Hobbs, L.M., Wallerstein, G., Hu, E.M., 1982, ApJ., 252, L17
\bibitem[\protect\citeauthoryear{Author}{2012}]{Author2012} 
Hunter, I., Smoker, J.V., Keenan, F.P., et al.,  2006, MNRAS, 367, 1478
\bibitem[\protect\citeauthoryear{Author}{2012}]{Author2012} 
Jenkins, E.B., Tripp, T.M., Fitzpatrick, E.L. et al., 1998, ApJ, 492, 147L
\bibitem[\protect\citeauthoryear{Author}{2012}]{Author2012} 
Jenkins, E.B., Wallerstein, G., Silk, J. 1976, ApJS, 32, 681
\bibitem[\protect\citeauthoryear{Author}{2012}]{Author2012} 
Jenkins, E.B.,  Wallerstein, G., Silk, J., 1984, 278, 649
\bibitem[\protect\citeauthoryear{Author}{2012}]{Author2012} 
Jenkins, E.B., Wallerstein, G., 1995, ApJ, 440, 227
\bibitem[\protect\citeauthoryear{Author}{2012}]{Author2012} 
Lef\`{e}vre, L., Marchenko, S.V., Moffat, A.F.J., Acker, A., 2009, A\& A, 507, 1141
\bibitem[\protect\citeauthoryear{Author}{2012}]{Author2012} 
Mayer, P., Drechsel, H., Irrgang, A., 2014, A\&A, 565, 86
\bibitem[\protect\citeauthoryear{Author}{2012}]{Author2012} 
Nichols, J.S., Slavin, J.D., 2004, ApJ, 610, 285
\bibitem[\protect\citeauthoryear{Author}{2012}]{Author2012} 
Price, R.J., Crawford, I.A., Barlow, M.J., 2000, MNRAS, 312, L43
\bibitem[\protect\citeauthoryear{Author}{2012}]{Author2012} 
Rao, N.K., Sriram, S., Jayakumar, K., Gabriel, F., 2005, JApA,  26, 331
\bibitem[\protect\citeauthoryear{Author}{2012}]{Author2012} 
Rao, N.K., Muneer, S., Lambert, D.L., Varghese, B.A., 2016, MNRAS, 455, 2529 (paper I)
\bibitem[\protect\citeauthoryear{Author}{2012}]{Author2012} 
Rao, N.K., Lambert, D.L., Reddy, A.B.S., Gupta, R., Muneer, S., Singh, H.P., 2017, 467, 1186 (paper II)
\bibitem[\protect\citeauthoryear{Author}{2012}]{Author2012} 
Reichley, P.E., Downs, G.S., Morris, G.A., 1970, ApJL, 159, L35
\bibitem[\protect\citeauthoryear{Author}{2012}]{Author2012} 
Rollinde, E., Boiss\'{e}, P., Federman, S.R., et al., 2003, A\&A, 401, 215
\bibitem[\protect\citeauthoryear{Author}{2012}]{Author2012} 
Sembach, K.R., Danks, A.C., 1994, A\&A, 289, 589
\bibitem[\protect\citeauthoryear{Author}{2012}]{Author2012} 
Slavin, J.D., Nichols, J.S., Blair, W.P., 2004, ApJ, 606, 900
\bibitem[\protect\citeauthoryear{Author}{2012}]{Author2012} 
Smith, K.T., Fossey, S.J., Cordiner, M.A., et al., 2013, MNRAS, 429, 939
\bibitem[\protect\citeauthoryear{Author}{2012}]{Author2012} 
Sushch, I., Hnatyk, B., Neronov, A., 2011, A\&A, 525, A154
\bibitem[\protect\citeauthoryear{Author}{2012}]{Author2012} 
Tokovinin, A., Mason, B.D., Hartkopf, W.I., 2010, A.J, 139, 743
\bibitem[\protect\citeauthoryear{Author}{2012}]{Author2012} 
Wallerstein, G., Silk, J., 1971, ApJ, 170, 289
\bibitem[\protect\citeauthoryear{Author}{2012}]{Author2012} 
Wallerstein, G., Silk, J., Jenkins, E.B., 1980, ApJ, 240, 834
\end{thebibliography}
\end{document}